\documentclass{aa}
\usepackage{txfonts}
\usepackage{graphicx}
\usepackage{subcaption}
\usepackage{wasysym}
\usepackage{sidecap}
\usepackage[normalem]{ulem}
\sidecaptionvpos{figure}{t}

\usepackage{xcolor}

\usepackage[bottom]{footmisc}
\makeatletter
\renewcommand*{\@fnsymbol}[1]{\ensuremath{\star}}
\makeatother

\begin{document} 

%--------------------------------------------------------------------------
\title{Ocean signatures in the total flux and polarization spectra of Earth-like exoplanets\thanks{The calculated spectra of the total fluxes, polarized fluxes and degree of polarization will be made available through the CDS.}}

\author{V.J.H.\ Trees  \inst{\ref{inst1}\ref{inst2}}\and D.M.\ Stam \inst{\ref{inst3}}}
         
\institute{Royal Netherlands Meteorological Institute (KNMI), 
           Utrechtseweg 297, 3731 GA, De Bilt, The Netherlands 
           \label{inst1} \and 
           Faculty of Civil Engineering and Geosciences, 
           Delft University of Technology, Stevinweg 1, 2628 CN Delft, 
           The Netherlands \label{inst2} \and 
           Faculty of Aerospace Engineering, Delft University of Technology, Kluyverweg 1, 2629 HS Delft, The Netherlands 
           \label{inst3}}

\date{Received 21 March 2022 / Accepted 10 May 2022}

%--------------------------------------------------------------------------
% \abstract{}{}{}{}{} 
% 5 {} token are mandatory
 
\abstract
% context heading (optional)
{Numerical simulations of starlight that is reflected by Earth-like exoplanets predict habitability signatures that can be searched for with future telescopes. }
% aims heading (mandatory)
{We explore signatures of water oceans in the flux and polarization spectra of this reflected light.}
% methods heading (mandatory)
{With an adding-doubling algorithm, we compute the total flux $F$, polarized flux $Q$ and degree of polarization $P_\mathrm{s}$ of starlight 
reflected by dry and ocean model planets with Earth-like atmospheres and patchy clouds. The oceans consist of Fresnel reflecting surfaces 
with wind-ruffled waves, foam and wave shadows, above natural blue seawater. Our results are presented as functions of wavelength (from 300 to 2500~nm with 
1~nm resolution) and as functions of the planetary phase angle from 
90$^\circ$ to 170$^\circ$.}
% results heading (mandatory)
{The ocean glint increases $F$, $|Q|$ and $P_\mathrm{s}$ with increasing 
phase angle at non-absorbing wavelengths, and causes the spectra of 
$F$ and $|Q|$ for the various phase angles to intersect. In the near-infrared, $Q$\ is negative, i.e. the direction of polarization is perpendicular to the plane through the star, planet, and observer. In the $P_\mathrm{s}$-spectra, the glint leaves 
dips (instead of peaks) in gaseous absorption bands. All those signatures 
are missing in the spectra of dry planets.}
% conclusions heading (mandatory)
{The dips in $P_\mathrm{s}$, and the negative $Q$ in the near-infrared, can 
be searched for at a phase angle of $90^\circ$, where the planet-star
separation is largest. Those ocean signatures in polarized light do not 
suffer from false positive glint signals that could be due to clouds or 
reflecting dry surfaces. For heavily cloudy planets, ocean detection 
is possible when the glint is (partially) cloud-free. 
When modelling signals of planets with oceans, using horizontally 
inhomogeneous cloud covers is thus crucial. Observations spread
over time would increase the probability of catching a 
cloud-free glint and detecting an ocean.}


%--------------------------------------------------------------------------
\titlerunning{Ocean signatures in the spectra of Earth-like exoplanets} 

\maketitle

%%%%%%%%%%%%%%%%%%%%%%%%%%%%%%%%%%%%%%%%%%%%%%%%%%%%%%%%%%%%%%%%%%%%%%%%%%%
\section{Introduction}

The discovery of liquid water on the surface of an extrasolar planet 
would be a milestone in the search for life beyond our solar system, 
as liquid water is considered to be a requirement for life as we know it. 
Water oceans on the surfaces of rocky planets can be searched for in 
planetary systems' habitable zones, which are defined as the regions 
around stars where liquid surface water could be stable \citep{Kasting1993}. 
The actual capability of a planet to maintain liquid surface water is 
difficult to determine because that depends on, amongst others, the planet’s 
atmospheric composition, and its vertical and horizontal pressure and 
temperature profiles, surface albedo, mass, size and rotational period, 
and the activity of its host star \citep[see e.g.][]{Kopparapu2020}.

Water vapor has already been detected in the atmospheres of hot 
gaseous extrasolar planets \citep[e.g.,][]{Tinetti2007, Deming2013,Fraine2014,Kreidberg2014,Kreidberg2015,Line2016} and 
of super Earth K2-18b which resides in the habitable zone of its
parent star \citep{Tsiaras2019}. Those detections were done using 
spectroscopy of the transits of these planets and/or of their 
secondary eclipses (when the planet moves behind the star). 
Although future transit or eclipse spectroscopy measurements 
of water vapor abundances on rocky planets, such as planned with
the James Webb Space Telescope (JWST), could give hints 
about habitable surface conditions, the light that is being detected
with such measurements only traveled through the exoplanet's (upper) 
atmosphere, and actual observations of liquid water 
oceans are only possible using a direct detection of the starlight 
that is reflected by the planet. 

The signal of starlight that is reflected by an Earth-like exoplanet 
and that reaches a (near-future) dedicated direct detection telescope 
will not provide any spatial resolution across an exoplanet, because 
of the enormous distances between the telescope and these planets. 
All information of the illuminated and visible part of the planetary 
disk will thus be stored in one pixel. However, if the signal is 
measured at multiple wavelengths, the pixel contains information 
about the planet's color and e.g.\ absorbing gases. 
The signal will also vary with the planetary phase angle (i.e. the 
angle between the observer, the host star and the planet orbiting 
its host star), as the optical properties of the surface and 
atmosphere usually depend on the scattering angle, and in time
as the planet rotates around its axis.

Light from the star can be assumed to be unpolarized \citep{Kemp1987}, 
and gets polarized upon scattering by the planet's atmosphere and/or 
reflection by the surface. 
The state of polarization of the light thus enhances the contrast 
between the star and the exoplanet, but it also contains information 
about the optical properties of the planet's atmosphere and surface. 
The dependence of the degree of polarization on scattering angle is 
in general much more significant than that of the total flux 
\citep{Hansen1974}, making polarimetry a strong tool for exoplanet 
characterization \citep[see e.g.][]{2000ApJ...540..504S,Stametal04,Stam08}.
Indeed, also for detecting and characterizing oceans on exoplanets, 
the polarization of light provides an extra dimension. Fresnel reflection 
of direct starlight by an ocean results in a mirror reflection of the star 
in the water, which is called the glint, which broadens with increasing 
ocean surface roughness (i.e., wind speed). In the total reflected flux,
the glint is strongest at the largest phase angles 
\citep[see e.g.][]{Robinsonetal2010,Robinson2014}, where unfortunately
the separation between the planet and its star is relatively small 
and which is therefore not an optimal angle for resolving the 
planetary signal from the direct starlight. The degree of polarization, 
however, peaks at the Brewster angle, which is at larger and thus
easier accessible planet-star 
separations \citep[see][]{Zugger2010,TreesStam2019}.

Measuring the degree of polarization also has practical advantages, 
because instrumental measurement errors in the total and polarized fluxes 
may cancel out upon division. Also, as a relative measure, the degree of
polarization of an exoplanet is generally insensitive to extinction 
along the optical path between the exoplanet and the observer. 
That is, when taking measurements of an exoplanet in the night sky with 
a telescope on the Earth's surface, absorption of the light within 
the Earth's atmosphere affects the total and polarized fluxes equally, 
leaving the degree of polarization of the exoplanet unaffected.

%\textit{In this paper, we extend the spectra towards the near-infrared (NIR), from 1000 nm to 2500 nm, which is particularly interesting for studying potential habitable planets orbiting M-dwarfs \citep[see e.g. ][for the spectra of Proxima Centauri and Trappist-1]{LinKaltenegger2020}. ...The planet Proxima Centauri b has an angular separation of 37 milliarcsecond (mas) from its host star Proxima Centuari, and can in the future potentially be resolved from Proxima Centuari with the Extremely Large Telescope which has an inner working angle of 6 mas at visible wavelengths. ...\\
%\textit{Ryan and Robinson 2022:  Detecting Oceans on Exoplanets with Phase-dependent Spectral Principal Component Analysis}

Current ground-based and space-based telescopes are not capable of 
measuring the polarized light that is reflected by Earth-like exoplanets. 
Examples of instruments that are designed for measuring polarized 
light of exoplanets are EPOL \citep{Keller2010}, which is the imaging 
polarimeter of the Exoplanet Imaging Camera and Spectrograph (EPICS) 
for the European Southern Observatory's (ESO) 
Extremely Large Telescope (ELT) \citep{Kasper2010}, 
and POLLUX, a UV polarimeter that is envisioned for NASA’s LUVOIR (Large UV 
Optical Infrared Surveyor) space telescope concept 
\citep{Bouret2018}. The choices in the design of those and other next 
generation instruments, their telescopes, and observational strategies 
(i.e.\ integration times and temporal coverage) depend on  
exoplanetary signals from numerical models. 

Recently, numerical simulations of total flux spectra of Earth-like planets 
with rough ocean surfaces have been presented by \citet{RyanRobinson2022}, 
who suggested to retrieve the glint signal with a principal component analysis. 
Another suggested technique for ocean detection is to retrieve a surface 
type map from rotationally resolved phase curves of the total flux at a 
single or multiple wavelength(s) 
\citep{Cowanetal2009,OakleyandCash2009,Lustig-Yaeger2018}. 
None of the papers mentioned above studied the state of polarization
of the reflected light. 
Without the polarization signal, false positive glint signatures may 
appear if in the crescent phase, reflecting surfaces or clouds have
a relatively strong contribution.
For example, \cite{Cowanfalseglint2012} proposed that the larger coverage 
fraction of poles on the planetary disk in the crescent phase could mimic 
the glint. 
%Furthermore, it should be noted neglecting polarization in 
%the numerical radiative transfer simulations of scattering atmospheres 
%results in %errors in the modeled reflected total flux 
%\citep[see][]{StamHovenier2005}. 

Numerical simulations of starlight that is reflected by exoplanets 
with oceans that take polarization into account have first been 
presented by \citet{Stam08}. \citet{Stam08} showed the spectra of the 
total flux and degree of polarization of Earth-like model exoplanets
with oceans, vegetated surfaces and water clouds. The ocean was
described by a flat Fresnel reflecting surface on top of a black water
body, without waves, wave shadows, foam, and reflecting water.
Due to the flat ocean surface and the uni-directional incident 
sunlight, the glint was infinitely narrow and its signature was 
lost in the numerical 
integration over the planetary disk. Simulations of planets with 
rough black ocean surfaces as functions of planetary phase angle 
showed that the glint in fact gives a strong peak in the degree of 
polarization \citep[see][who did not include the scattering by 
atmospheric gas]{Williams2008}. 
The glint shifts the peak of the degree of polarization at a 
phase angle of 90$^\circ$, due to Rayleigh scattering by the gaseous 
molecules, to the Brewster angle at a larger phase angle 
\citep[see][who included scattering by gas in the atmosphere]{Zugger2010}. 
\citet{Zugger2011IR} proposed that for ocean detection absorption-free 
spectral windows in the infrared could be more convenient than visible 
wavelengths as there the scattering optical thickness of the atmosphere 
is smaller which improves the visibility of the glint. 

The spectra of \citet{Stam08} and the phase curves of 
\citet{Zugger2011IR,Zugger2010} are for horizontally homogeneous planets:
the local atmosphere-surface systems do not vary in the latitudinal and 
longitudinal directions. The spectra of partly cloudy planets were 
mimicked by weighted sums of completely cloudy and completely 
cloud-free planets, 
which results in fading of the ocean signatures. 
\citet{Zugger2010,Zugger2011IR} therefore concluded that the detectability of 
an exo-ocean is strongly limited by the presence of clouds. 
\citet{TreesStam2019}, however, simulated the total and polarized fluxes and 
the degree of polarization of light reflected by planets with rough ocean 
surfaces and with horizontally inhomogeneous clouds, and found that the 
glint appearing and hiding behind patchy clouds results in a variability 
of the phase curves. Moreover, \citet{TreesStam2019} combined the phase 
curves at various wavelengths between 350 and 865 nm, to show that the 
glint increasingly reddens the degree of polarization and polarized flux with 
increasing phase angle. This color change in polarized light does not happen 
for dry planets and can thus identify an exo-ocean, also in the mean signal 
of heavily cloudy ocean planets with random patchy cloud patterns, and 
when surface pressures are higher than on Earth.

Modeled signals of Earth-like exoplanets can be validated with measurements of 
reflected sunlight by the Earth as a whole  
obtained from a sufficiently large distance. 
As far as we know, no such measurements of polarized light have yet been 
done, but 
an instrument that is in development to do so, is the Lunar 
Observatory of Unresolved Polarimetry of Earth 
\citep[LOUPE;][]{2021RSPTA.37990577K,Hoeijmakers2016,Karalidietal2012}.
Sunlight that is reflected by Earth can also be measured with an Earth-based telescope observing the night-side of the lunar disk in the sky, often referred to as Earth-shine measurements. 
Polarimetric Earth-shine measurements have been presented and analyzed by 
\citet{Bazzonetal2013}, \citet{MilesPaez2014}, 
\citet{Sterzik2012,Sterzik2019,Sterzik2020} and 
\citet{Takahashi2013,Takahashi2021}. 
The visible broadband measurements by \citet{Sterzik2019} indeed show 
a reddening of the degree of polarization with increasing 
phase angle, and \citet{Takahashi2021} found a shift of the peak degree of 
polarization towards phase angles larger than 90$^\circ$ in the near-infrared,
which could be explained by the glint. It should be noted that, 
in Earth-shine measurements, the unknown, and depolarizing reflectance 
properties\footnote{The Moon's reflectance properties of incident sunlight can
rather straightforwardly be measured, but its reflectance properties
of incident polarized light are unknown.} of the lunar surface remain a 
considerable source of uncertainty.

Attempts to fit model simulation results to measured Earth-shine spectra have 
been made by \cite{Emdeetal2017}. They used a model planet with a rough black 
ocean surface and patchy clouds at four wavelengths (469, 555, 645 and 858.5 nm), 
and concluded that the glint could explain the enhancement of the degree of 
polarization in the red part of the measured spectrum by \citet{Sterzik2012}.
\cite{Emdeetal2017} also analyzed the influence of aerosols, water clouds and 
ice clouds on the degree of polarization spectra, but without an ocean below. 
Note that all spectra (i.e. at a 1~nm wavelength resolution) presented by 
\cite{Emdeetal2017} were of (weighted sums) of horizontally homogeneous planets.
\cite{Emdeetal2017} did not show spectra of patchy cloudy planets with oceans, 
and did not study the phase angle variation of the spectropolarimetric glint signature.

In this paper, we present the computed spectra of the total and polarized fluxes and 
the degree of polarization of starlight that is reflected by dry and ocean planets, 
having Earth-like atmospheres with clouds consisting of spherical, liquid water 
droplets and horizontally inhomogeneous cloud patterns. Absorption of light in the 
planet's atmosphere by gaseous H$_2$O, O$_2$, O$_3$, CO$_2$ and CH$_4$ is taken 
into account. The wind over our oceans generates surface waves which broaden 
the ocean glint, wave shadows and sea foam, and the seawater has a natural blue 
color. The computed spectra range from the ultraviolet (300~nm) to the infrared 
(2500~nm) at a 1~nm wavelength resolution, and we present the results for 
planetary phase angles between 90$^\circ$ and 170$^\circ$.

This paper is structured as follows. In Sect.~\ref{sect_numerical}, we state our 
definitions of the total and polarized fluxes and the degree of polarization of 
light. We explain our numerical method to compute those quantities for 
starlight reflected by our model planets, and we describe our atmosphere-ocean 
planet models. In Sect.~\ref{sect_results}, we present the computed spectra of 
the reflected total and polarized fluxes and the degree of polarization 
for dry and ocean planets. In Sect.~\ref{sec:h2o}, we discuss the influence 
of various atmospheric parameters on the ocean signature that we find in gaseous 
absorption bands in the degree of polarization. In Sect.~\ref{sect_comparison}, 
we compare our ocean signatures with earlier published 
results. Finally, in Sect.~\ref{sect_conclusion}, we summarize the 
spectropolarimetric ocean signatures and state our conclusions.

%%%%%%%%%%%%%%%%%%%%%%%%%%%%%%%%%%%%%%%%%%%%%%%%%%%%%%%%%%%%%%%%%%%%%%%%%%%%%%%
\section{Numerical method}
\label{sect_numerical}

%------------------------------------------------------------------------------
\subsection{Definitions of fluxes and polarization}
\label{sect_definitions}

The radiance and polarization state of a quasi-monochromatic light beam 
can be described by the following (column) vector ${\bf I}$ 
\begin{equation}
   \mathbf{I} = \left[I, Q, U, V\right],
\label{eq_stokes1}
\end{equation}
where $I$ is the total radiance, $Q$ and $U$ the linearly polarized 
radiances, and $V$ the circularly polarized radiance
\citep[see, e.g.,][]{Hansen1974,2004Hovenier}. 
The units of $I$, $Q$, $U$ and $V$ are W m$^{-2}$ m$^{-1}$ sr$^{-1}$, 
where m$^{-1}$ indicates the dependence on wavelength $\lambda$ and 
sr$^{-1}$ are the units of solid angle $\Omega$. 
In this paper, we also use the flux (column) vector $\pi \bf{F}$, 
described by 
\begin{equation}
   \pi\mathbf{F} = \pi\left[F, Q, U, V\right],
\label{eq_stokes2}
\end{equation}
for which the units are W m$^{-2}$ m$^{-1}$.

Elements $Q$ and $U$ are defined with respect to a reference 
plane. Because the distance between the observer and the exoplanet 
is much larger than the radius of the exoplanet, our reference 
plane is the planetary scattering plane, i.e., the plane through the 
centers of the planet, the host star and the observer. 
The linearly polarized fluxes, $\pi Q$ and $\pi U$, can then in 
principle be obtained from the fluxes measured through a linear polarization 
filter perpendicular to the propagation direction of the light:
\begin{equation}
    \pi Q = \pi F_{0^\circ} - \pi F_{90^\circ},
    \label{eq:piQ}
\end{equation}
\begin{equation}
    \pi U = \pi F_{45^\circ} - \pi F_{135^\circ},
    \label{eq:piU}    
\end{equation}
where the subscripts indicate the rotation angles of the filter's 
optical axis with respect to the planetary scattering plane, 
measured in the anti-clockwise direction when looking into the 
propagation direction of the light 
\citep[see][]{Hansen1974,hovenier2004transfer}. 
The total flux, $\pi F$, is measured according to
\begin{equation}
\pi F = \pi F_{0^\circ} + \pi F_{90^\circ} = 
        \pi F_{45^\circ} + \pi F_{135^\circ}
\label{eq:piF}
\end{equation}
Elements $Q$ and $U$ can be converted to $Q$ and $U$ w.r.t. another 
reference plane, such as the optical plane of the polarimeter, 
using a rotation matrix \citep[see][]{HoveniervanderMee83}. 

We define the degree of polarization $P$, which is independent of 
the reference plane, as 
\begin{equation}
    P = \frac{\sqrt{Q^2 + U^2 + V^2}}{F}.
    \label{eq:dop}
\end{equation}
The starlight that is incident on a planet is assumed to be unpolarized
($P = 0$). This assumption is based on the very small
degree of polarization of the disk-integrated sunlight \citep[][]{Kemp1987} 
(on the order of 10$^{-6}$) and of FGK-stars \citep[][]{Cotton2017} 
(23.0 $\pm$ 2.2 ppm for active stars and 2.9 $\pm$ 1.9 ppm for inactive stars), 
and on the small degree of polarization ($\sim$ $10^{-6}$) expected 
from stellar spots and flares 
breaking the symmetry of the stellar disk 
\citep{Berdyugina2011,kostogryz2015polarization}. 
We thus write the stellar flux vector that is incident on the exoplanet 
as $\pi \mathbf{F_0} = \pi F_0 \mathbf{1}$ where $\mathbf{1}$ is the 
unit column vector and $\pi F_0$ is the incident stellar flux measured 
perpendicular to the direction of propagation. 

With unpolarized incident starlight and if the planet is mirror-symmetric 
with respect to the reference plane, the integrated $\pi U$  
and $\pi V$ over the planetary disk will equal zero. 
In this paper, all cloud-free and fully cloudy planets are mirror-symmetric
while the planets with patchy clouds are near mirror-symmetric 
with respect to the planetary scattering plane resulting in 
$\pi U$ and $\pi V$ being a factor 10$^3$ and 10$^5$, respectively, 
smaller than $\pi F$. Therefore, we may use an alternative definition of the 
degree of polarization, namely
\begin{equation}
    P_{\rm s} = -\frac{Q}{F} = -\frac{F_{0^\circ} - F_{90^\circ}}{F_{0^\circ} + F_{90^\circ}}
    \label{eq:dops}
\end{equation}
For $P_{\rm s} > 0$ ($P_{\rm s} < 0$), the light is polarized perpendicular 
(parallel) to the reference plane.

%%%%%%%%%%%%%%%%%%%%%%%%%%%%%%%%%%%%%%%%%%%%%%%%%%%%%%%%%%%%%%%%%%%%%%%%%%
% FIGURE 1
%%%%%%%%%%%%%%%%%%%%%%%%%%%%%%%%%%%%%%%%%%%%%%%%%%%%%%%%%%%%%%%%%%%%%%%%%%
\begin{figure}[t]
\centering
\includegraphics[width=0.8\linewidth]{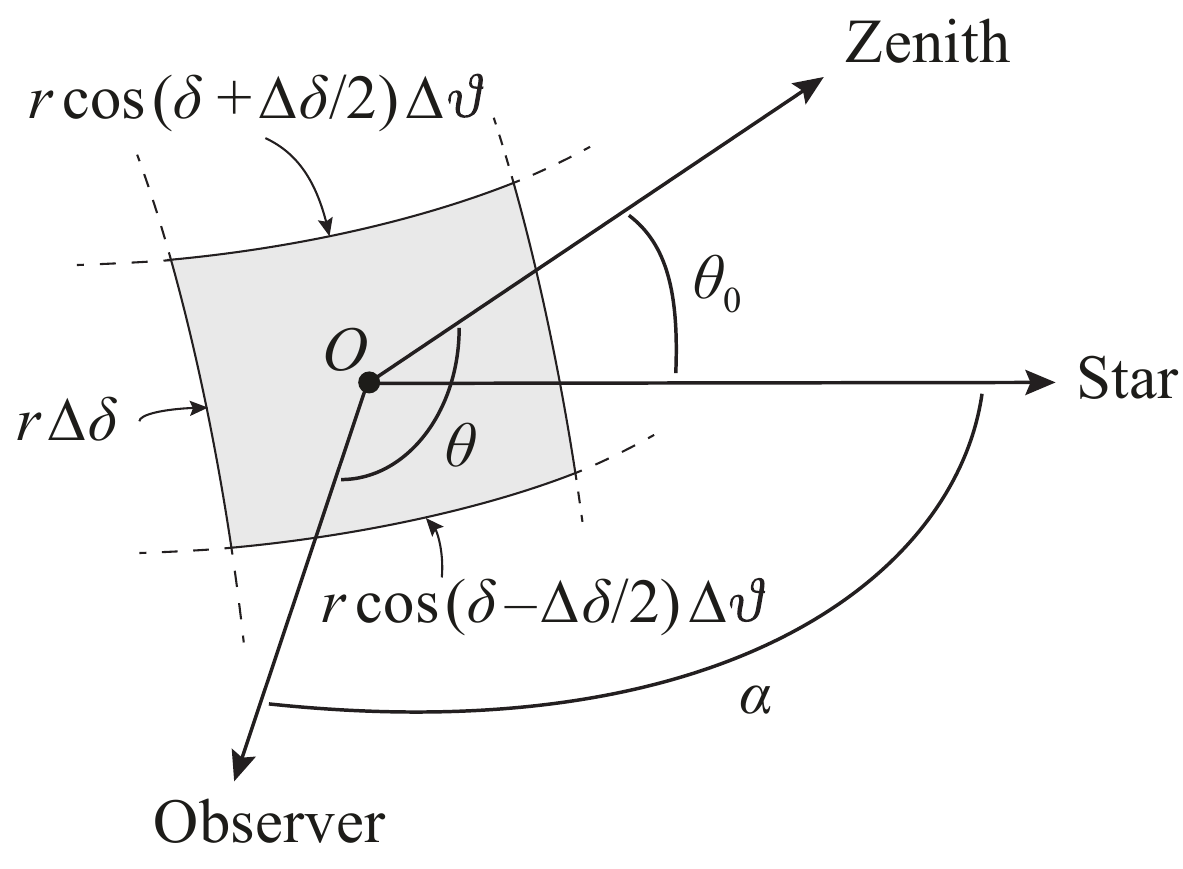}
\caption{Sketch of the geometry of the reflection at location $O$ 
	on the exoplanet, with latitude $\delta$ and longitude $\vartheta$.
	Quantities $r$, $\theta$, $\theta_0$ and $\alpha$ are the planetary 
	radius, viewing zenith angle, stellar zenith angle, and planetary 
	phase angle, respectively. The grey area element represents the 
	area on the planet subtended by latitude element 
	$\Delta \delta$ and longitude element $\Delta \vartheta$.}
\label{fig:element}
\end{figure}
%%%%%%%%%%%%%%%%%%%%%%%%%%%%%%%%%%%%%%%%%%%%%%%%%%%%%%%%%%%%%%%%%%%%%%%%%%

%-----------------------------------------------------------------------------
\subsection{Computing the reflected starlight}
\label{sect_calculating_reflected_starlight}

Starlight with wavelength $\lambda$ 
that is reflected by an orbiting planet arrives at 
an observer at a large distance from the planet-star system. 
Consider a point $O$ on the planet's surface (see Fig.~\ref{fig:element}) 
at latitude $\delta$ and longitude $\vartheta$ which are defined with 
respect to the vector from the planet's center towards the observer. 
If stellar flux $\pi F_0$ is incident at $O$, 
the reflected intensity vector towards the observer can be written as 
\citep[see][]{Stam2006,Hansen1974}:
\begin{equation}
   {\bf I}(\lambda,\mu,\mu_0,\phi-\phi_0)= 
           \mu_0 \hspace{0.1cm}    {\bf L}(\beta) \hspace*{0.05cm}
           {\bf R}(\lambda,\mu,\mu_0,\phi-\phi_0) 
           {\bf 1} \hspace*{0.05cm} F_0(\lambda),
\label{eq:reflected_vector}
\end{equation}
where $\mu_0 = \cos \theta_0$ with $0^\circ \leq \theta_0 \leq 90^\circ$ 
the local stellar zenith angle, $\mu = \cos \theta$ with 
$0^\circ \leq \theta \leq 90^\circ$ the local viewing zenith angle, 
and $\phi-\phi_0$ the azimuth difference angle between the incident 
and reflected beam ($\phi-\phi_0=0^\circ$ when the propagation
directions of the incident and the reflected beams are in the same
vertical half plane).
Furthermore, $\bf{R}$ is the $4 \times 4$ reflection matrix of the 
atmosphere-surface system at $O$. 
Since the optical properties of the atmosphere and surface 
are wavelength dependent, $\bf{R}$ also depends on wavelength. 
The $4 \times 4$ rotation matrix $\bf{L}(\beta)$ converts $Q$ and $U$
that are defined with respect to the local meridian plane (through the 
local zenith and the direction to the observer) directly 
after the reflection at $O$, to $Q$ and $U$ defined with respect to 
the planetary scattering plane \citep[see][]{HoveniervanderMee83}. 
Angle $\beta$ is the rotation angle between the two reference planes, 
which is positive in the anti-clockwise direction when looking in the 
direction of light propagation. 
Quantities $\mu$, $\mu_0$, $\phi-\phi_0$ and $\beta$ depend on location 
($\delta$,$\vartheta$). 
Quantities $\mu_0$ and $\phi-\phi_0$ also depend on the phase angle 
$\alpha$, which is the angle between the star and the observer 
measured from the planet's center (at $\alpha = 0^\circ$, the planet 
is behind the star as seen from the observer, and at $\alpha = 180^\circ$, 
the planet is precisely in front of the star). 
We set the orbital plane equal to the planetary scattering plane 
(i.e.\, the orbit is seen 'edge-on'), and the tilt of the planet's 
rotation axis with respect to the orbital plane's normal is 
set equal to zero.  

The reflected flux vector by the planet as a whole, arriving at the 
observer at a distance $d$ from the planet can be obtained for each 
$\lambda$ and $\alpha$ by integrating $\bf{I}$ over the solid angle 
subtended by the planetary disk 
\citep[see e.g.][]{Stam2006}:
\begin{equation}
   \pi {\bf F}
   (\lambda,\alpha) =
   \frac{1}{d^2}
   \int_{\rightmoon} \mu \hspace*{0.05cm}
   {\bf I}(\lambda,\mu,\mu_{0},\phi-\phi_{0}) \hspace*{0.05cm}
   dO.
   \label{eq:reflected_flux}
\end{equation}
We normalize $\pi {\bf F}$, indicated by $\overline{\textbf{F}}$, 
such that at $\alpha = 0^\circ$,
its first element $\overline{F}$ equals the planet's geometric albedo
as follows
\begin{equation}
 \overline{\textbf{F}}(\lambda,\alpha) = 
     \frac{\pi {\bf F}(\lambda,\alpha)}{\pi F_0 (\lambda)} \frac{d^2}{r^2}.
 \label{eq:normalization}
\end{equation}
Here, $r$ is the radius of the planet. 
We compute $\overline{\textbf{F}}$ independent of $\pi F_0$, $r$ and $d$. 
By specifying those parameters for a given exoplanet-star system, 
the values of $\overline{\textbf{F}}$ presented in this paper can 
straightforwardly be converted to $\pi {\bf F}$ of the given system 
using Eq.~\ref{eq:normalization}. 
The degree of polarization $P$ (or $P_{\rm s}$) is 
independent of $\pi F_0$, $r$ and $d$.

%%%%%%%%%%%%%%%%%%%%%%%%%%%%%%%%%%%%%%%%%%%%%%%%%%%%%%%%%%%%%%%%%%%%%%%%%%
% FIGURE 2
%%%%%%%%%%%%%%%%%%%%%%%%%%%%%%%%%%%%%%%%%%%%%%%%%%%%%%%%%%%%%%%%%%%%%%%%%%
\begin{figure}
\centering
\includegraphics[width=0.99\linewidth]{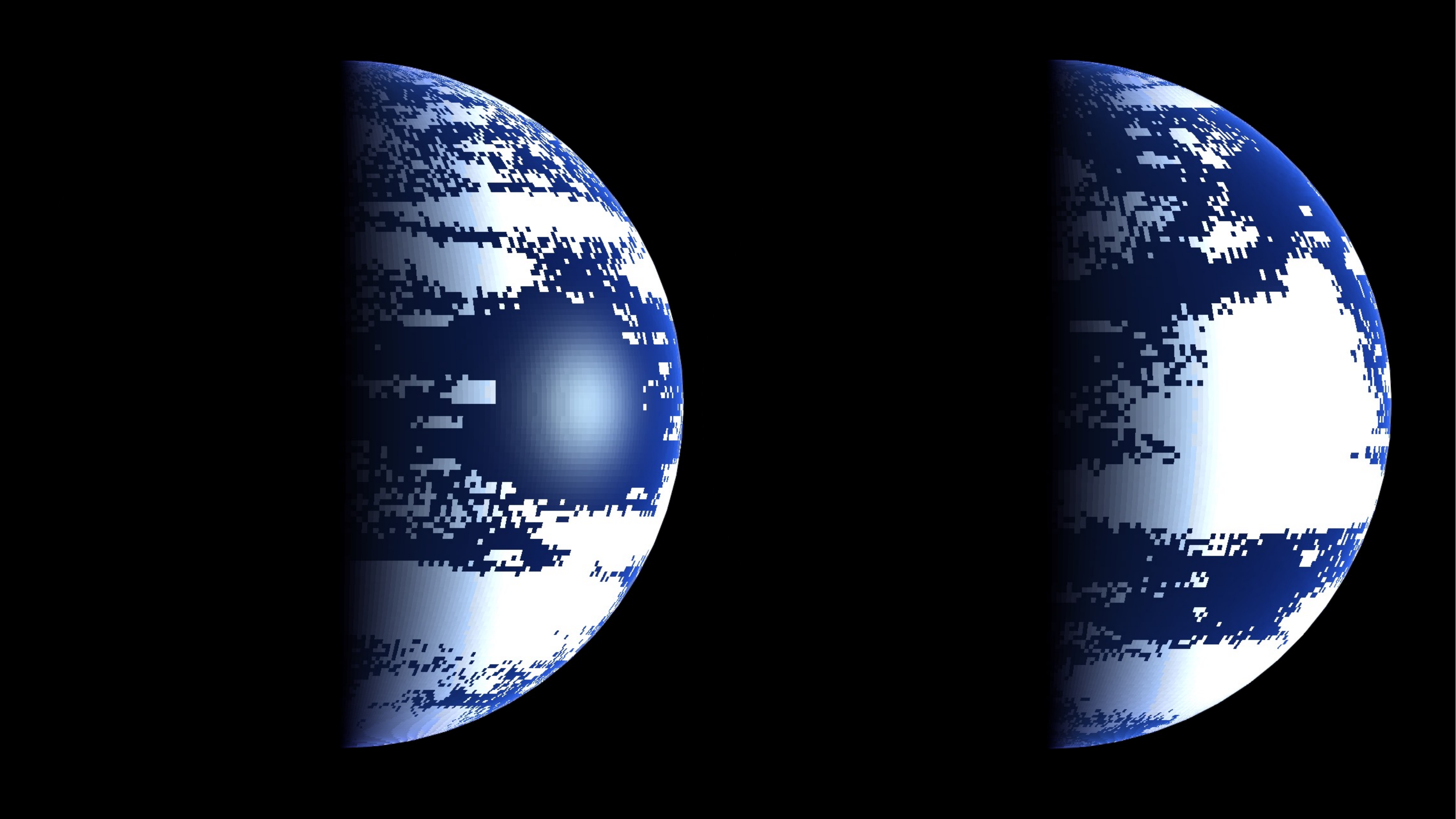}
\caption{Disk-resolved computed RGB images using weighted 
	     additive color mixing of the pixel intensity 
	     $\overline{F}_{ij}\pi r^2/(\mu_{ij}\Delta O_{ij})$ at 
	     $\lambda=450$, 550, and 650~nm, at $\alpha=90^\circ$ of 
	     ocean planets with patchy clouds and a cloud-free glint 
	     (left image) and a cloud-covered glint (right image). 
	     For both images, the cloud fraction $f_{\rm c}$ equals 0.5, 
	     and the surface wind speed $v$ is 7~m/s.}
\label{fig:disks}
\end{figure}
%%%%%%%%%%%%%%%%%%%%%%%%%%%%%%%%%%%%%%%%%%%%%%%%%%%%%%%%%%%%%%%%%%%%%%%%%%

To evaluate Eq.~\ref{eq:reflected_flux}, we divide the planet's surface 
into elements $\Delta O_{ij}$ centered at points $O_{ij}$ located at
$(\delta_i,\vartheta_j)$, with angular dimensions $\Delta \delta$ and 
$\Delta \vartheta$ such that
\begin{align}
    \Delta O_{ij} & = \int_{\vartheta_j-\Delta \vartheta /2}^{\vartheta_j+\Delta \vartheta / 2} \int_{\delta_i-\Delta 
    \delta / 2}^{\delta_i+\Delta \delta / 2} r^2 \cos\delta 
    \hspace*{0.05cm}d\delta \hspace*{0.05cm} d\vartheta \notag
       \\
      & = 2 r^2 \Delta \vartheta \cos \delta_{i} \sin(\Delta\delta/2)
\label{eq:Oi}
\end{align}
Combining Eqs.~\ref{eq:reflected_vector}, \ref{eq:reflected_flux}, and
\ref{eq:normalization}, and replacing the integral in 
Eq.~\ref{eq:reflected_flux} by a summation, we arrive at
\begin{equation}
\overline{\textbf{F}}(\lambda,\alpha) \approx
 \sum_{j=1}^{M}\sum_{i=1}^{N} \mu_{0ij} \hspace*{0.1cm}
   {\bf L}(\beta_{ij})
   {\bf R}_{ij}(\lambda,\mu_{ij},\mu_{0ij},\phi_{ij}-\phi_{0ij}) {\bf 1} \hspace*{0.1cm}
   \frac{\mu_{ij}\Delta O_{ij}}{\pi r^2},
\label{eq:summation}
\end{equation}
where $N$ and $M$ are the number of latitudes and longitudes, 
respectively. At locations that are out of view ($\theta_{ij} > 90^\circ$) 
or not illuminated ($\theta_{0ij} > 90^\circ$), ${\bf R}_{ij}=0$.
The projections of $\Delta O_{ij}$ on a plane perpendicular to the 
observer's viewing direction are approximately $\mu_{ij}\Delta O_{ij}$. 
Hence, the integration error at $\alpha = 0^\circ$ can be expressed 
by the deviation of 
$\sum_{j=1}^{M}\sum_{i=1}^{N} \mu_{ij}\Delta O_{ij}/(\pi r^2)$ from $1$. We use $\Delta \delta = \Delta \vartheta = 1^\circ$, resulting in an
integration error at $\alpha = 0^\circ$ of about $2\cdot10^{-15}$.

%%%%%%%%%%%%%%%%%%%%%%%%%%%%%%%%%%%%%%%%%%%%%%%%%%%%%%%%%%%%%%%%%%%%%%%%%%
% FIGURE 3
%%%%%%%%%%%%%%%%%%%%%%%%%%%%%%%%%%%%%%%%%%%%%%%%%%%%%%%%%%%%%%%%%%%%%%%%%%
\begin{figure*}
\centering
\includegraphics[width=0.75\linewidth]{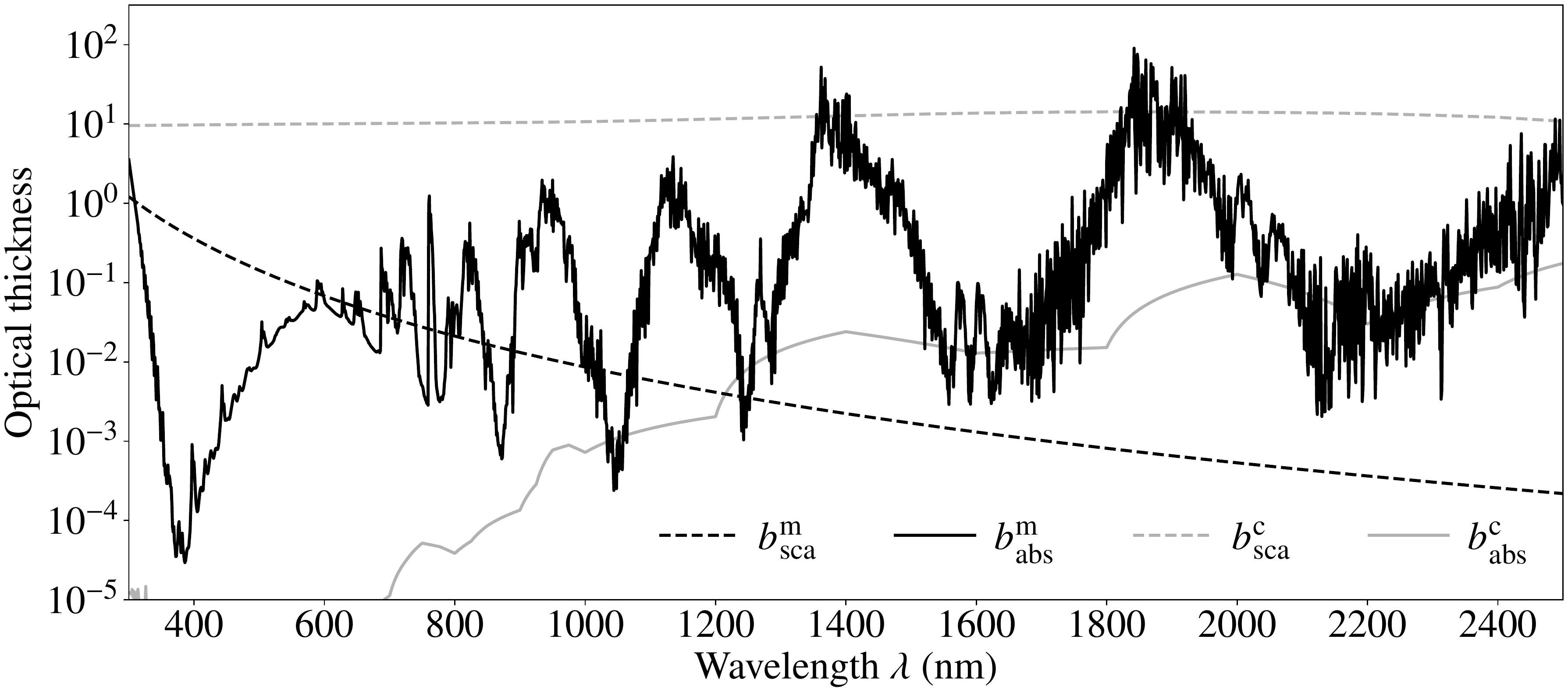}
\caption{Optical thickness of the model atmosphere for scattering by 
         gas $b_\mathrm{sca}^\mathrm{m}$ (black dashed line), 
         absorption by gas $b_\mathrm{abs}^\mathrm{m}$ (black solid line,
         see Fig.~\ref{fig:opticalthicknessgases} for the values per
         gas species),
         scattering by water cloud droplets $b^\mathrm{c}_\mathrm{sca}$ 
         (grey dashed line) and absorption by water cloud droplets
         $b^\mathrm{c}_\mathrm{abs}$ (grey solid line).}
\label{fig:opticalthickness}
\end{figure*}
%%%%%%%%%%%%%%%%%%%%%%%%%%%%%%%%%%%%%%%%%%%%%%%%%%%%%%%%%%%%%%%%%%%%%%%%%%

%%%%%%%%%%%%%%%%%%%%%%%%%%%%%%%%%%%%%%%%%%%%%%%%%%%%%%%%%%%%%%%%%%%%%%%%%%
% FIGURE 4
%%%%%%%%%%%%%%%%%%%%%%%%%%%%%%%%%%%%%%%%%%%%%%%%%%%%%%%%%%%%%%%%%%%%%%%%%%
\begin{figure*}
\centering
\includegraphics[width=0.75\linewidth]{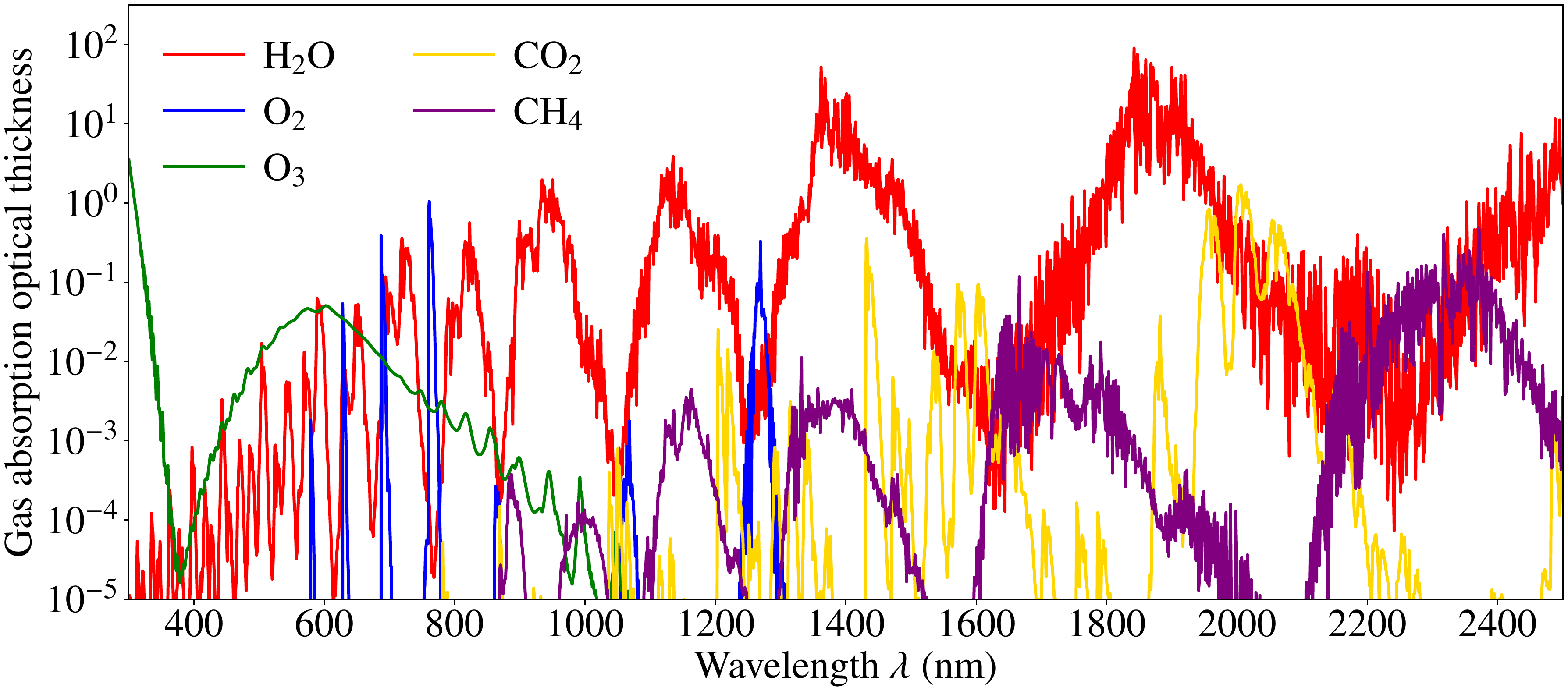}
\caption{Absorption optical thicknesses of the various gases in our
         model atmosphere, 
         computed by taking the natural logarithm of the 1~nm bin 
         averaged direct line-by-line transmittance of each 
         absorbing gas \citep[see e.g.\ Eq.~1 of][]{Irwinetal2008}. 
         In our computations, we do not use these values directly,
         instead we use 10~$k$-distribution 
         Gauss points in each 1~nm wide wavelength bin (see the text 
         for more details).}
\label{fig:opticalthicknessgases}
\end{figure*}
%%%%%%%%%%%%%%%%%%%%%%%%%%%%%%%%%%%%%%%%%%%%%%%%%%%%%%%%%%%%%%%%%%%%%%%%%%

The computation of $\overline{\textbf{F}}$ is performed at wavelengths 
$\lambda$ from 300 nm to 2500 nm with a step size of 1 nm. To compute
$\overline{\textbf{F}}$ for a given phase angle $\alpha$ and locations
$(\delta_i,\vartheta_j)$, we first compute $\Delta O_{ij}$, $\mu_{ij}$, 
$\mu_{0ij}$, $\phi_{ij}-\phi_{0ij}$ and $\beta_{ij}$, and we select a local 
atmosphere-surface model for each location. 
We pre-compute $\mathbf{R}$ of the local atmosphere-surface models 
(see Sect.~\ref{sect_atmosphere_surface_models}) for each $\lambda$ on 
a grid of 100 $\mu$-points, 100-$\mu_0$ points and a Fourier-series 
representation for the azimuth direction 
\citep[see][for details about the Fourier series expansion]{deHaan87,Rossietal2018}. 
We interpolate between those grids to obtain $\mathbf{R}_{ij}$ at 
$(\delta_i,\vartheta_j)$ and $\lambda$ for locally reflected light in the 
direction of the observer, and evaluate Eq.~\ref{eq:summation}. 

For our horizontally homogeneous planets the pre-computed $\mathbf{R}$ can be
used for each location on the planet. For our horizontally inhomogeneous
planets,
i.e. that have cloudy and cloud-free locations, one $\mathbf{R}$ is 
pre-computed 
for the cloudy areas and another one for the cloud-free areas. 
We generate the patchy cloud locations randomly on the planet for each 
$\alpha$ and in a zonally oriented pattern. 
Figure~\ref{fig:disks} shows two RGB images of our ocean planet models with
horizontally inhomogeneous cloud patterns at $\alpha = 90^\circ$.  
The cloud fraction $f_\text{c}$ on the visible and illuminated part of the 
planetary disk is 0.5 in both images, but in the image on the left 
the ocean glint is cloud-free while it is fully covered by clouds on the right.

%------------------------------------------------------------------------------
\subsection{The planetary surface--atmosphere models}
\label{sect_atmosphere_surface_models}

The local reflection matrix $\mathbf{R}$ (Eq.~\ref{eq:summation}) depends 
on the optical properties of the local atmosphere and surface. 
It is computed with a multi-stream adding-doubling radiative transfer 
algorithm taking into account polarization of light for all orders of 
scattering \citep[see][]{deHaan87, Stam08},
extended with reflection by and transmission through a rough ocean surface 
and scattering in the water body \citep[see][]{TreesStam2019}. 
Below, we describe our atmosphere and surface models. 

%------------------------------------------------------------------------------
\subsubsection{The atmosphere}
\label{sect_atmosphere_model}

The atmospheres of our local atmosphere-surface models consist of a stack 
of 16 horizontally homogeneous layers. For each layer, the optical 
thickness $b$, the single-scattering albedo $a$ and the scattering matrix 
$\boldsymbol{F}_\mathrm{sca}$ \citep[see][]{hovenier2004transfer} of the 
mixture of molecules and cloud particles need to be specified to perform 
the radiative transfer calculations. We refer to \citet{Stam08} and the 
references therein for the relevant equations. We use 
cloud-free atmospheres, in which light is scattered and absorbed 
only by gaseous molecules, and cloudy atmospheres, that also have 
cloud particles that scatter and absorb light. 

In our model atmospheres, gas molecules scatter as anisotropic Rayleigh 
scatterers \citep[see][]{Hansen1974}, with the wavelength-dependent 
depolarization factor of \citet{Bates1984} between 300 and 1000~nm which we
extrapolate to 2500~nm, the refractive index of dry air varies with 
wavelength \citep{PeckandReader1972} 
and the mean molecular mass is 29~g~mole$^{-1}$.
The acceleration of gravity is 9.81~m~s$^{-2}$.
The molecular density in a layer depends on the pressure difference 
across each layer. We use the pressure-temperature profile 
throughout the 16 layers as specified in Table~1 of \citet{Stam08}, 
which is based on the mid-latitude summer profile of \citet{1972McClatchey}.

Our mix of atmospheric gases is Earth-like. It consists of ozone
(O$_3$), water vapor (H$_2$O), oxygen (O$_2$), methane (CH$_4$), 
and carbon dioxide (CO$_2$). The volume mixing ratios (VMRs) of O$_3$ 
and H$_2$O in each layer are taken from Table~1 of \citet{Stam08}. 
The VMRs of O$_2$, CH$_4$ and CO$_2$ are assumed constant with 
height: 0.21, 1.7~$\cdot 10^{-6}$ and 410 $\cdot 10^{-6}$, respectively. 
The molecular absorption cross sections of O$_3$ are from 
\cite{Gorshelev2014} and \cite{Serdyuchenko2014}, and depend on the 
layer's temperature. The molecular absorption cross sections of O$_2$,
H$_2$O, CH$_4$, and CO$_2$ are from the HITRAN database \citep{HITRAN2016} 
with the HAPI package \citep{HAPIKOCHANOV201615}, and depend on the 
pressure and temperature of the layer.

Figures~\ref{fig:opticalthickness} and~\ref{fig:opticalthicknessgases}
show the optical thickness of the
model atmosphere for molecular scattering ($b_\text{sca}^\text{m}$) and
molecular absorption ($b_\text{abs}^\text{m}$) ranging from 300~nm to 
2500~nm at a 1~nm wavelength resolution. The value of 
$b_\text{sca}^\text{m}$ decreases with increasing wavelength according 
to the $\lambda^{-4}$-dependence of the Rayleigh scattering cross section. 
The value of $b_\text{abs}^\text{m}$ varies strongly with $\lambda$ in 
Fig.~\ref{fig:opticalthickness}, but also within the 1-nm wide bins. 
Therefore, we do not use $b_\text{abs}^\text{m}$ directly 
but convert the molecular absorption cross sections of the gas mixture 
to so-called $k$-distributions \citep{LacisOinas1991,2000JQSRT..64..131S},
using 10~Gaussian quadrature points per wavelength bin. 
The absorbing molecules responsible for enhanced values of 
$b_\text{abs}^\text{m}$ in Fig.~\ref{fig:opticalthickness} can be 
identified using Fig.~\ref{fig:opticalthicknessgases}. It should be noted 
that while the concentrations of CH$_4$, O$_2$ and CO$_2$ are assumed 
to be altitude independent, H$_2$O mainly resides and thus absorbs in 
the bottom layers 
of our model atmosphere while O$_3$ mainly resides and thus
absorbs in the top layers 
\citep[see Table 1 of][]{Stam08}.

In a cloudy atmosphere, the clouds also contribute to the scattering 
and absorption optical thicknesses, the single-scattering albedo, and the 
scattering matrix \citep[see][]{Stam08}. For consistence with 
\cite{Stam08}, we assume spherical homogeneous water cloud droplets with 
a standard size distribution \citep{Hansen1974}, having an effective 
radius of 2.0~$\mu$m and an effective variance of 0.1. The cloud base 
altitude is 2~km ($p$= 802~hPa) and the cloud top altitude is 4~km 
($p$= 628~hPa). 
The cloud droplet single scattering albedo, scattering matrix and 
wavelength dependence of the droplet extinction cross section are 
computed using Mie-theory \citep{de1984expansion}. The real and imaginary 
parts of the refractive index of the water droplets vary with wavelength 
between 300 and 2500~nm from $1.3490 + (1.60\cdot10^{-8})i$ to 
$1.2605 + (2.06\cdot10^{-3})i$ \citep{hale1973optical}. 
The cloud optical thickness $b^\text{c}$ is 10 at $\lambda=550$~nm, 
and varies with wavelength between 300 and 2500~nm from 9.57 to 11.01 
(and peaks with 14.35 at 1887~nm) according to the wavelength 
dependence of the droplets' extinction cross--section. 
The cloud droplets absorb radiation in the infrared too, 
although their scattering contribution is much larger at all 
wavelengths as can be seen in Figs.~\ref{fig:opticalthickness}
and~\ref{fig:opticalthicknessgases}.

%%%%%%%%%%%%%%%%%%%%%%%%%%%%%%%%%%%%%%%%%%%%%%%%%%%%%%%%%%%%%%%%%%%%%%%%%%
% FIGURE 5
%%%%%%%%%%%%%%%%%%%%%%%%%%%%%%%%%%%%%%%%%%%%%%%%%%%%%%%%%%%%%%%%%%%%%%%%%%
\begin{figure}
\centering
\includegraphics[width=0.9\linewidth]{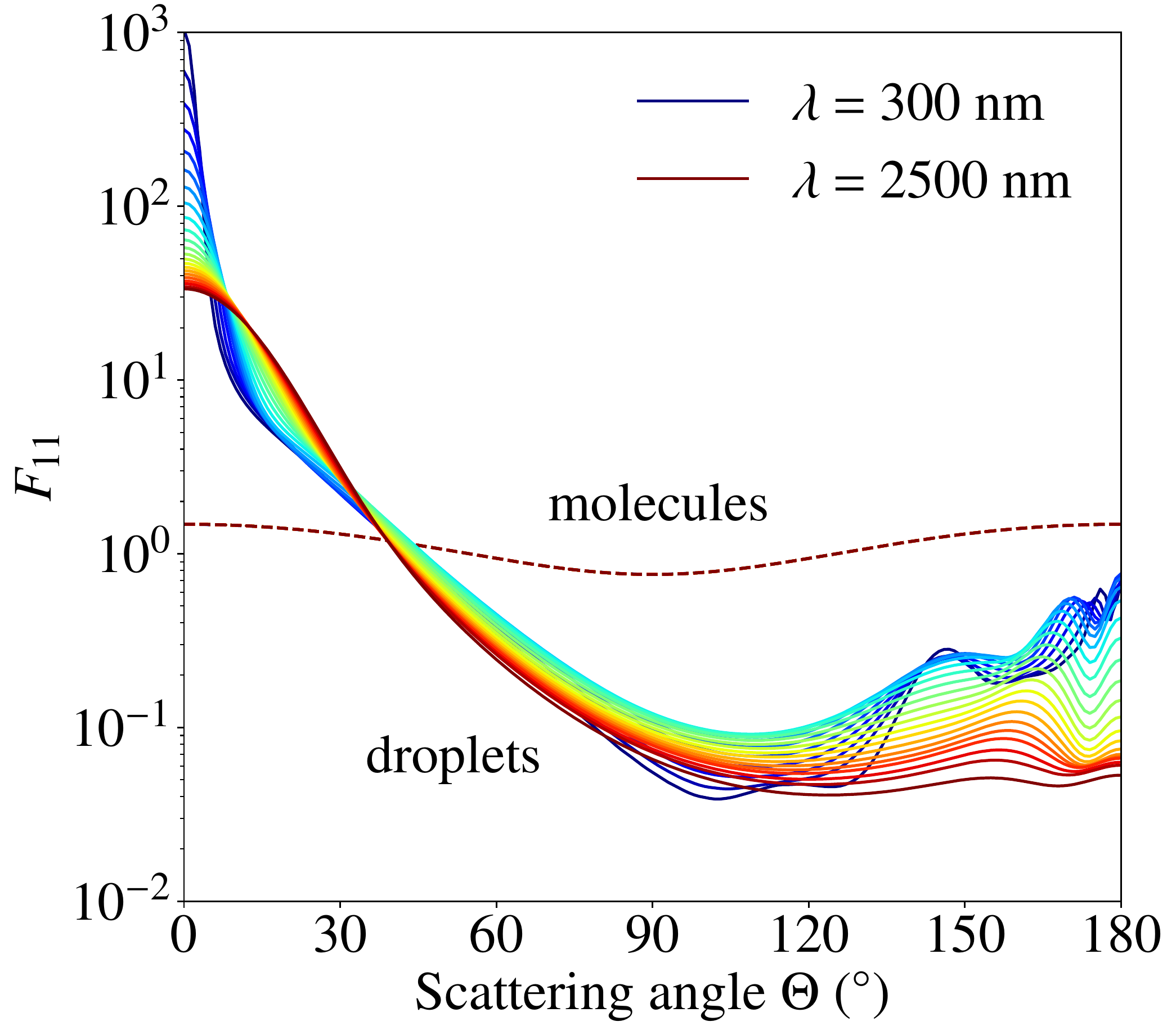}
\includegraphics[width=0.9\linewidth]{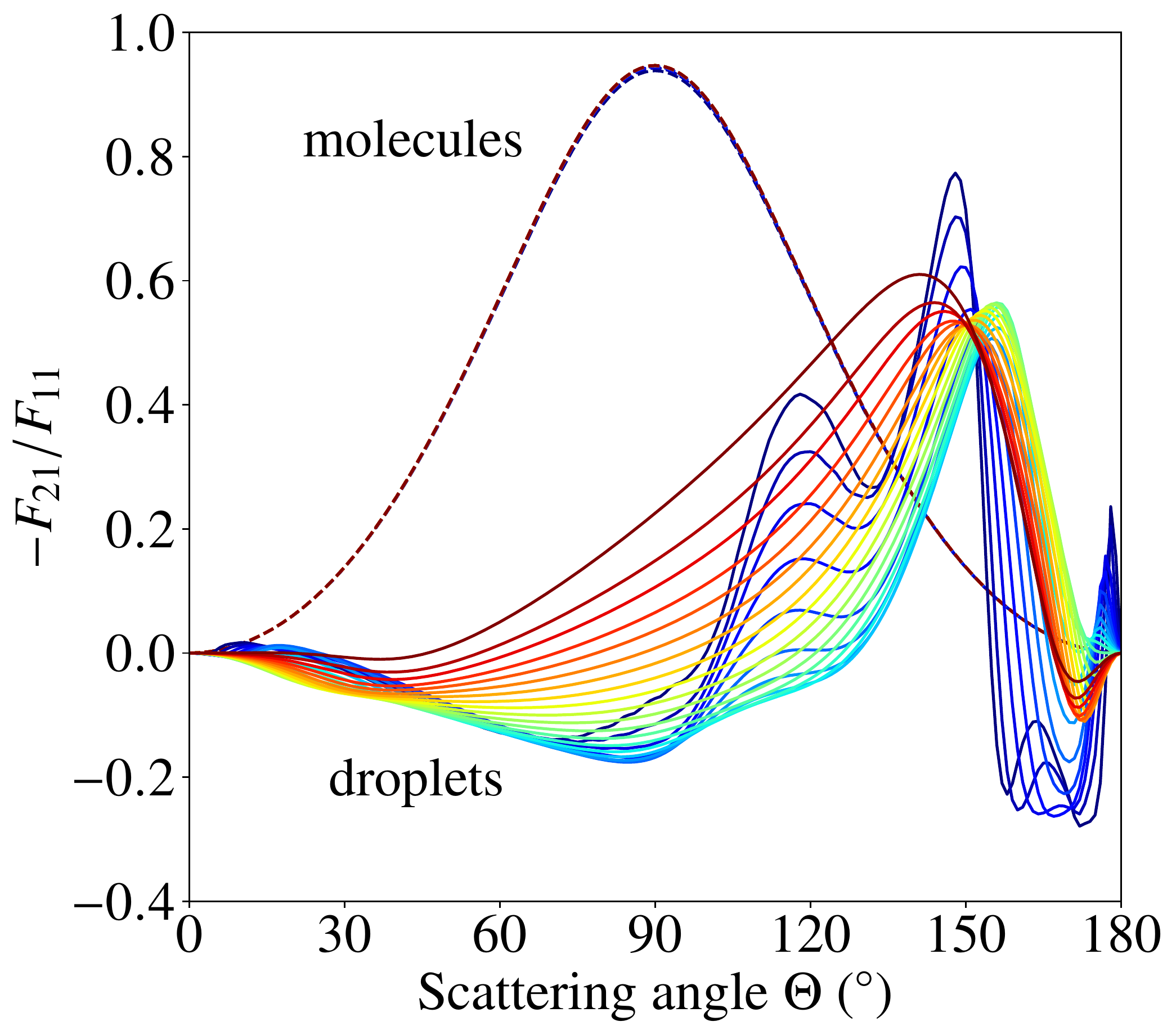}
\caption{The phase function $F_{11}$ (element (1,1) of scattering 
         matrix $\boldsymbol{F}_\mathrm{sca}$) (top) and
         degree of linear polarization ($-F_{21}/F_{11}$) (bottom) 
         as functions of the scattering angle $\Theta$, for singly 
         scattered light by gaseous molecules (dashed lines) and cloud 
         droplets (solid lines). 
         Scattering angle $\Theta$ is the angle between the directions
         of propagation of the incident and the scattered light 
         ($\Theta = 180^\circ - \alpha$).
         The wavelengths range from 300~nm (dark blue line) to 2500~nm 
         (dark red line), in steps of 100~nm. 
         The lines for the gaseous molecules at the different 
         wavelengths virtually overlap.}
\label{fig:phasefunction}
\end{figure}
%%%%%%%%%%%%%%%%%%%%%%%%%%%%%%%%%%%%%%%%%%%%%%%%%%%%%%%%%%%%%%%%%%%%%%%%%%

Figure~\ref{fig:phasefunction} shows the phase functions and the degree
of linear polarization $P_\mathrm{s}$
of unpolarized incident light that is singly scattered by the gaseous 
molecules and the cloud droplets. 
%Similar curves can be seen in Fig.~1 of \citet{Stam08} but only for 
%$\lambda= 440$, $550$, and $870$~nm. 
From this figure it is clear that the single 
scattering properties of the cloud droplets depend on wavelength 
across the entire range between 300 and 2500~nm. 
For example, the diffraction peak in the phase function near 
$\Theta= 0^\circ$ weakens with increasing $\lambda$. 
Also, the sign changes of $P_\mathrm{s}$ depend on $\lambda$ and 
the secondary rainbow feature, which is near $\Theta=120^\circ$ for 
$\lambda=300$~nm, is unobservable for $\lambda \gtrsim 1000$~nm. 
The phase function and $P_\mathrm{s}$ of the gaseous molecules 
also depend on $\lambda$ according to the wavelength dependence 
of the depolarization factor of air, but as can be seen in 
Fig.~\ref{fig:phasefunction}, this dependence is very small.

%%%%%%%%%%%%%%%%%%%%%%%%%%%%%%%%%%%%%%%%%%%%%%%%%%%%%%%%%%%%%%%%%%%%%%%%%%
% FIGURE 6
%%%%%%%%%%%%%%%%%%%%%%%%%%%%%%%%%%%%%%%%%%%%%%%%%%%%%%%%%%%%%%%%%%%%%%%%%%
\begin{figure*}[t!]
\centering
\includegraphics[width=0.795\linewidth]{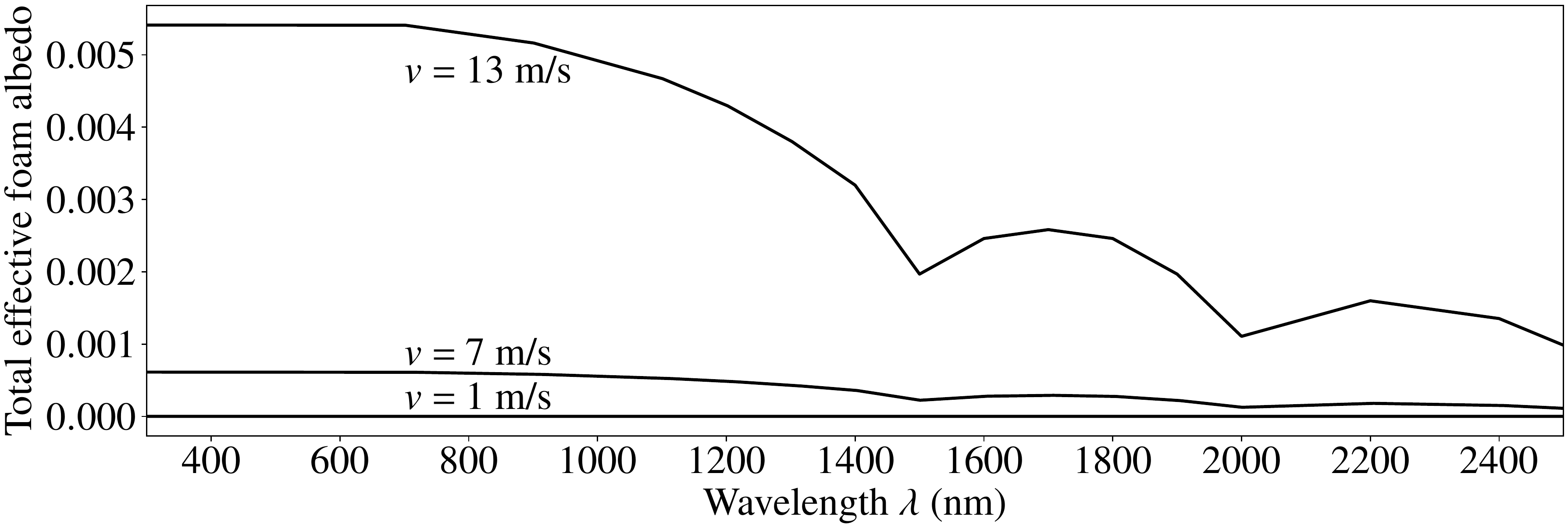}
\caption{The total effective sea foam albedo, which is the effective foam albedo of \citet{koepke1984effective} taking into account the aging of the whitecaps on the seawater multiplied by the wind-speed dependent fraction of the ocean surface that is covered with whitecaps of \citet{Monahan1980}: $2.95\cdot10^{-6} v^{3.52}$, with $v$ the wind speed in m/s.}
\label{fig:foamspectrum}
\end{figure*}
%%%%%%%%%%%%%%%%%%%%%%%%%%%%%%%%%%%%%%%%%%%%%%%%%%%%%%%%%%%%%%%%%%%%%%%%%%

%%%%%%%%%%%%%%%%%%%%%%%%%%%%%%%%%%%%%%%%%%%%%%%%%%%%%%%%%%%%%%%%%%%%%%%%%%
% FIGURE 7
%%%%%%%%%%%%%%%%%%%%%%%%%%%%%%%%%%%%%%%%%%%%%%%%%%%%%%%%%%%%%%%%%%%%%%%%%%
\begin{figure*}[t!]
\centering
\hspace{0.12cm}
\includegraphics[width=0.783\linewidth]{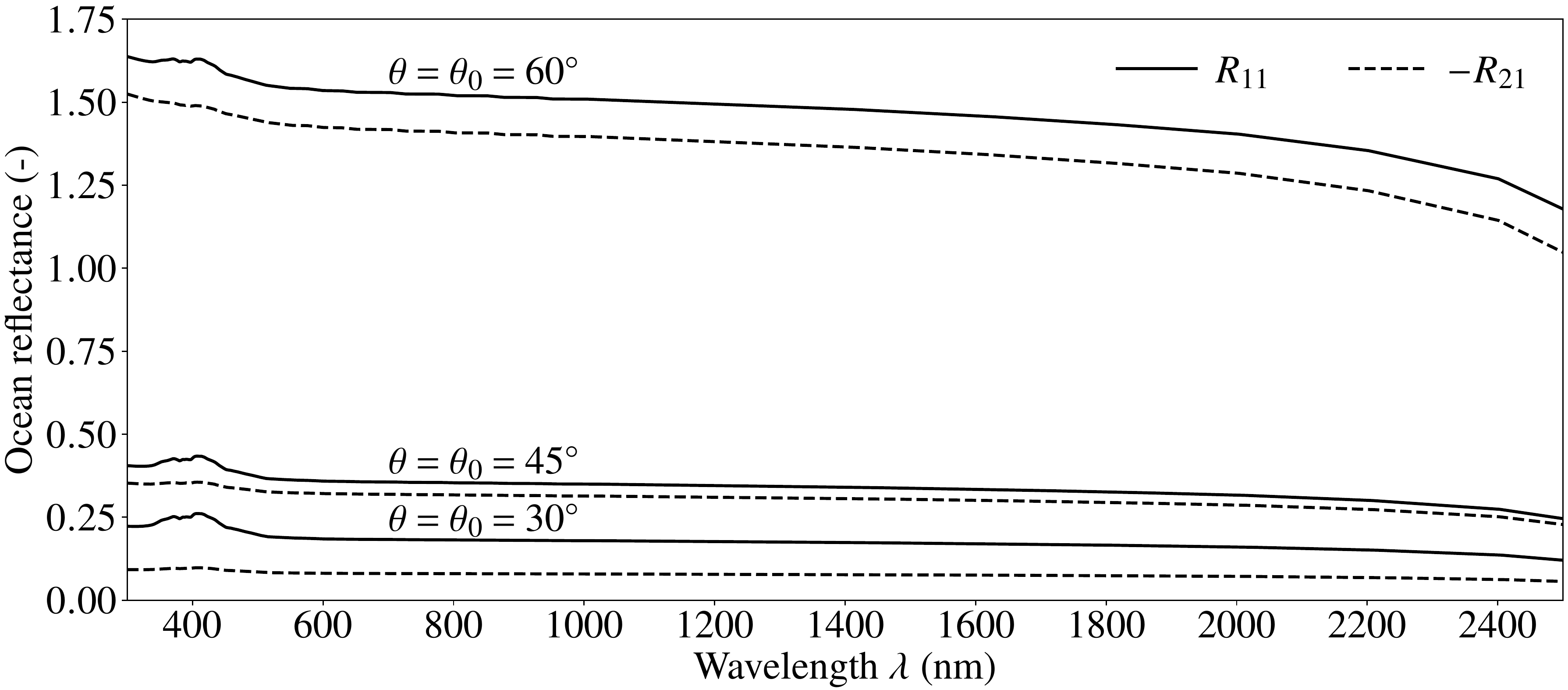}
\caption{Reflectance of the model ocean 
         from 300~to 2500~nm for 
         $\phi-\phi_0=0^\circ$ and $\theta=\theta_0=30^\circ$, 
         $45^\circ$ and $60^\circ$. 
         $R_{11}$ is the total reflectance (element (1,1) of 
         matrix $\mathbf{R_\text{ocean}}$), and -$R_{21}$ is the linearly 
         polarized reflectance (element (2,1) of $\mathbf{R_\text{ocean}}$). 
         The wind speed $v$ is 7~m/s.}
\label{fig:oceanspectrum}
\end{figure*}
%%%%%%%%%%%%%%%%%%%%%%%%%%%%%%%%%%%%%%%%%%%%%%%%%%%%%%%%%%%%%%%%%%%%%%%%%%

%-----------------------------------------------------------------------------
\subsubsection{The surface}
\label{sect_surface_model}

Below the atmosphere, the surface of our model planets is either 
a Lambertian reflector (i.e. isotropically reflecting and completely
depolarizing), or an ocean. 
The ocean consists of a Fresnel reflecting and 
transmitting air-water interface, a water body below the interface 
and a surface below the water body 
\cite[see Appendix A of][for the relevant equations]{TreesStam2019}. 
Other than in \citet{Stam08}, where the Fresnel reflecting ocean is
flat, our ocean surface can be rough as it can be ruffled by the wind.  
The Fresnel reflection by the air-water interface of incident 
starlight that is directly transmitted through the exoplanet's 
atmosphere results in the so--called ocean glint, 
whose size depends on the surface roughness. 
The surface roughness depends on the wind speed according to the 
empirical relation between wind speed and wave slope inclination 
distribution of \cite{CoxandMunk1954}. 
The width of the glint pattern increases with increasing wind speed, 
while its brightness gradually decreases 
\citep[see Fig.~7 of][]{TreesStam2019}.
We use wind speeds of 1~m/s (light breeze), 7~m/s (moderate breeze), 
and 13~m/s (strong breeze).

With increasing wind speed the contribution of sea foam in our 
ocean model also increases \citep[according to][]{Monahan1980}. 
We assume that the foam reflects Lambertian (isotropic and 
depolarizing). 
Although the foam albedo is approximately constant across the 
visible ($\sim$ 0.22), for wavelengths longer than 1000~nm, the albedo 
decreases significantly, see Fig.\ref{fig:foamspectrum}.
Therefore we use the effective wavelength-dependent albedo of
\citet[][]{koepke1984effective} that decreases from 0.22 to 0.04 
between 300 and 2500~nm. 
This albedo takes into account the aging of the whitecaps on the waves. 
The net contribution of the foam to the total ocean reflectance across 
our wavelength range is shown in Fig.~\ref{fig:foamspectrum}.
In the following, we will call this contribution the 'total effective 
foam albedo'. For $v=13$~m/s, 2.45\% of the ocean surface is covered 
with foam, while for $v=7$~m/s and $v=1$~m/s, it is 0.28\% and 
2.96 $\cdot 10^{-4}$\%, respectively, which explains the strong 
dependence of the total effective foam albedo on $v$ 
\citep[see also Eq.~A.15 of][]{TreesStam2019}. 
With increasing wind speed, the influence of wave shadows 
(i.e., the blockage of incident and/or reflected light by waves)
also increases, which is taken into account using 
the shadowing function of \cite{Smith1967} and \cite{Sancer1969}. 
The reflection and transmission matrices of the air-water interface 
are corrected for the energy deficiency caused by neglecting reflections 
of light between different wave facets \citep{NakajimaandTanaka1983}. 

Light that is transmitted through the air-water interface is scattered 
and/or absorbed by the water or absorbed by the surface below the water 
body. For consistence with \cite{TreesStam2019}, we assume a 100~m deep 
ocean and a black surface.
The water body is modeled as a stack of horizontally homogeneous layers 
of pure seawater, and its reflectance is computed using an adding-doubling 
program taking into account multiple scattering, absorption and polarization 
\citep[]{deHaan87}. The single scattering properties of the seawater are 
approximated by the scattering matrix for anisotropic Rayleigh scattering 
\citep{Hansen1974} with a depolarization factor of 0.09 
\citep{Chowdharyetal2006,morel1974optical}. The absorption and 
scattering coefficients are taken from \cite{SogandaresandFry1997}, 
\cite{popeandfry1997} and \cite{SmithandBaker1981}, resulting in a blue 
color across the visible. 
For wavelengths longer than 800~nm, all radiation that is transmitted through 
the air-water interface is absorbed by the water. For more details on 
our ocean model, we refer to \citet{TreesStam2019}.

The Fresnel reflection (and transmission) by (and through) the 
air-water interface depends on the real refractive index of water 
\citep[see Appendix A of][]{TreesStam2019}. 
The real refractive index of air is set equal to 1.0 in the interface 
computations of our ocean model. The real refractive index of water
\citep{hale1973optical} is approximately constant across the visible 
(it varies from 1.339 at 400~nm to 1.330 at 750~nm), but decreases more
significantly in the near-infrared to 1.2605 at 2500~nm (see also 
Sect.~\ref{sect_atmosphere_model}). 
Therefore, unlike in the ocean model of \citet{TreesStam2019}, 
we use the wavelength-dependent water refractive index of 
\citet{hale1973optical} when computing the reflection and 
transmission matrices of the air-water interface 
\cite[Eqs. A.1 and A.6 of][]{TreesStam2019}.

Figure~\ref{fig:oceanspectrum} shows elements (1,1) and (2,1)
of $\mathbf{R_\text{ocean}}$, the reflection matrix of the ocean 
(thus without the atmosphere), from 300 to 2500~nm, for three 
geometries in the forward reflection direction. 
These elements represent the total and the linearly polarized 
reflectances of the ocean, respectively. For a non-isotropic reflecting surface, such as a
Fresnel reflecting surface, the reflectance can be larger than 
1.0 because it is directional (for an isotropically reflecting
surface, such as a Lambertian reflecting surface, the reflectance
is equal to the surface albedo and thus always smaller than or 
equal to 1.0). With increasing $\theta$ and $\theta_0$, the reflection angle at the 
air-water interface increases and the Fresnel reflection strengthens, 
resulting in a larger reflectance, i.e., a brighter glint. 
The spectral signature of the ocean glint is relatively flat: 
there is only a slight decrease with wavelength, except for 
$\lambda \lesssim 400$ nm and $\lambda \gtrsim 2000$~nm where the 
spectral slopes are steeper as a result of the wavelength dependence 
of the water refractive index. The bump in the visible,  
between 300 and 500~nm, is responsible for the blue color of the 
seawater. 

Figure~\ref{fig:oceanspectrum} also shows the linearly polarized 
reflectance of the ocean. The spectral shape of this 
reflectance follows the spectral shape of the total reflectance. 
The degree of polarization increases with increasing $\theta$ and 
$\theta_0$ from $30^\circ$ to $45^\circ$, because the number of randomly 
oriented wave facets that experience local incidence angles equal or 
close to the Brewster angle (which is 53$^\circ$ in the visible)
increases.\footnote{At the Brewster angle, the degree of polarization 
of light reflected by a clean and flat air-water interface reaches its 
maximum of 1.0.} For $\theta$ and $\theta_0$ larger than 53$^\circ$, 
the degree of polarization decreases again, but the absolute linearly 
polarized reflectance increases as the total ocean reflectance strengthens 
(as shown by the spectrum for $\theta=\theta_0=60^\circ$). 
The linearly polarized ocean reflectance between 300 and 
500~nm slightly increases because of the higher ocean reflectance 
and because the light that leaves the water 
gets partly polarized upon scattering inside the water.

As explained in \citet{TreesStam2019}, our energy balance, reflection 
and transmission matrices for the rough air-water interface have been 
verified against \citet{NakajimaandTanaka1983}, 
\citet{MishchenkoandTravis1997} and \citet{Zhaietal2010}. 
More recently, \citet{Chowdhary2020} published results for the 
(polarized) reflectance by locally plane-parallel atmosphere-ocean 
models, as a benchmark for state-of-the-art spectropolarimetric 
radiative transfer models, as a preparation for NASA's upcoming 
Plankton, Aerosol, Cloud, ocean Ecosystem (PACE) Earth observation 
satellite mission \citep{Chowdhary2019PACEperspective,Werdell2019PACE}. 
We compared against \citet{Chowdhary2020}, for the rough ocean model 
without atmosphere (their model AOS-2) and with molecular atmosphere (their model AOS-3). The absolute differences of $R_{11}$ and $|R_{21}|$ with respect to those of \citet{Chowdhary2020} were smaller than $3.7\cdot10^{-4}$ for all scattering geometries of the test results, and were on average of the order $10^{-5}$.

%%%%%%%%%%%%%%%%%%%%%%%%%%%%%%%%%%%%%%%%%%%%%%%%%%%%%%%%%%%%%%%%%%%%%%%%%%
% FIGURE 8
%%%%%%%%%%%%%%%%%%%%%%%%%%%%%%%%%%%%%%%%%%%%%%%%%%%%%%%%%%%%%%%%%%%%%%%%%%
\begin{figure}[b]
\vspace*{0.5cm}
\centering
\includegraphics[width=0.99\linewidth]{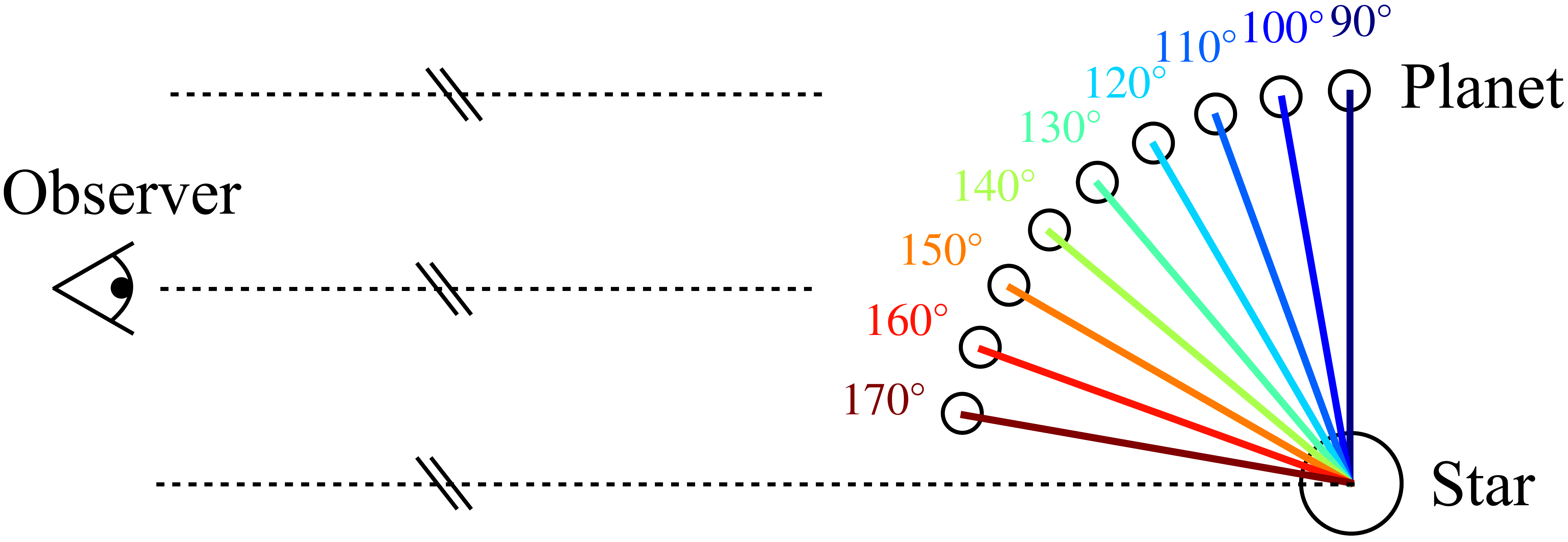}
	\caption{Sketch of the observing geometries used in our computations, 
	         for various phase angles $\alpha$. The $\alpha$-dependent 
	         colors correspond to the colors of the $\alpha$-dependent 
	         spectra in Figs.~\ref{fig:cloudfreeF}-\ref{fig:CloudyP}, \ref{fig:Stam08} and \ref{fig:quasi}.}
	\label{fig:phaseangles}
\end{figure}
%%%%%%%%%%%%%%%%%%%%%%%%%%%%%%%%%%%%%%%%%%%%%%%%%%%%%%%%%%%%%%%%%%%%%%%%%%

%%%%%%%%%%%%%%%%%%%%%%%%%%%%%%%%%%%%%%%%%%%%%%%%%%%%%%%%%%%%%%%%%%%%%%%%%%
% FIGURE 9
%%%%%%%%%%%%%%%%%%%%%%%%%%%%%%%%%%%%%%%%%%%%%%%%%%%%%%%%%%%%%%%%%%%%%%%%%%
\begin{figure*}[t!]
\captionsetup[subfigure]{aboveskip=-1pt,belowskip=-1pt}
\centering
\begin{subfigure}{0.518\textwidth}
\centering
\caption{\large{  Cloud-free dry planets}}
\label{fig:cloudfreelambF}
\includegraphics[width=1\linewidth]{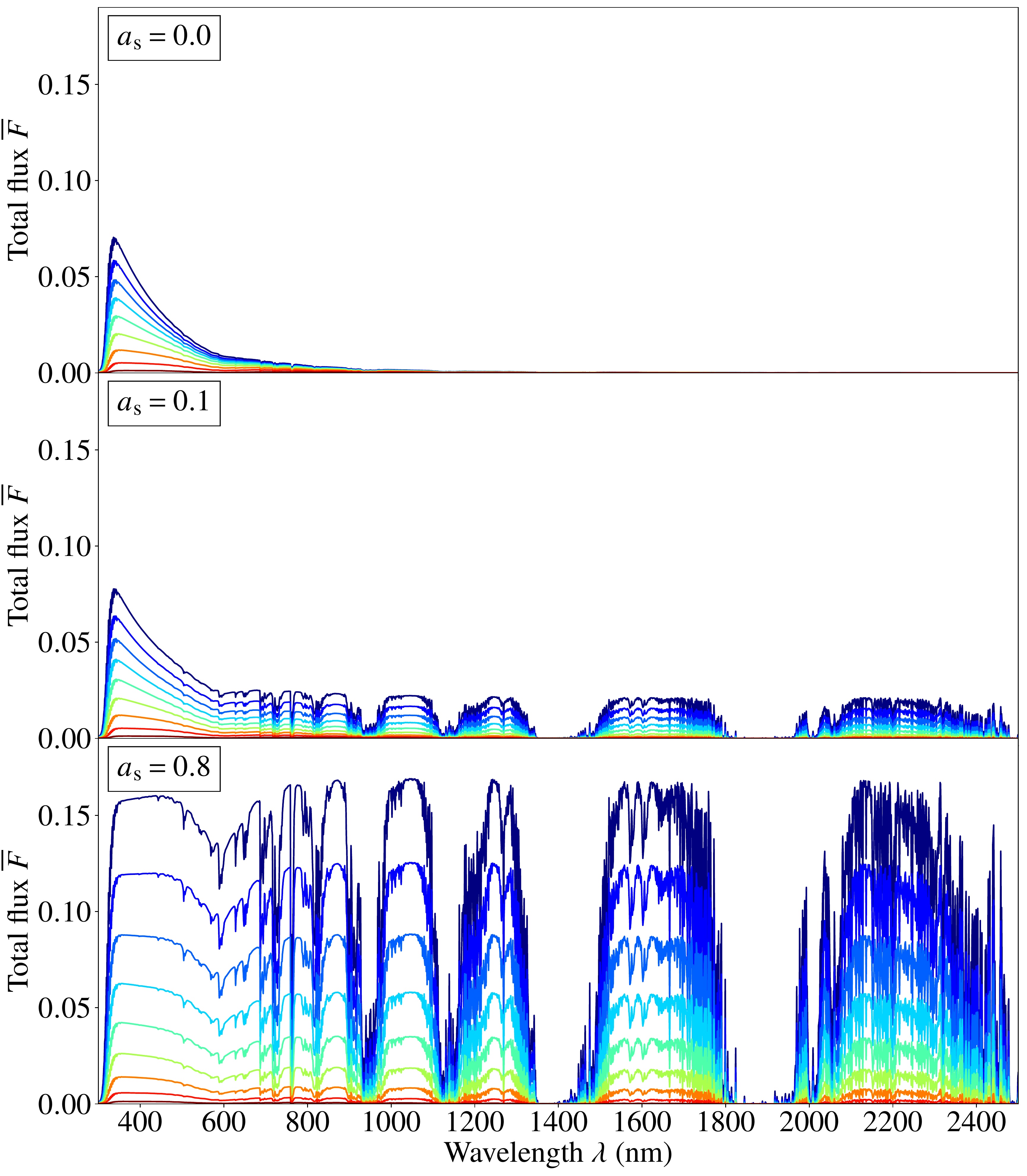}
\end{subfigure}
%\hfill
\begin{subfigure}{0.477\textwidth}
\centering
\caption{\large{  Cloud-free ocean planets}}
\label{fig:cloudfreeoceanF}
\includegraphics[width=1\linewidth]{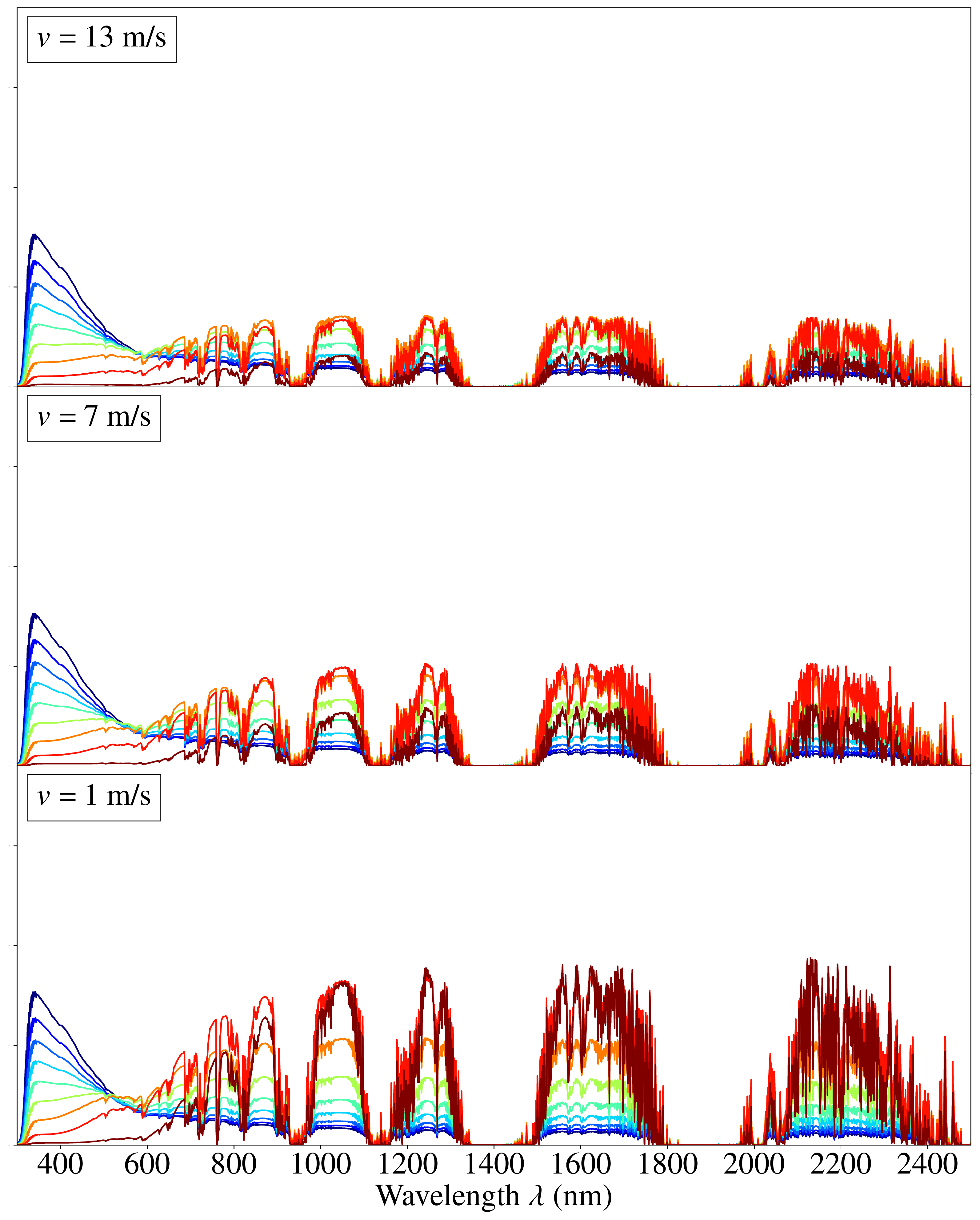}
\end{subfigure}
\caption{The total flux $\overline{F}$ of starlight that is reflected by 
         cloud-free, dry model planets (left column) and cloud-free 
         ocean planets (right column) as functions of $\lambda$ 
         at 1~nm resolution, for nine phase angles $\alpha$ from 
         90$^\circ$ (dark blue) to 170$^\circ$ (dark red) with steps 
         of 10$^\circ$ (see Fig.~\ref{fig:phaseangles} for the colors 
         of the different $\alpha$'s). The surface albedo $a_\text{s}$ 
         of the dry planets is 0.0 (top left), 0.1 (middle left), and 
         0.8 (bottom left). The wind speed $v$ over the oceans is 13~m/s 
         (top right), 7~m/s (middle right), and 1~m/s (bottom right).}
\label{fig:cloudfreeF}
\end{figure*}
%%%%%%%%%%%%%%%%%%%%%%%%%%%%%%%%%%%%%%%%%%%%%%%%%%%%%%%%%%%%%%%%%%%%%%%%%%

%%%%%%%%%%%%%%%%%%%%%%%%%%%%%%%%%%%%%%%%%%%%%%%%%%%%%%%%%%%%%%%%%%%%%%%%%%
% FIGURE 10
%%%%%%%%%%%%%%%%%%%%%%%%%%%%%%%%%%%%%%%%%%%%%%%%%%%%%%%%%%%%%%%%%%%%%%%%%%
\begin{figure*}[t!]
\captionsetup[subfigure]{aboveskip=-1pt,belowskip=-1pt}
\centering
\begin{subfigure}{0.518\textwidth}
\centering
\caption{\large{  Cloudy dry planets: $a_\mathrm{s} = 0.0$}}
\label{fig:CloudydryF}
\includegraphics[width=1\linewidth]{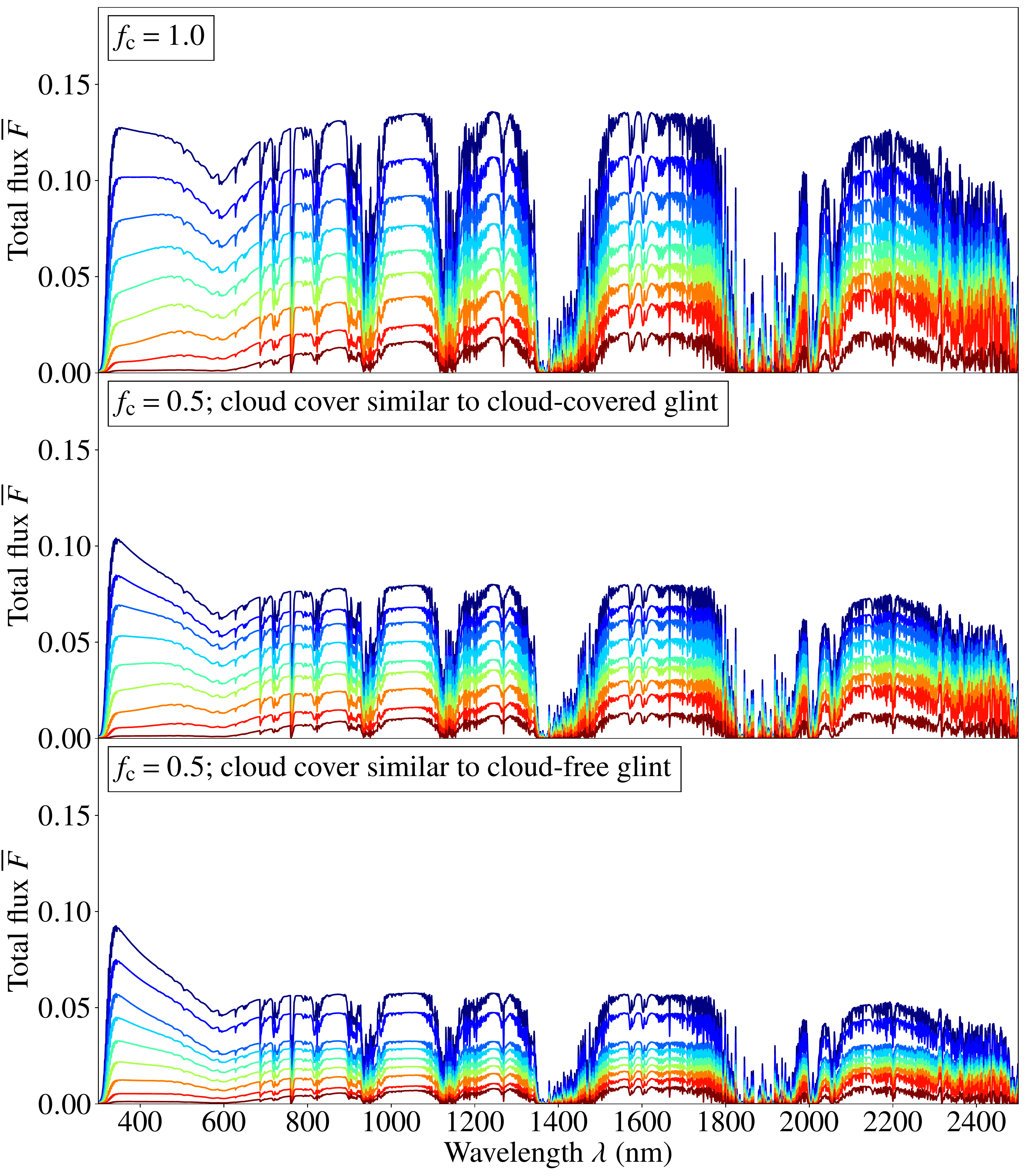}
\end{subfigure}
\hfill
\begin{subfigure}{0.477\textwidth}
\centering
\caption{\large{  Cloudy ocean planets; $v = 7$ m/s}}
\label{fig:cloudyoceanF}
\includegraphics[width=1\linewidth]{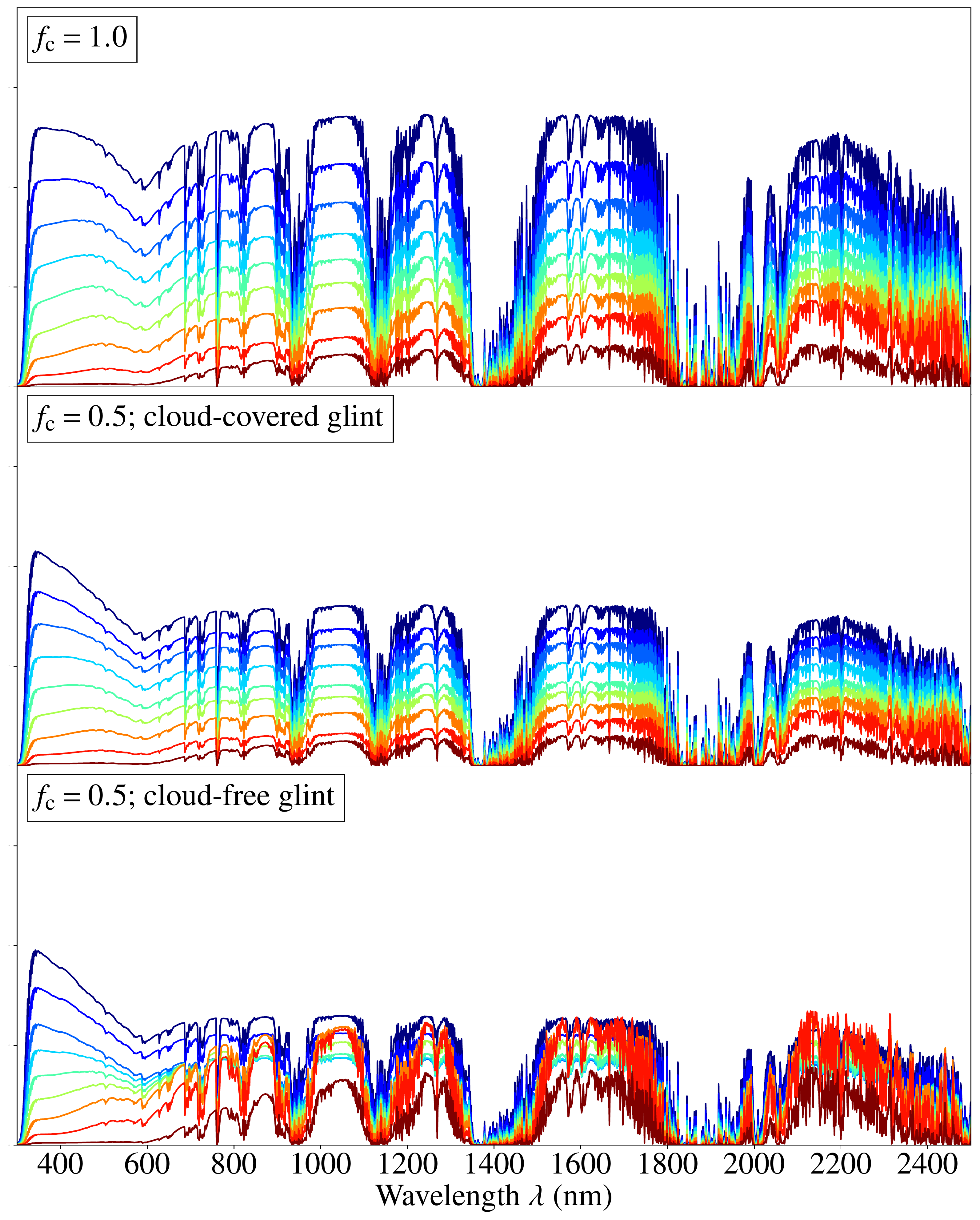}
\end{subfigure}
\caption{Similar to Fig.~\ref{fig:cloudfreeF} except with clouds. 
         The cloud coverage fraction $f_\text{c}$=1.0 (top) or
         0.5 (middle and bottom). The middle figures have similar 
         patchy cloud patterns, chosen such that the glint of the 
         ocean planet (right) is fully covered. 
         The bottom figures also have similar patchy cloud patterns, 
         but chosen such that the glint of the ocean planet is cloud-free.}
\label{fig:cloudyF}
\end{figure*}
%%%%%%%%%%%%%%%%%%%%%%%%%%%%%%%%%%%%%%%%%%%%%%%%%%%%%%%%%%%%%%%%%%%%%%%%%%

%-----------------------------------------------------------------------------
\section{Results}
\label{sect_results}

Here, we present the computed spectra of the total flux $\overline{F}$ 
(Sect.~\ref{sec:totalflux}) and polarized flux $\overline{Q}$ 
(Sect.~\ref{sec:polarizedflux}), normalized using the geometric albedo as explained in Sect. \ref{sect_calculating_reflected_starlight}, and the degree of polarization $P_\mathrm{s}$ (Sect.~\ref{sec:dop}) of starlight that is reflected by our model planets.  

%-----------------------------------------------------------------------------
\subsection{Total flux $\overline{F}$}
\label{sec:totalflux}

%-----------------------------------------------------------------------------
\subsubsection{$\overline{F}$ of cloud-free planets}

Figure~\ref{fig:cloudfreelambF} shows the computed total flux $\overline{F}$
that is reflected by cloud-free model planets with dry, Lambertian reflecting
surfaces, for $\lambda$ between 300 and 2500~nm and at nine phase angles
$\alpha$ from 90$^\circ$ to 170$^\circ$ (see Fig.~\ref{fig:phaseangles}). 
The surface albedo $a_\mathrm{s}$ increases in the sub-figures from top 
to bottom: $a_\text{s} = 0.0$, 0.1 and 0.8. 

The $\overline{F}$-spectra at $\alpha=90^\circ$ that are presented in 
Fig.~\ref{fig:cloudfreelambF}, agree well with published model computations 
between 300 and 1000~nm by \citet{Stam08}: for planets with dark surfaces 
($a_\text{s}=0.0$ and 0.1) and for $\lambda >$ 340~nm, 
$\overline{F}$ decreases with increasing $\lambda$ because of the 
$\lambda^{-4}$ dependence of the Rayleigh scattering efficiency by the gas. 
With increasing $a_\text{s}$, $\overline{F}$ increases at all $\lambda$ 
outside the absorption bands because of the contribution of light that 
is reflected by the surface. With increasing $\alpha$, $\overline{F}$ 
decreases at all $\lambda$ outside the absorption bands for all surface 
albedos. This decrease of $\overline{F}$ for larger $\alpha$ is a direct 
consequence of the smaller illuminated and visible fraction of the 
planetary disk \citep[see e.g.\ Fig.~4 of][for the planetary phase curves of $\overline{F}$ for dry planets with various surface albedos]{Stam08}.

Superimposed on the continua of the surface reflectance and the 
scattering gas, are gaseous absorption features (cf.\ the gaseous 
absorption optical thicknesses in Fig.~\ref{fig:opticalthicknessgases}). 
For example, O$_3$ absorbs the reflected light in the so-called Huggins 
absorption band at $\lambda <$ 340~nm, and in the so-called Chappuis 
absorption band between 500 and 700~nm. The absorption in the Chappuis 
band is more apparent when the surface albedo is higher
(see the results for $a_\text{s}=$ 0.8). 
For $a_\text{s}=$ 0.1 and 0.8, the O$_2$-A band (around $\lambda \sim$ 760~nm)
can clearly be identified, whereas the other O$_2$ absorption bands 
are more easily confused with H$_2$O absorption bands 
(cf.\ Fig.~\ref{fig:opticalthicknessgases}). 
For a more detailed discussion of the absorption features in the 
$\overline{F}$-spectra of Earth-like planets between 300 and 1000~nm, 
see e.g.\ Sect.~4.1.1 of \cite{Stam08}.

Here, we extend the spectra towards the near-infrared (NIR): from 1000 
to 2500~nm. In this spectral range, the main gaseous absorber is H$_2$O, 
with absorption bands near 1150, 1400, 1900, and 2500~nm 
(cf. Fig.~\ref{fig:opticalthicknessgases}). 
It should be noted that O$_2$ has an absorption band at about 
1270~nm, thus well-separated from H$_2$O absorption bands, which can also 
be identified in Fig.~\ref{fig:cloudfreelambF}. This O$_2$ absorption 
feature has already been observed in measured disk-integrated IR 
$\overline{F}$-spectra of the Earth during the EPOXI mission 
\citep[see][]{Livengood2011,Fujii2013}. Small absorption features 
of CO$_2$ near 1600~nm can also be identified. 
Absorption by CH$_4$ (e.g.\ near 2300~nm) is more difficult to 
distinguish by eye because of its less distinctive absorption bands
and nearby H$_2$O vapor absorption features 
(cf.\ Fig.~\ref{fig:opticalthicknessgases}).

Figure~\ref{fig:cloudfreeoceanF} shows $\overline{F}$ that is reflected 
by cloud-free planets with ocean surfaces over the same spectral range and
at the same phase angles. The wind speed $v$ decreases in the sub-figures 
from 13~m/s at the top, to 1~m/s at the bottom. At $v= 1$~m/s, the glint
is the narrowest but also the brightest (directionally reflecting and 
polarizing) \citep[see Fig.~7 of][]{TreesStam2019}.

At $\alpha=90^\circ$, ocean planets are relatively dark and their 
$\overline{F}$-spectrum is thus comparable to that of a dry planet 
with a low surface albedo (cf.\ Fig.~\ref{fig:cloudfreelambF} for 
$a_\text{s}=0$ and 0.1). Compared to the dry planet with 
$a_\text{s}=0$, $\overline{F}$ at $\alpha=90^\circ$ of an ocean planet
is slightly larger at $\lambda \lesssim 550$~nm, with a small bump 
between 300 and 500~nm, because of the natural blue color of our model 
ocean (cf.\ Fig.~\ref{fig:oceanspectrum}). For a discussion of the 
influence of the ocean color on $\overline{F}$ at various $\alpha$, 
we refer to Fig.~5 in \citet{TreesStam2019}. At longer wavelengths and 
outside the absorption bands, the ocean glint enhances $\overline{F}$. 
At $\alpha=90^\circ$, the contribution of the glint to $\overline{F}$ 
is virtually independent of the wind speed: the contribution of a narrow, 
bright glint is approximately equal to that of a wide, dim glint
when integrated over the cloud-free planetary disk for 
$\alpha \lesssim 135^\circ$ \citep[see also Sect.~3.1.2 of][]{TreesStam2019}. 

Wind speed dependent differences in $\overline{F}$ could also be due 
to the larger contribution of the foam on the surface for higher wind 
speeds (see Fig.~\ref{fig:foamspectrum}). Because the visibility of the
ocean surface decreases with increasing $\alpha$ because of the
increasing average optical path lengths through the
atmosphere, we expect the signature of foam to be most apparent 
at $\alpha=90^\circ$ in Fig.~\ref{fig:cloudfreeoceanF}. 
Indeed, for example at 865~nm, where there is relatively little 
absorption, a small increase of $\overline{F}$ is seen for 
$\alpha=90^\circ$ and $v=13$~m/s ($\overline{F}= 0.0111$,~0.0103, and
0.0103 for $v=13$, 7, and 1~m/s, respectively). 
However, this weak foam signature in $\overline{F}$ can easily be 
confused with other signatures that shape $\overline{F}$ of a planet 
in the continuum, such as surface reflection 
(Fig.~\ref{fig:cloudfreelambF}) and clouds (see Sect.~\ref{sec:Fcloudy}).

With increasing $\alpha$ from 90$^\circ$ to $160^\circ$, $\overline{F}$ 
in Fig.~\ref{fig:cloudfreeoceanF} increases for $\lambda \gtrsim 400$~nm 
outside the absorption bands compared to $\overline{F}$ of a dry planet with a black surface ($a_\mathrm{s} = 0$). This $\alpha$-dependent increase is the result of increasing Fresnel reflection of direct starlight (i.e., a brighter glint), 
due to the increasing local reflection angles on the waves (see 
Sect.~\ref{sect_surface_model}). When $\lambda \gtrsim$~900~nm 
and outside absorption bands, this contribution of the glint is 
spectrally more or less flat (cf.\ Fig.~\ref{fig:oceanspectrum}). 
The spectrally flat glint serves as a continuum with superimposed gaseous
absorption bands: 
in the bands, direct starlight is absorbed and does not reach the 
ocean surface or is absorbed after being reflected by the ocean, thus 
decreasing the visibility of the glint. 

For $\lambda \lesssim$~900 nm outside absorption bands, the
$\alpha$-dependent glint contribution disappears gradually with 
decreasing $\lambda$ (see Fig.~\ref{fig:cloudfreeoceanF}) as the atmospheric
optical thickness increases and less light reaches the surface.
Consequently, the $\overline{F}$-spectra of an ocean planet 
at the different values of $\alpha$ intersect between about 500 and 700~nm. 
These results are consistent with the intersecting planetary phase curves 
of cloud-free ocean planets at various (roughly absorption-free) wavelengths 
between 350 and 865~nm in Fig.~1 of \citet{TreesStam2019}.

%-----------------------------------------------------------------------------
\subsubsection{$\overline{F}$ of cloudy planets}
\label{sec:Fcloudy}

Figures~\ref{fig:CloudydryF} and b show the spectra of 
$\overline{F}$ that is reflected by dry and ocean planets, respectively, 
for cloud coverage fractions, $f_\mathrm{c}$, equal to 1.0, i.e., the 
planet is completely covered by clouds (top), and 0.5 (middle and
bottom). 
On the cloudy dry planets, $a_\mathrm{s}=0.0$, and on the cloudy ocean 
planets, $v=7$~m/s. Compared to their cloud-free equivalents 
(the top of Figs.~\ref{fig:cloudfreelambF} and b), 
the clouds increase $\overline{F}$ at all wavelengths outside the 
absorption bands. 
Because for $f_\mathrm{c}=1$, the ocean glint is fully covered by 
the optically thick clouds, the spectra for the cloudy dry and 
cloudy ocean planets are virtually the same at each $\alpha$:
there is no ocean-glint-related increase in $\overline{F}$ with 
increasing $\alpha$, nor a small bump in $\overline{F}$ between 300 
and 500~nm related to the blue water color. 

%%%%%%%%%%%%%%%%%%%%%%%%%%%%%%%%%%%%%%%%%%%%%%%%%%%%%%%%%%%%%%%%%%%%%%%%%%
% FIGURE 11
%%%%%%%%%%%%%%%%%%%%%%%%%%%%%%%%%%%%%%%%%%%%%%%%%%%%%%%%%%%%%%%%%%%%%%%%%%

\begin{figure*}[pt!]
\captionsetup[subfigure]{aboveskip=-1pt,belowskip=-1pt}
\centering
\begin{subfigure}{.523\textwidth}
\centering
\caption{\large{  Cloud-free dry planets}}
\label{fig:cloudfreelambQ}
\includegraphics[width=1\columnwidth]{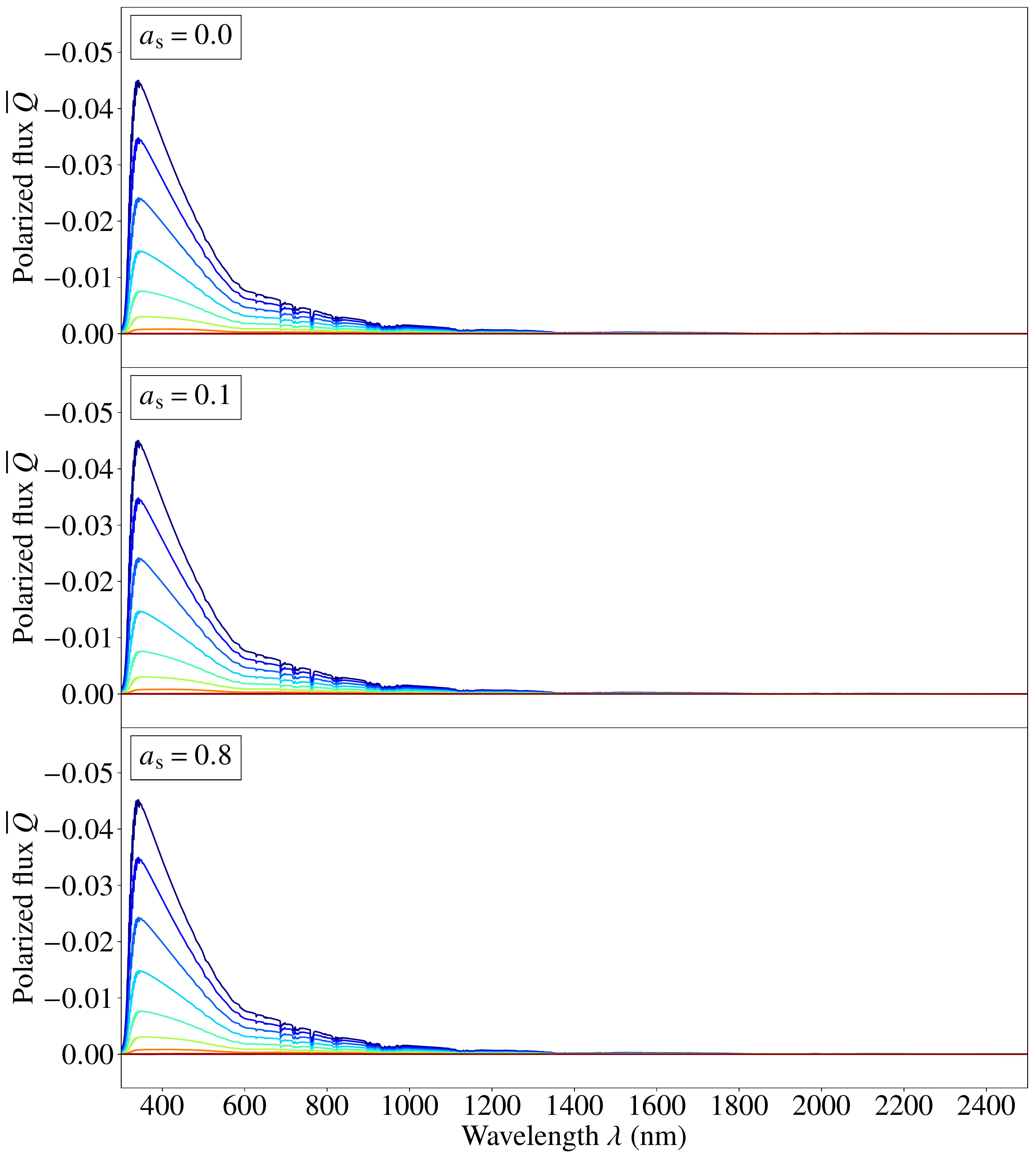}
\end{subfigure}
%\hfill
\begin{subfigure}{.47\textwidth}
\centering
\caption{\large{  Cloud-free ocean planets}}
\label{fig:cloudfreeoceanQ}
\includegraphics[width=1\columnwidth]{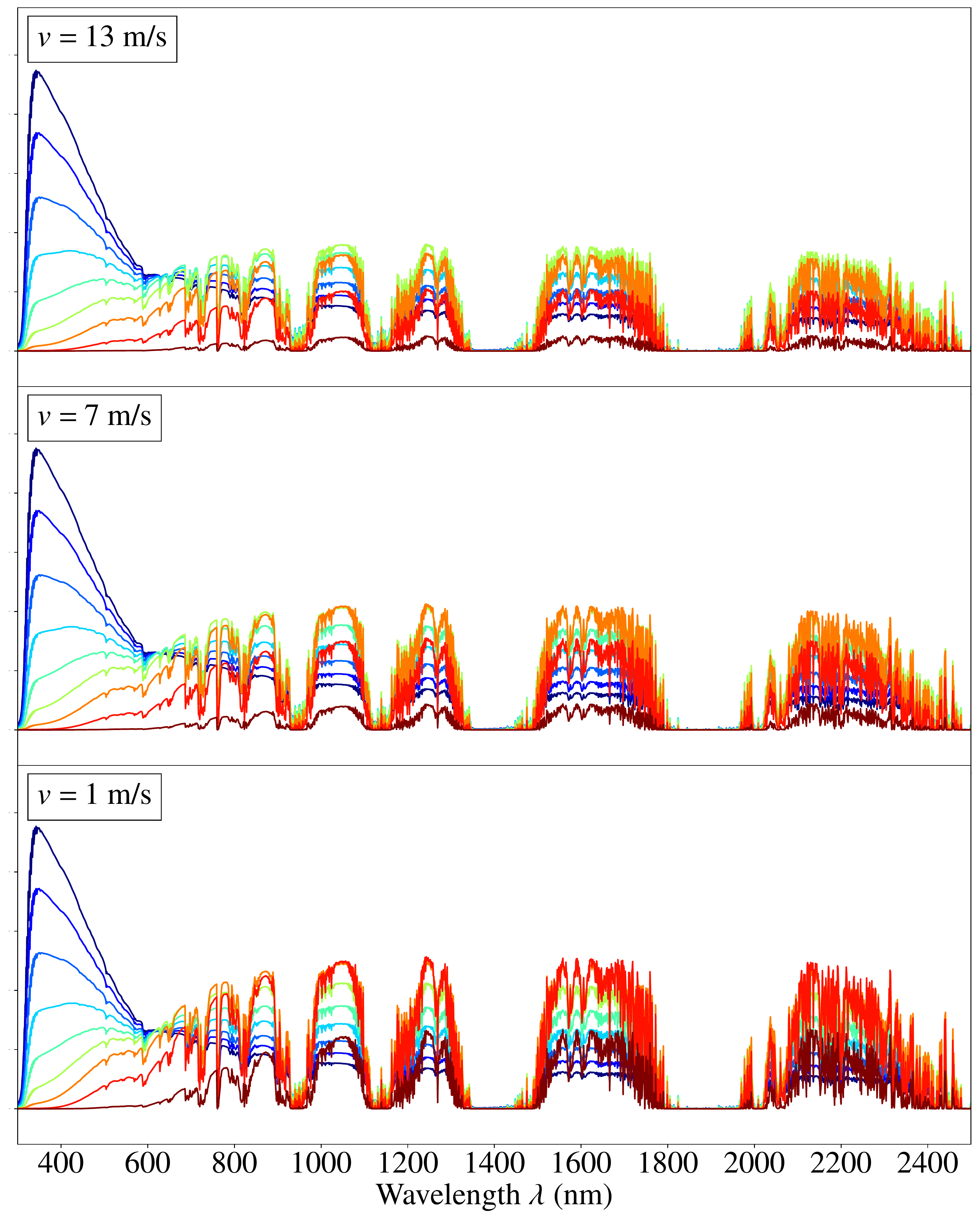}
\end{subfigure}
\caption{Similar to Fig.~\ref{fig:cloudfreeF}, except for the linearly 
         polarized flux $\overline{Q}$. Note the inverted vertical axis!}
\label{fig:cloudfreeQ}
\end{figure*}
%%%%%%%%%%%%%%%%%%%%%%%%%%%%%%%%%%%%%%%%%%%%%%%%%%%%%%%%%%%%%%%%%%%%%%%%%%

%%%%%%%%%%%%%%%%%%%%%%%%%%%%%%%%%%%%%%%%%%%%%%%%%%%%%%%%%%%%%%%%%%%%%%%%%%
% FIGURE 12
%%%%%%%%%%%%%%%%%%%%%%%%%%%%%%%%%%%%%%%%%%%%%%%%%%%%%%%%%%%%%%%%%%%%%%%%%%
\begin{figure*}[pt!]
\captionsetup[subfigure]{aboveskip=-1pt,belowskip=-1pt}
\centering
\begin{subfigure}{.523\textwidth}
\centering
\caption{\large{  Cloudy dry planets; $a_\mathrm{s} = 0.0$}}
\label{fig:cloudylambQ}
\includegraphics[width=1\columnwidth]{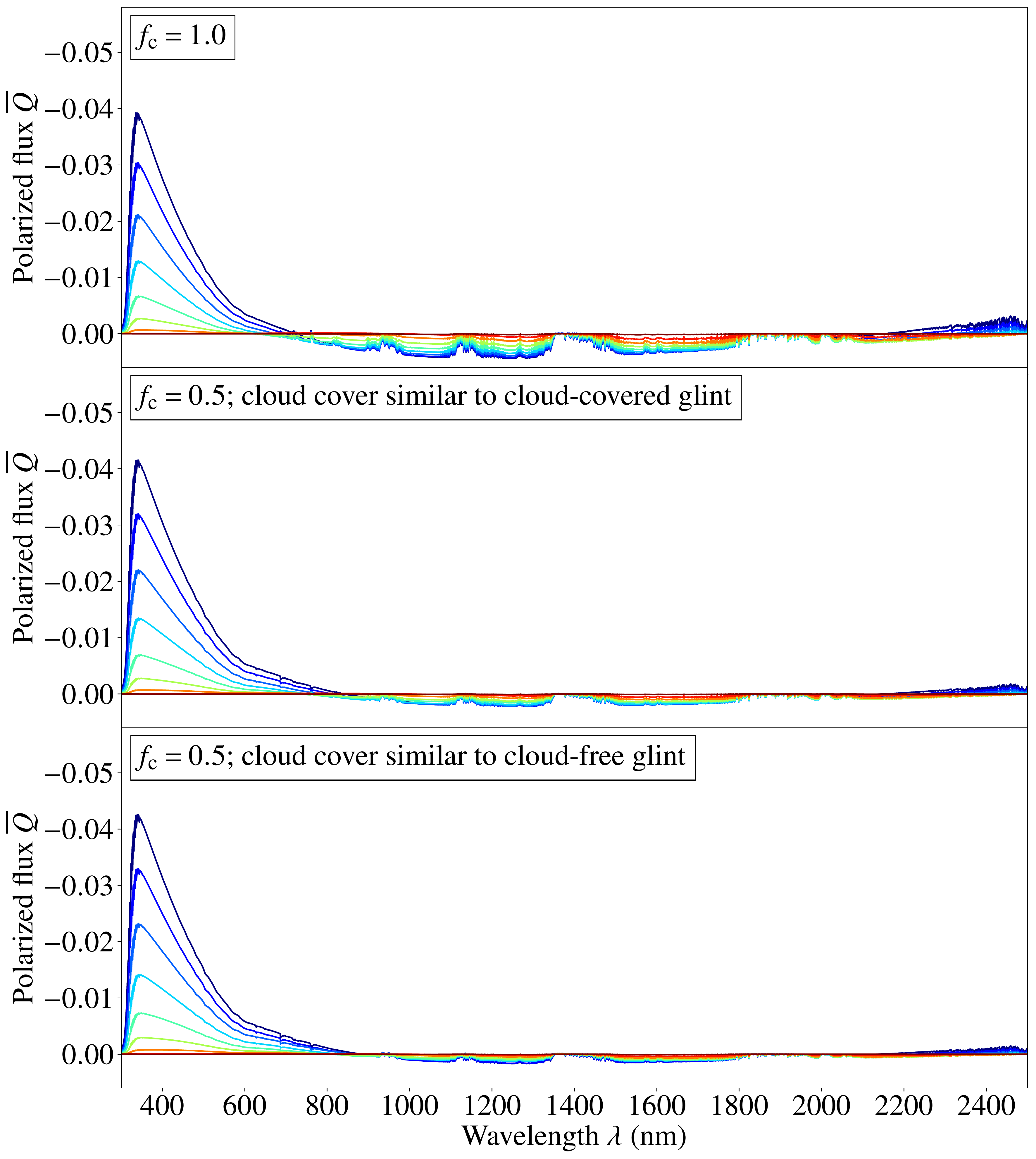}
\end{subfigure}
%\hfill
\begin{subfigure}{.47\textwidth}
\centering
\caption{\large{  Cloudy ocean planets; $v = 7$ m/s}}
\label{fig:cloudyoceanQ}
\includegraphics[width=1\columnwidth]{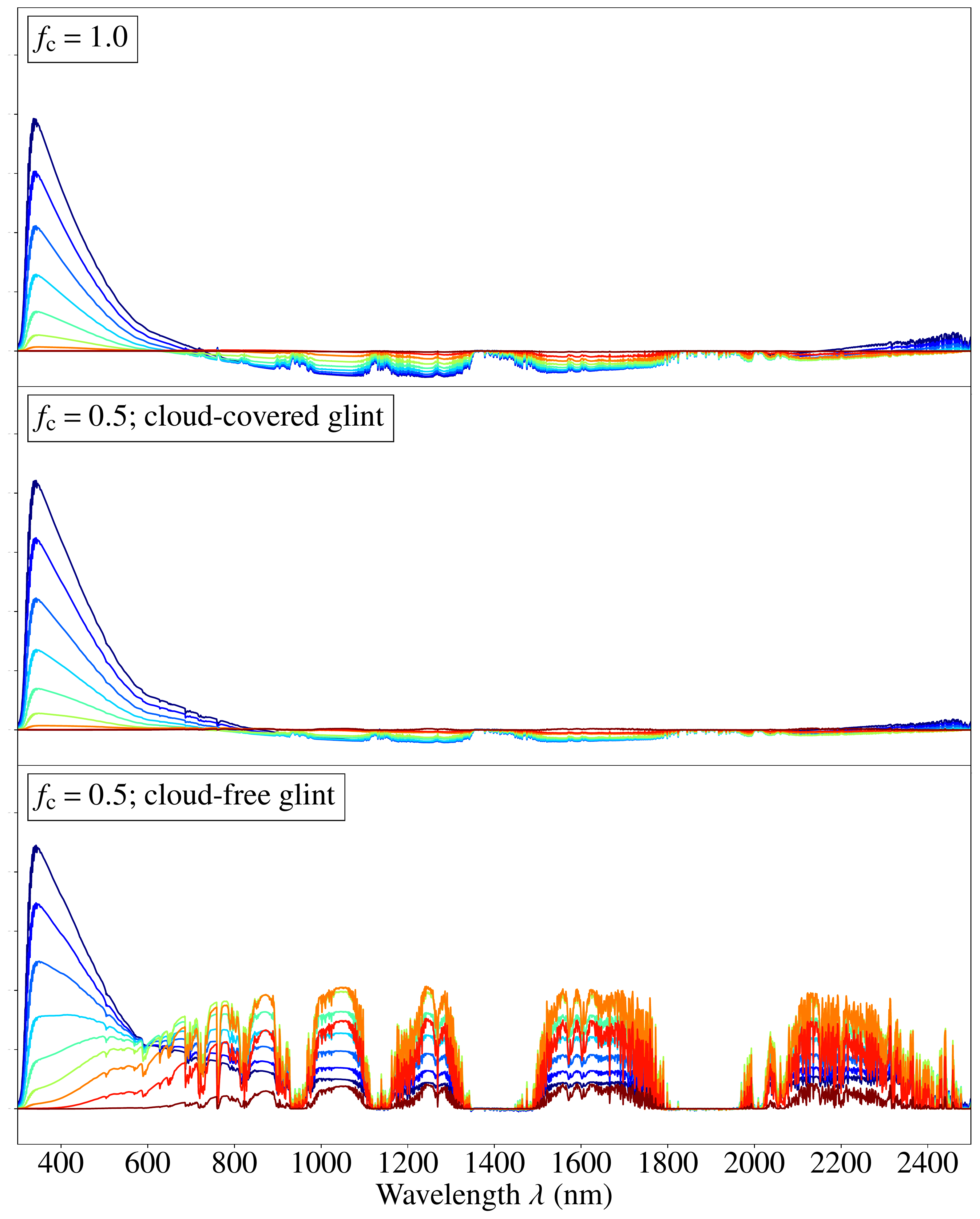}
\end{subfigure}
\caption{Similar to Fig.~\ref{fig:cloudyF}, except for the linearly polarized flux $\overline{Q}$.}
\label{fig:cloudyQ}
\end{figure*}
%%%%%%%%%%%%%%%%%%%%%%%%%%%%%%%%%%%%%%%%%%%%%%%%%%%%%%%%%%%%%%%%%%%%%%%%%%

In the middle row of Figs.~\ref{fig:CloudydryF} and b, 
$f_\mathrm{c}=0.5$, and the patchy cloud pattern is such that on the ocean 
planets, the glint is fully covered by clouds (see the right image in 
Fig.~\ref{fig:disks}). For comparison, the cloud pattern on the dry planet 
is similar to that on the ocean planet. When the glint is covered by clouds, 
there is no increase of $\overline{F}$ with increasing $\alpha$. 
Only a very subtle remnant of the ocean color bump near 
$\lambda \sim 400$~nm is visible at $\alpha=90^\circ$, which is caused by 
light that is reflected by the ocean in between the cloud patches.

In the bottom row of Figs.~\ref{fig:CloudydryF} and b, 
$f_\mathrm{c}$ is also 0.5, but here the patchy cloud pattern leaves
the ocean glint uncovered (see the left image in Fig.~\ref{fig:disks}). 
The uncovered glint significantly enhances $\overline{F}$ with increasing
$\alpha$ outside the absorption bands. 
Because of the decreasing optical thickness of the gaseous atmosphere 
with increasing $\lambda$, some $\overline{F}$-spectra for various $\alpha$ 
intersect. The different cloud patterns also influences the spectra of the 
dry planets: the 'glint region' is at the well-illuminated, 
well-visible equator, so with a dark and dry planet, a cloud-free 
'glint-region' results in a smaller $\overline{F}$. 

From Figs.~\ref{fig:CloudydryF} and b, we conclude 
that the detectability of an ocean glint in $\overline{F}$ is 
not limited by the cloud coverage fraction $f_\mathrm{c}$, but depends
strongly on the spatial distribution of the clouds, in particular 
on the coverage of the glint.

%-----------------------------------------------------------------------------
\subsection{Polarized flux $\overline{Q}$}
\label{sec:polarizedflux}

%-----------------------------------------------------------------------------
\subsubsection{$\overline{Q}$ of cloud-free planets}

Figures~\ref{fig:cloudfreelambQ} and b are
similar to Figs.~\ref{fig:cloudfreelambF} and~b
except for the polarized flux $\overline{Q}$.
The obvious difference in $\overline{Q}$ between dry and ocean planets is
that for all wind speeds $v$ and for $\alpha$ between $90^\circ$ to 
$150^\circ$, ocean planets show an enhancement of $|\overline{Q}|$ 
with increasing $\alpha$ at all $\lambda$ outside absorption bands and intersections of the $\overline{Q}$-spectra at various values of $\alpha$
between $\sim 600$ and $\sim 900$~nm, while for the dry planets, there is
only a noticeable polarized flux for $\lambda \lesssim 1000$~nm, 
and the $\overline{Q}$-spectra do not intersect at any $\alpha$, 
regardless of the surface albedo $a_\mathrm{s}$.

The enhancement of $|\overline{Q}|$ with increasing $\alpha$ outside 
absorption bands for cloud-free ocean planets is due to the increasing 
polarized Fresnel reflection of direct starlight with increasing local 
reflection angles on the ocean waves. The intersections of the 
$\overline{Q}$-spectra for various $\alpha$'s is caused by the spectral 
dependence of the scattering by the gas. At the shortest wavelengths, the polarized radiance is determined by Rayleigh scattering and 
hence is strongly $\alpha$ dependent, while with increasing $\lambda$,
the atmospheric scattering decreases and the visibility of the 
polarized ocean glint increases.
The very small values of $|\overline{Q}|$ of the dry, cloud-free planets 
for $\lambda \gtrsim 1000$~nm are due to the absence of a significant 
polarizing scatterer or reflector at those wavelengths, as the 
Rayleigh scattering optical thickness is very small and as the 
surface reflection is unpolarized. 

For the cloud-free ocean planets, $\overline{Q}$ outside the 
absorption bands seems virtually independent of the wind speed $v$ 
for $\alpha=90^\circ$ to $120^\circ$. At $\alpha=90^\circ$, the suppression
of the Fresnel reflection by the open ocean surface due to foam leaves
a negligibly small decrease of $|\overline{Q}|$ for $v=13$~m/s with 
respect to $v=7$~m/s (e.g.\ $1.04 \cdot 10^{-4}$ at $\lambda = 865$ nm).
Indeed, Fig.~4 of \citet{TreesStam2019} shows that $|\overline{Q}|$ only
depends on $v$ for $\alpha \gtrsim 123^\circ$ as there the glint is 
increasingly dominant in the signal. 
In Fig.~\ref{fig:cloudfreeoceanQ}, the strongest enhancement of 
$|\overline{Q}|$ is at $\alpha = 140^\circ$ for $v= 13$~m/s, and 
at 160$^\circ$ for $v= 1$~m/s. 
Figure~4 of \citet{TreesStam2019} shows that with decreasing $v$, 
$|\overline{Q}|$ at a NIR wavelength ($\lambda = 865$ nm) peaks at 
a larger $\alpha$, which is explained by the brighter but narrower glint 
for smaller $v$, that is also less affected by wave shadows.

Similar to $\overline{F}$ of the cloud-free ocean planet, the 
glint in the near-infrared $\overline{Q}$ has a spectrally approximately 
flat continuum with superimposed gaseous absorption features.
Inside absorption bands, less direct light reaches the ocean and/or 
the observer after being reflected, resulting in a
smaller $|\overline{Q}|$. For example, absorption features of O$_2$ and 
CO$_2$ can be identified in $|\overline{Q}|$ at 1270~and 1600~nm,
respectively.

%-----------------------------------------------------------------------------
\subsubsection{$\overline{Q}$ of cloudy planets}

Figures~\ref{fig:cloudylambQ} and b are
similar to Figs.~\ref{fig:CloudydryF} and b
except for the polarized flux $\overline{Q}$.
For completely cloudy planets ($f_\mathrm{c}=1.0$), $\overline{Q}$
is, like the total flux, virtually similar for dry and ocean planets: 
the thick clouds cover the ocean signatures and
at $\lambda \lesssim 700$~nm, $\overline{Q}$ is determined by the Rayleigh 
scattering of the gas above (and to a lesser extent inside and below) 
the clouds. 
At $\lambda \sim 700$~nm, the direction of polarization can be seen to 
change: the sign of $\overline{Q}$ changes from negative to positive
because at the longer wavelengths the scattering by the cloud 
particles overtakes the Rayleigh scattering, and the single scattering 
matrix element $F_{21}$ of the water cloud droplets is positive at
small to intermediate scattering angles $\Theta$, thus at
intermediate to large phase angles $\alpha$ (see
Fig.~\ref{fig:phasefunction}).  
The sign change in Figs.~\ref{fig:cloudylambQ} and~b 
is most apparent at $\alpha=90^\circ$, because with increasing $\alpha$ 
from $90^\circ$, the observable fraction of the planetary disk decreases, 
thus the absolute amount of flux decreases, and in addition, 
the degree of polarization of the light that is singly scattered by the 
cloud droplets generally decreases with decreasing $\Theta$
(Fig.~\ref{fig:phasefunction}). For $\lambda \gtrsim 2100$~nm, 
the sign of $F_{21}$ of the droplets, and consequently the sign of 
$\overline{Q}$, changes again.

For the planets with patchy clouds ($f_\mathrm{c}=0.5$) that cover
the glint (the middle row of Figs.~\ref{fig:cloudylambQ} and b), 
the sign changes of $\overline{Q}$ 
are somewhat less apparent because there are simply less scattering 
cloud particles across the planet.
With the glint covered by patchy thick clouds, the ocean planets have 
virtually similar $\overline{Q}$-spectra as the dry planets: no ocean
signatures can be identified.

In case the glint is not covered by clouds (the bottom row of
Figs.~\ref{fig:cloudylambQ} and b), the
$\overline{Q}$-spectra of the ocean planet differ significantly from 
those of the similarly clouded dry planet: with increasing $\alpha$ up 
to $150^\circ$, the glint increases $|\overline{Q}|$ outside the absorption bands as compared to 
$|\overline{Q}|$ of the dry planet.  
The latter spectra are almost unaffected by a different cloud pattern: 
$|\overline{Q}|$ at $\lambda \lesssim 800$~nm is only slightly 
increased because the equatorial regions are mostly cloud-free, 
such that relatively more signal is received from regions where 
Rayleigh scattering gas dominates the local signals.

Comparing Figs.~\ref{fig:cloudfreeQ} and~\ref{fig:cloudyQ} at NIR 
wavelengths ($\lambda > 800$~nm) outside absorption bands shows that, 
for all dry planets, $\overline{Q}$ is close to zero or slightly positive, 
while for all ocean planets with a cloud-free glint, $\overline{Q}$ 
is negative. Moreover, for the latter planets, $|\overline{Q}|$ increases 
with increasing $\alpha$, causing the $\overline{Q}$-spectra for the 
different values of $\alpha$ to intersect. 
The negative $\overline{Q}$ at NIR wavelengths and the intersecting  
$\overline{Q}$-spectra for various $\alpha$ could reveal 
the presence of an ocean.

%-----------------------------------------------------------------------------
\subsection{Degree of polarization $P_\mathrm{s}$}
\label{sec:dop}

%%%%%%%%%%%%%%%%%%%%%%%%%%%%%%%%%%%%%%%%%%%%%%%%%%%%%%%%%%%%%%%%%%%%%%%%%%
% FIGURE 13
%%%%%%%%%%%%%%%%%%%%%%%%%%%%%%%%%%%%%%%%%%%%%%%%%%%%%%%%%%%%%%%%%%%%%%%%%%
\begin{figure*}[pt!]
\captionsetup[subfigure]{aboveskip=-1pt,belowskip=-1pt}
\centering
\begin{subfigure}{.5145\textwidth}
\centering
\caption{\large{  Cloud-free dry planets}}
\label{fig:cloudfreelambP}
\includegraphics[width=1\linewidth]{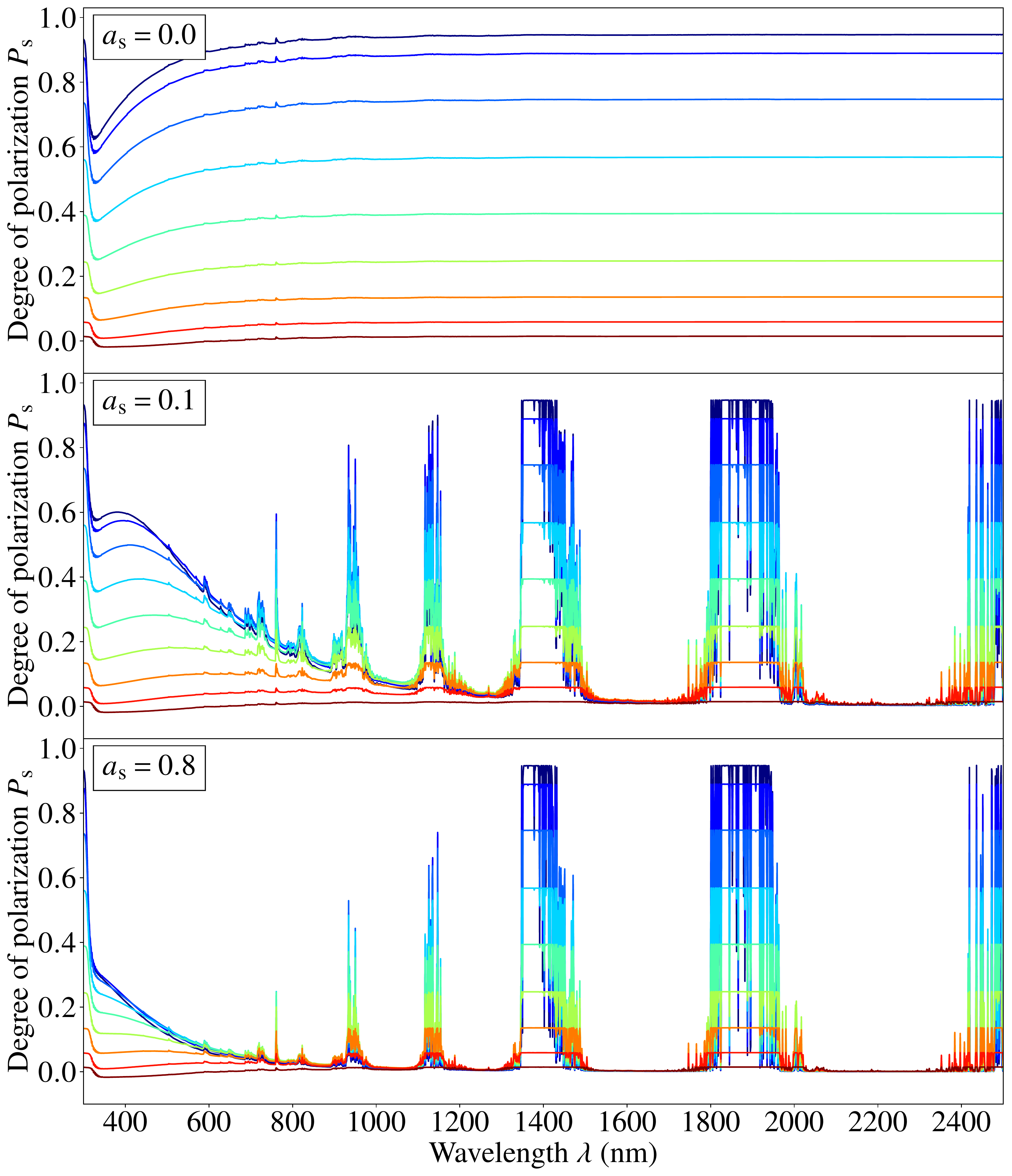}
\end{subfigure}
\hfill
\begin{subfigure}{.48\textwidth}
\centering
\caption{\large{  Cloud-free ocean planets}}
\label{fig:cloudfreeoceanP}
\includegraphics[width=1\linewidth]{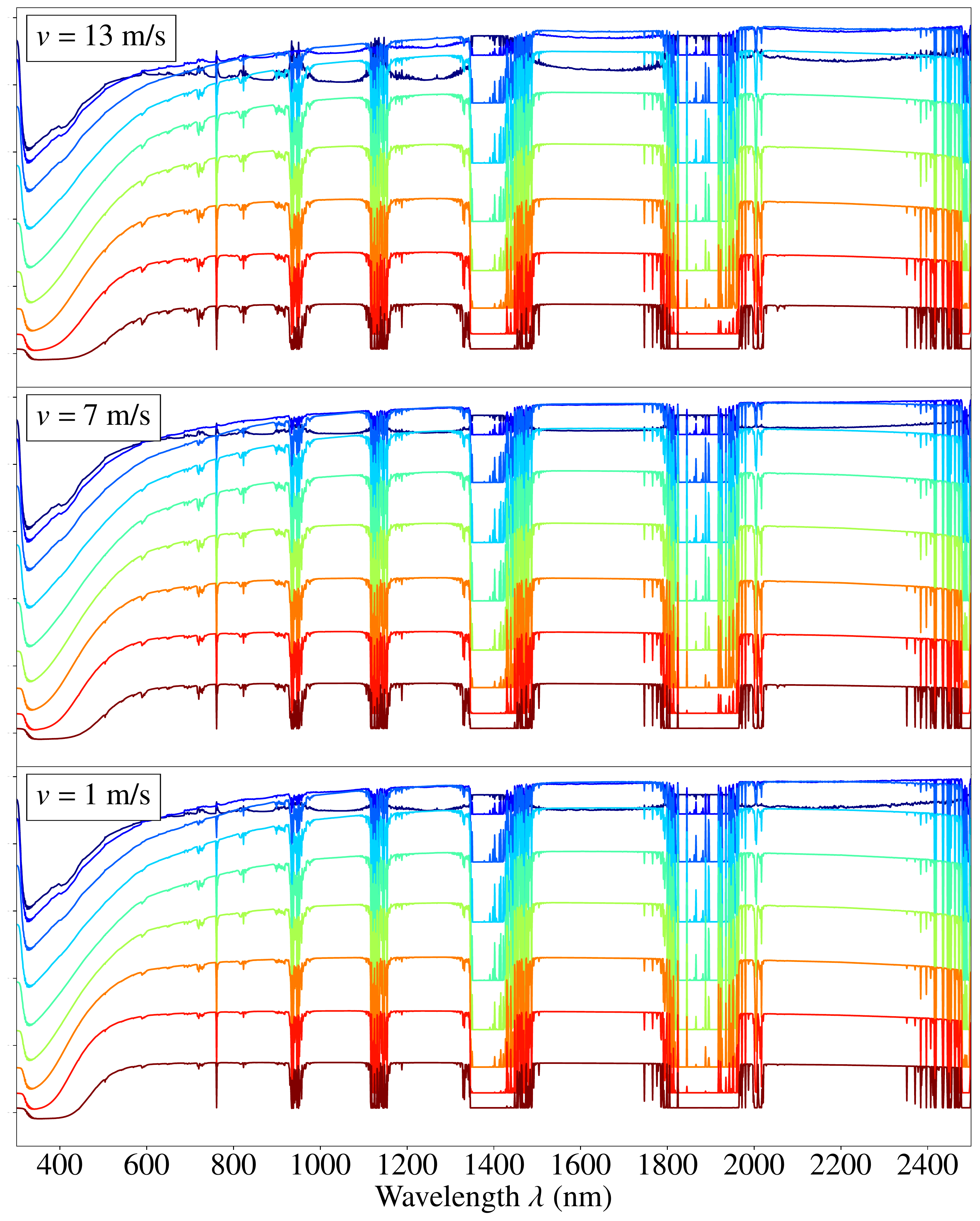}
\end{subfigure}
\caption{Similar to Fig.~\ref{fig:cloudfreeF}, except for the degree of polarization $P_\mathrm{s}$.}
\label{fig:cloudfreeP}
\end{figure*}
%%%%%%%%%%%%%%%%%%%%%%%%%%%%%%%%%%%%%%%%%%%%%%%%%%%%%%%%%%%%%%%%%%%%%%%%%%

%%%%%%%%%%%%%%%%%%%%%%%%%%%%%%%%%%%%%%%%%%%%%%%%%%%%%%%%%%%%%%%%%%%%%%%%%%
% FIGURE 14
%%%%%%%%%%%%%%%%%%%%%%%%%%%%%%%%%%%%%%%%%%%%%%%%%%%%%%%%%%%%%%%%%%%%%%%%%%
\begin{figure*}[pt!]
\captionsetup[subfigure]{aboveskip=-1pt,belowskip=-1pt}
\centering
\begin{subfigure}{.5145\textwidth}
\centering
\caption{\large{  Cloudy dry planets; $a_\mathrm{s} = 0.0$}}
\label{fig:CloudydryP}
\includegraphics[width=1\linewidth]{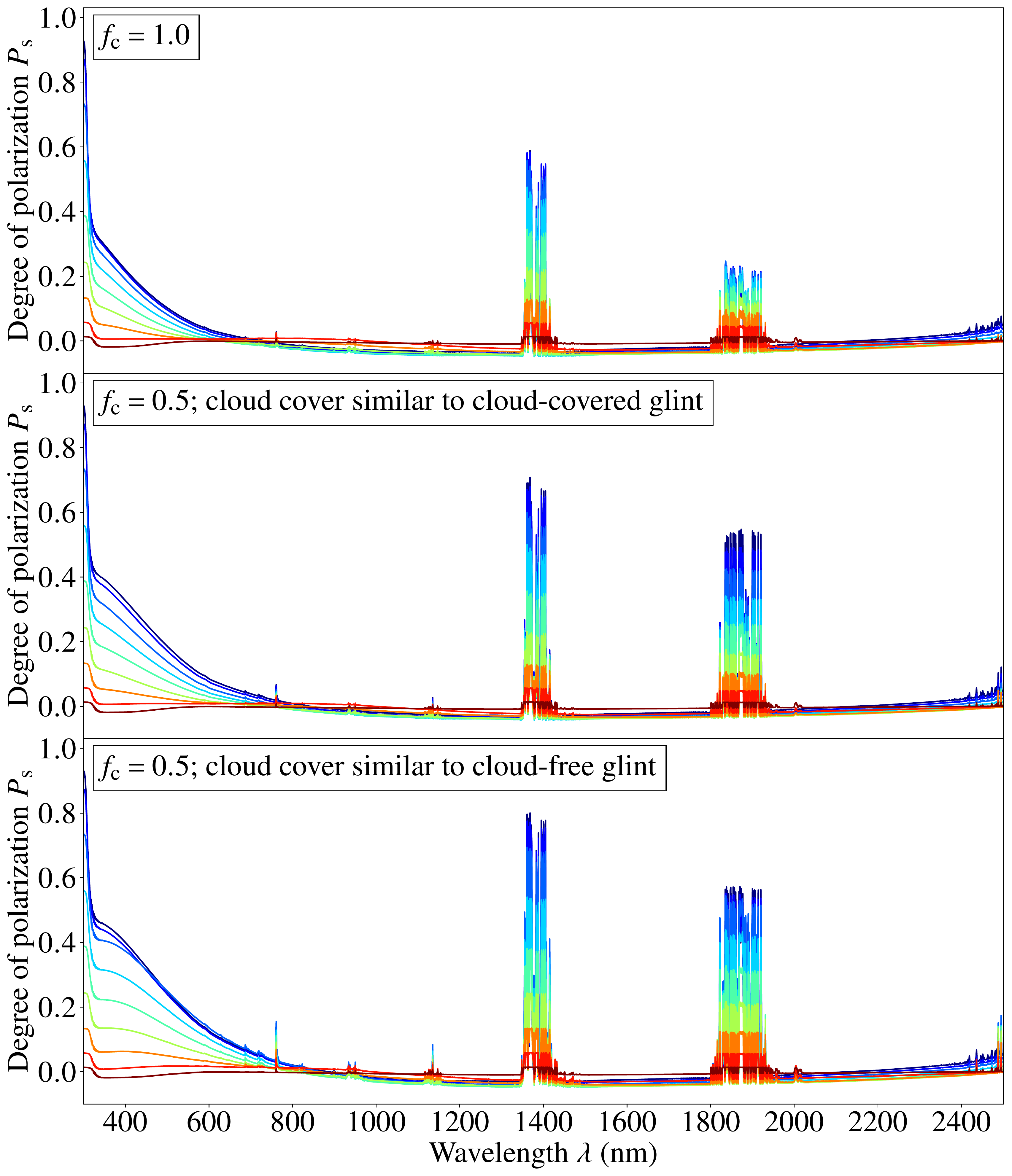}
\end{subfigure}
\hfill
\begin{subfigure}{.48\textwidth}
\centering
\caption{\large{  Cloudy ocean planets; $v = 7$ m/s}}
\label{fig:cloudyoceanP}
\includegraphics[width=1\linewidth]{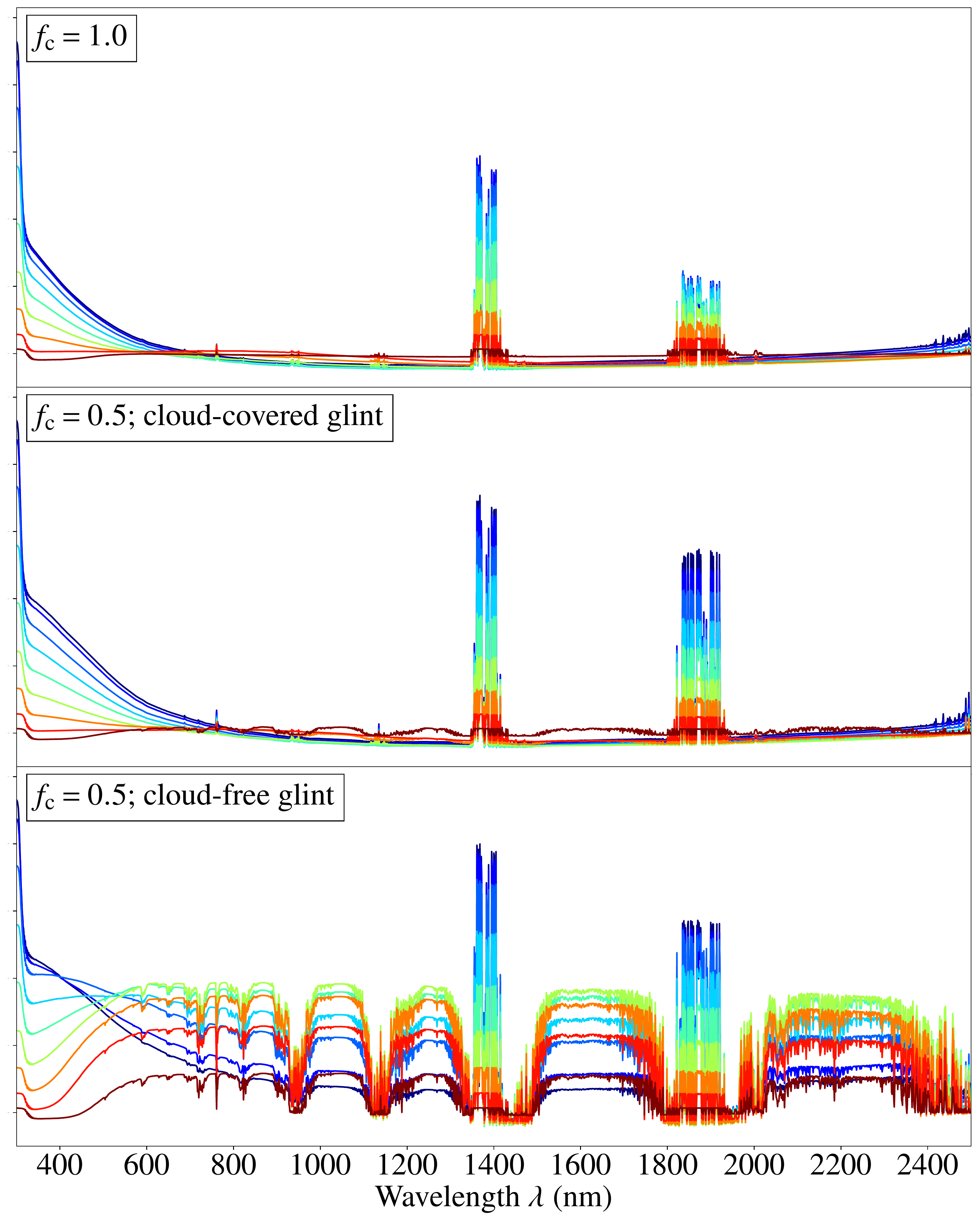}
\end{subfigure}
\caption{Similar to Fig.~\ref{fig:cloudyF}, except for the degree of polarization $P_\mathrm{s}$.}
\label{fig:CloudyP}
\end{figure*}
%%%%%%%%%%%%%%%%%%%%%%%%%%%%%%%%%%%%%%%%%%%%%%%%%%%%%%%%%%%%%%%%%%%%%%%%%%

%-----------------------------------------------------------------------------
\subsubsection{$P_\mathrm{s}$ of cloud-free planets}

Figure~\ref{fig:cloudfreelambP} is similar to Figs.~\ref{fig:cloudfreelambF}
and~\ref{fig:cloudfreelambQ}, except for $P_\mathrm{s}$. Between 300
and 1000~nm, the results agree well with earlier published 
polarization spectra by \citet{Stam08} (the latter do not
go beyond 1000~nm). For dry planets with a black surface 
($a_\text{s}=0.0$), $P_\mathrm{s}$ is highest around $\alpha= 90^\circ$
due to Rayleigh scattering. 
Because of multiple scattering of light, $P_\mathrm{s}$ decreases with 
decreasing $\lambda$ and increasing gaseous scattering optical thickness
from the visible towards the ultraviolet, reaching a local minimum at 
$\sim 320$~nm \citep[see also Fig.~3 of][]{Stam08}.
At shorter wavelengths, absorption by O$_3$ (the Huggins band) 
suppresses multiple scattering, which leads to an increase of $P_\mathrm{s}$
\citep[see, e.g.][]{Stam08,1999GeoRL..26..591A}.
Although $P_\mathrm{s}$ in the near-infrared or in the Huggins band is
relatively high due to single scattering of light by the atmospheric gas
above the black surface, the polarized flux $|\overline{Q}|$ is very small
at those wavelengths due to the small gaseous scattering optical thickness
and the absorption (cf.\ Fig.~\ref{fig:cloudfreelambQ}). 

Even a little bit of reflection by the surface ($a_s=0.1$) influences
$P_\mathrm{s}$ significantly, as unpolarized light is added
to the planet's signal, decreasing $P_\mathrm{s}$ outside the 
absorption bands. In particular at the longer wavelengths, this decrease 
is significant, because the polarized flux $|\overline{Q}|$ from light
that has been singly scattered by the atmospheric gas is very small 
(see Fig.~\ref{fig:cloudfreelambQ}). A similar, but stronger decrease
of $P_\mathrm{s}$ is found for the dry planet with a bright 
surface ($a_s= 0.8$).

The $P_\mathrm{s}$-spectra of the cloud-free dry planets with non-black 
surfaces show absorption band features, mostly due to O$_2$ and H$_2$O 
(see Fig.~\ref{fig:opticalthicknessgases}). 
Inside those absorption bands, $P_\mathrm{s}$ is high compared to 
outside the bands, because the absorption prevents light to reach 
the reflecting surface and any unpolarized light that is reflected
by the surface to reach the observer, leaving the strongly polarized 
signal of single Rayleigh scattering in the $P_\mathrm{s}$-spectra. 
Indeed, in the bands, $P_\mathrm{s}$ reaches its single scattering 
value, as if the surface were black (top row). 
The absorption of light by gases in the cloud-free atmospheres of 
dry exoplanets with reflecting surfaces thus shows up as 
peaks in the $P_\mathrm{s}$-spectra
\citep[e.g.][and references therein]{Stam08}.

Figure~\ref{fig:cloudfreeoceanP} shows $P_\mathrm{s}$ of 
cloud-free ocean planets. At first sight, and outside the absorption
bands, these spectra look rather similar to those of the cloud-free
dry planet with a black surface, but there are some profound differences,
three of which we will discuss below.

Firstly, at the shortest wavelengths, the optically thick atmosphere 
prevents Fresnel reflected light from the ocean surface to reach the
observer directly, but weakly polarized light that is diffusely reflected by 
the ocean still slightly decreases $P_\mathrm{s}$ 
\citep[see also Fig.~5 of ][]{TreesStam2019}. 
The bump in the ocean reflection between 300 and 500~nm (see 
Fig.~\ref{fig:oceanspectrum}) adds somewhat more unpolarized light to 
the planet signal, as is most apparent from the small dip in $P_\mathrm{s}$
for $\alpha=90^\circ$ (at that phase angle, the contribution of the blue ocean is larger than at larger phase angles, as the optical paths 
through the planet's atmosphere increase with increasing $\alpha$).

Secondly, for the cloud-free ocean planets, the maximum $P_\mathrm{s}$ 
does not occur at $\alpha=90^\circ$, like for the cloud-free dry planet,
but at larger values of $\alpha$. 
The reason is that with increasing $\lambda$, the influence of the 
Fresnel reflecting ocean increases, shifting the maximum 
$P_\mathrm{s}$ from that of the Rayleigh scattering to the 
phase angle corresponding to the local Brewster angle, that is given by
$\alpha_\text{B} = 2 \arctan( n_\text{water}/n_\text{air})$ 
\citep[see Fig. 1][]{TreesStam2019}. 
For example, at 1000~nm, $\alpha_\text{B}= 106^\circ$, while at 2500~nm
$\alpha_\text{B}=107^\circ$. Consequently, the $P_\mathrm{s}$-spectrum at 
$\alpha= 90^\circ$ intersects with the spectra at $\alpha= 100^\circ$ and 
$110^\circ$. Because $P_\mathrm{s}$ of the cloud-free ocean planet
peaks at larger $\alpha$, the $P_\mathrm{s}$-continua of cloud-free 
ocean planets are higher than those of the cloud-free dry black planet
for $\alpha > \alpha_\text{B}$.
As can be seen in Fig.~\ref{fig:cloudfreeoceanP}, a high wind speed
($v=13$ m/s) causes an extra decrease of the $P_\mathrm{s}$-spectrum 
for $\alpha=90^\circ$ and 100$^\circ$, due to the 
sea foam that covers 2.46\% of the ocean when $v=13$ m/s. 
The foam suppresses $P_\mathrm{s}$ because it
lowers the contribution of the (polarized) Fresnel reflection and it
diffusely reflects unpolarized light. As the sea foam albedo decreases 
towards longer wavelengths, the latter suppressing effect also decreases
with wavelength (at 1000~nm and $v=13$~m/s, $P_\mathrm{s}$ decreases 
with $0.081$ with respect to $v=7$~m/s, while at 2200~nm, it decreases with 
$0.037$). As the optical paths through the 
exoplanet's atmosphere increase with increasing $\alpha$, the influence 
of the foam on $P_\mathrm{s}$ is more apparent for smaller $\alpha$ 
\citep[see Appendix B of][]{TreesStam2019}.  

Thirdly, at wavelengths where the influence of the Fresnel reflecting 
ocean dominates that of the atmosphere ($\lambda \gtrsim 600$~nm), 
$P_\mathrm{s}$ is \textit{lower} in the absorption bands than outside
for all wind speeds. These dips of $P_\mathrm{s}$ (instead of 
the peaks for the dry planets) are explained as follows. 
Inside the absorption bands, $P_\mathrm{s}$ is close to its single 
scattering value for the given scattering geometry, which is similar
to $P_\mathrm{s}$ for the cloud-free dry planet with a black surface 
($a_s = 0$).
Outside the absorption bands, $P_\mathrm{s}$ for the ocean planet
is determined by the strongly polarized fluxes of the ocean glint 
(Fig.~\ref{fig:cloudfreeoceanQ}).
As discussed above, for $\alpha > \alpha_\text{B}$, this $P_\mathrm{s}$
is higher than the $P_\mathrm{s}$ of the cloud-free, dry planet with
a black surface. 
Thus, for $\alpha > \alpha_\text{B}$, the gaseous absorption bands show
up as \textit{dips} in the $P_\mathrm{s}$-spectra of cloud-free 
ocean planets.

%-----------------------------------------------------------------------------
\subsubsection{$P_\mathrm{s}$ of cloudy planets}

Figures~\ref{fig:CloudydryP} and b are similar to 
Figs.~\ref{fig:CloudydryF} and b, and 
\ref{fig:cloudylambQ} and b, except for $P_\mathrm{s}$.
As before, for completely cloudy planets ($f_\mathrm{c}=1.0$),
the $P_\mathrm{s}$-spectra of the dry and ocean planets are virtually
the same because the clouds cover the ocean.
The polarization for $\lambda \lesssim 700$~nm is due to Rayleigh
scattering (see the discussion of Figs.~\ref{fig:cloudylambQ} and b).
And, similar to the sign change of $\overline{Q}$, $P_\mathrm{s}$ 
changes sign near $\lambda \sim 700$~nm because of the influence of 
light that is scattered by the clouds (see Sect.~\ref{sec:polarizedflux}). 

%%%%%%%%%%%%%%%%%%%%%%%%%%%%%%%%%%%%%%%%%%%%%%%%%%%%%%%%%%%%%%%%%%%%%%%%%%
% FIGURE 15
%%%%%%%%%%%%%%%%%%%%%%%%%%%%%%%%%%%%%%%%%%%%%%%%%%%%%%%%%%%%%%%%%%%%%%%%%%
\begin{figure*}[t!]
\centering
\begin{subfigure}{.495\textwidth}
\centering
\caption{Varying H$_2$O VMR ($f_\mathrm{c} = 0.00$)}
\label{fig:h2o_05h2o}
\includegraphics[width=0.99\linewidth]{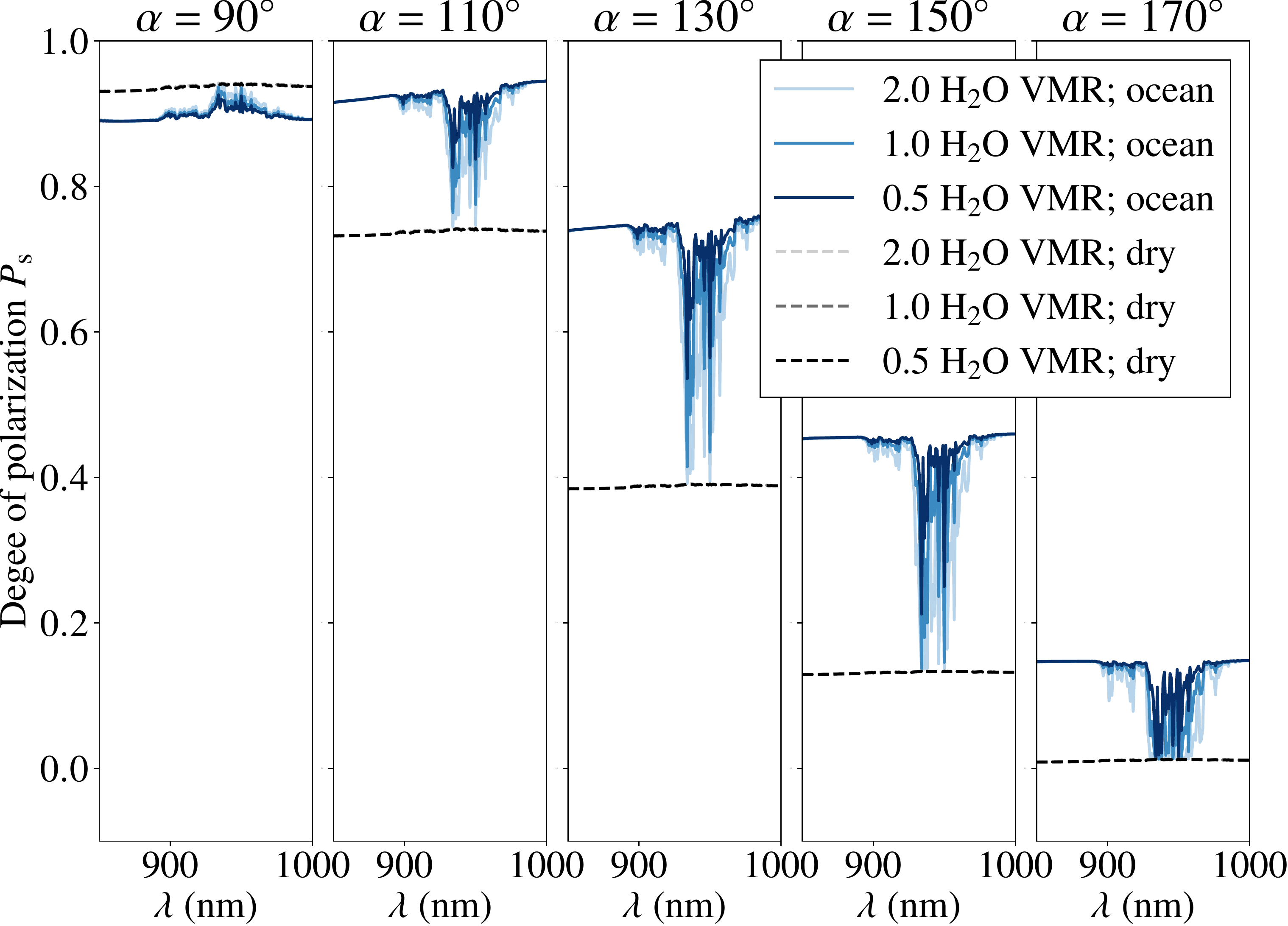}
\end{subfigure}
%%%%%%%%%%%%%%%%%%%%%%%%%%%%%%%%
\hfill
\vspace{0.5cm}
\begin{subfigure}{.495\textwidth}
\centering
\caption{Varying cloud optical thickness $b^\mathrm{c}$ ($f_\mathrm{c} = 1.00$; $h_\mathrm{c} = 2-4$ km)}
\label{fig:h2o_bc}
\includegraphics[width=0.99\linewidth]{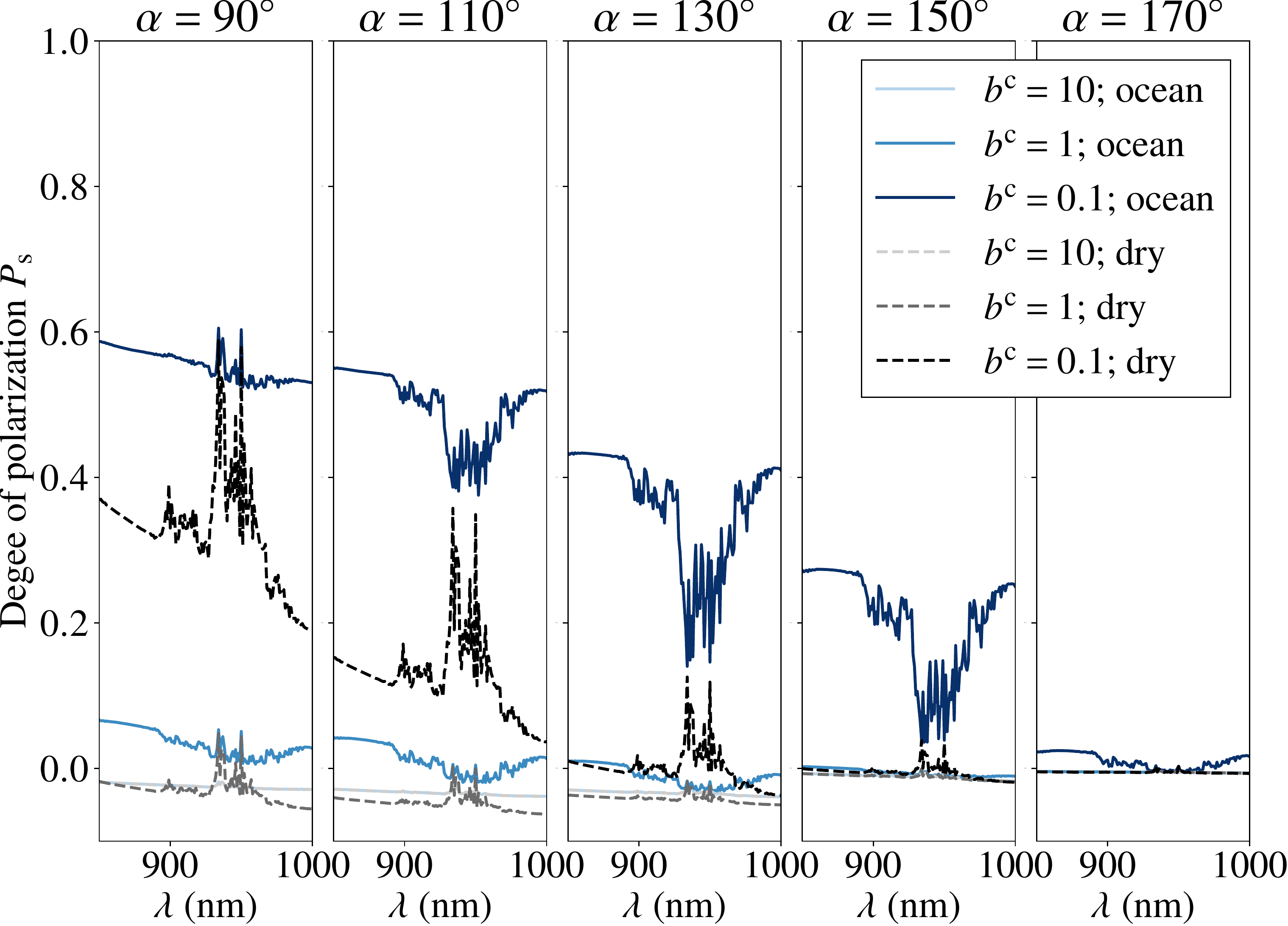}
\end{subfigure}
%%%%%%%%%%%%%%%%%%%%%%%%%%%%%%%%
\begin{subfigure}{.495\textwidth}
\centering
\caption{Varying cloud fraction $f_\mathrm{c}$ (cloud-free glint; $b^\mathrm{c} = 10$; $h_\mathrm{c} = 2-4$ km)}
\label{fig:h2o_fc}
\includegraphics[width=0.99\linewidth]{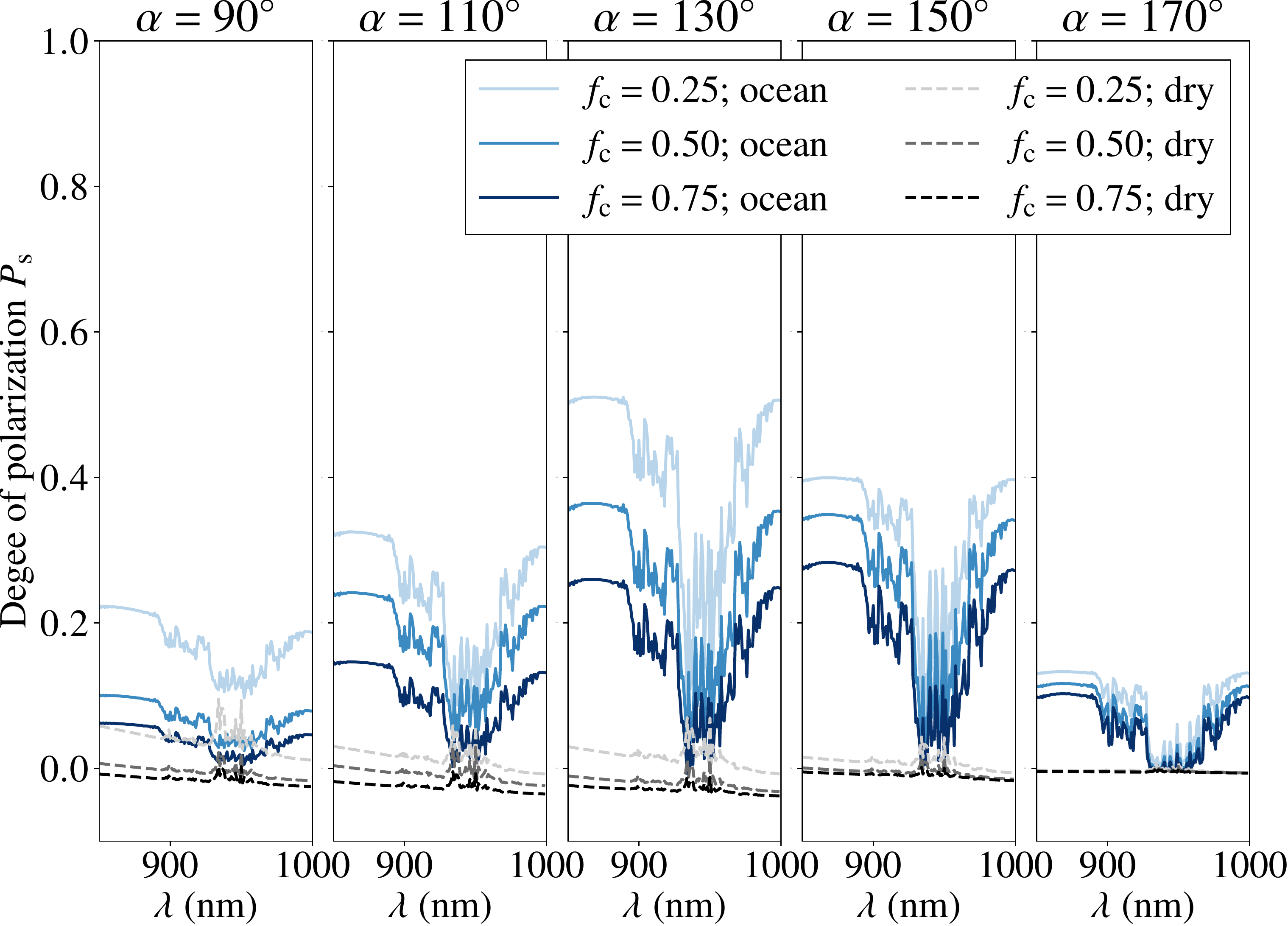}
\end{subfigure}
%%%%%%%%%%%%%%%%%%%%%%%%%%%%%%%%
\hfill
\begin{subfigure}{.495\textwidth}
\centering
\caption{Varying cloud height $h_\mathrm{c}$ ($f_\mathrm{c} = 0.50$; cloud-free glint; $b^\mathrm{c} = 10$)}
\label{fig:h2o_hc}
\includegraphics[width=0.99\linewidth]{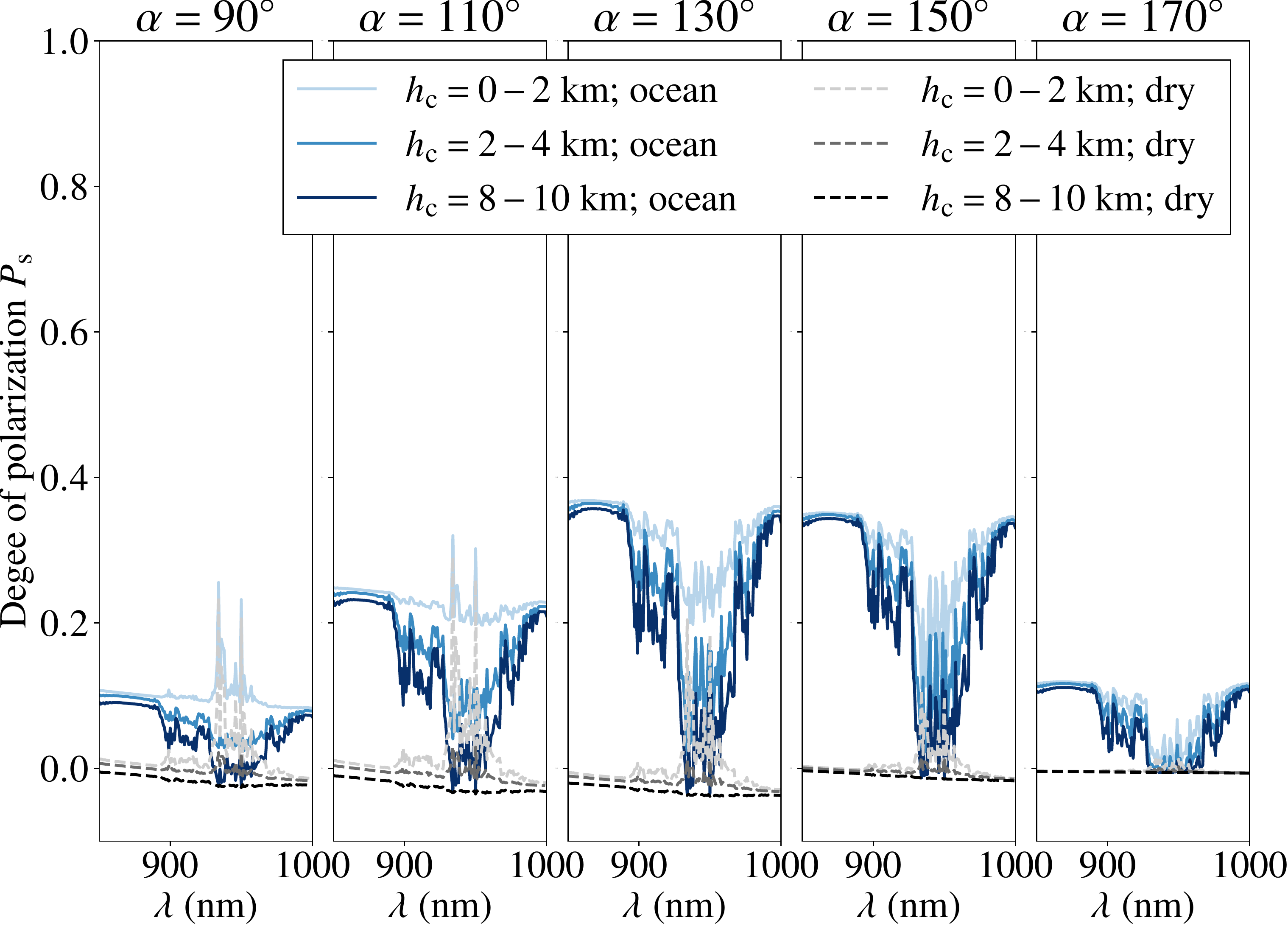}
\end{subfigure}
%%%%%%%%%%%%%%%%%%%%%%%%%%%%%%%%
\caption{The computed degree of polarization $P_\mathrm{s}$ of 
         starlight reflected by planets with ocean surfaces 
         (solid blue lines) or dry surfaces (dashed black lines), 
         and different atmospheres (different line color darkness),
         with $h_{\rm c}$ indicating the vertical extent of the 
         clouds, presented at five planetary phase angles $\alpha$ from 90$^\circ$ to 170$^\circ$. The wind speed $v$ over the oceans is 7~m/s. The dry planets
         have black surfaces ($a_\mathrm{s}$=0.0). In 
         Fig.~\ref{fig:h2o_05h2o}, the lines for the dry planets 
         with various H$_2$O VMRs virtually overlap.}
\label{fig:h2o_analysis}
\end{figure*}
%%%%%%%%%%%%%%%%%%%%%%%%%%%%%%%%%%%%%%%%%%%%%%%%%%%%%%%%%%%%%%%%%%%%%%%%%%

Because of multiple scattering of light between the cloud droplets,
the optically thick clouds reflect strongly depolarized light,
just like the surfaces with non-zero albedos 
(cf.\ $a_s=0.1$ and $a_s=0.8$ in Fig.~\ref{fig:cloudfreelambP}).  
The main difference between the effect of a bright surface and a 
thick cloud deck is due to the altitude at which the light is reflected: 
the cloud deck extends up to 4~km and thus covers a large fraction of 
the H$_2$O molecules as those are concentrated in the lowest atmospheric 
layers. 
Therefore, the H$_2$O absorption band features in $P_\mathrm{s}$ of 
cloudy planets are generally less strong than in $P_\mathrm{s}$ 
of cloud-free planets with non-zero $a_s$. 
Only in the strongest absorption bands (e.g.\ $\sim$~1400~nm 
and $\sim$~1900~nm), $P_\mathrm{s}$ reaches the single scattering 
values, resulting in spectral peaks.

The $P_\mathrm{s}$-spectra of half-cloudy ($f_\mathrm{c}=0.5$) 
dry and ocean planets with the glint on the latter covered 
by clouds, are virtually the same (middle row of Figs.~\ref{fig:CloudydryP} 
and b): no ocean signature can be identified. 

When the glint is cloud-free, however, the $P_\mathrm{s}$-spectra of 
the dry and ocean planets differ significantly as can be seen in the
bottom row of Figs.~\ref{fig:CloudydryP} and~b.
Outside the absorption bands, $P_\mathrm{s}$ is enhanced due to the 
Fresnel reflecting ocean. 
The enhancement increases with $\alpha$ up to $140^\circ$ 
just like the polarized flux $|\overline{Q}|$. 
The value of $P_\mathrm{s}$ outside the bands does not reach the 
value for the cloud-free ocean planet (Fig.~\ref{fig:cloudfreeoceanP})
because the cloud patches add unpolarized flux to the 
planetary signal. 
However, the values of $P_\mathrm{s}$ outside the bands are still 
larger than those for the cloud-free, dry planet with a black surface,
suggesting that the value of $P_\mathrm{s}$ outside absorption bands
could reveal the presence of an ocean.

Interestingly, when patchy clouds surround (but not cover) the glint, 
the dips of $P_\mathrm{s}$ at absorbing wavelengths
can still be identified (see Fig.~\ref{fig:cloudyoceanP}). 
In fact, it could be argued that patchy clouds enable the detection of 
this ocean signature because with clouds, the dips in $P_\mathrm{s}$ appear
already at smaller $\alpha$'s, such as 90$^\circ$ and 100$^\circ$, 
and the dips are wider and already apparent at shorter $\lambda$,
compared to the case for the cloud-free ocean planets 
(Fig.~\ref{fig:cloudfreeoceanP}).
In the H$_2$O absorption bands, the polarizing glint is 
invisible to the observer, but the clouds are not. In the bands, 
$P_\mathrm{s}$ attains similar values as for the dry planet with 
a black surface and similar patchy clouds (Fig.~\ref{fig:cloudfreelambP}).
In the deepest bands, the clouds are also invisible and for each $\alpha$,
$P_\mathrm{s}$ returns to the value of single Rayleigh scattering.

Comparing Figs.~\ref{fig:cloudfreelambP} and b, and
\ref{fig:CloudydryP} and b, we find that for all
dry planets, with black or bright surfaces, 
and with or without clouds, $P_\mathrm{s}$ in gaseous absorption
bands is either close to its continuum value, or higher.
For dry planets, $P_\mathrm{s}$ is thus either flat or 
shows peaks in absorption bands. 
Dips in $P_\mathrm{s}$ are only observed for ocean planets 
and only when the glint is cloud-free. 

%-----------------------------------------------------------------------------
\section{Discussion of absorption peaks and dips in $P_\mathrm{s}$}
\label{sec:h2o}

Here, we further analyze and discuss the influence of atmospheric 
parameters on the peaks and dips in $P_\mathrm{s}$ at wavelengths inside
gaseous absorption bands. We will focus on absorption by H$_2$O vapor 
between 850~and 1000~nm. The wind speed $v$ over the oceans is 7~m/s 
and the dry planets have black surfaces ($a_\mathrm{s}$=0.0).

Figure~\ref{fig:h2o_05h2o} shows the computed $P_\mathrm{s}$ of starlight 
that is reflected by cloud-free ($f_\mathrm{c}=0.0$) ocean planets and 
dry planets, similar as in Fig.~\ref{fig:cloudfreeP}, but here we have
added lines for halved and doubled H$_2$O vapor volume mixing ratios (VMRs). 
The results are presented at five phase angles $\alpha$: $90^\circ$, 
$110^\circ$, $130^\circ$, $150^\circ$, and $170^\circ$. 
For the cloud-free ocean planet, the dip in $P_\mathrm{s}$ decreases 
(increases) with a decreased (increased) H$_2$O VMR.
For the cloud-free dry planet, the $P_\mathrm{s}$-spectrum across the 
absorption band is virtually flat for all values of the H$_2$O VMR,
because for this planet and at these long wavelengths, $P_\mathrm{s}$
is determined by single Rayleigh scattering both outside and inside the 
band.

Figure~\ref{fig:h2o_bc} is similar to Fig.~\ref{fig:h2o_05h2o}, except
with the standard H$_2$O VMR and here the planets are completely covered
by clouds ($f_\mathrm{c} = 1.0$) with an optical thickness 
$b^\mathrm{c}=10$ (at 550~nm), similar as in Fig.~\ref{fig:CloudyP}, 
but also for $b^\mathrm{c}=0.1$, and 1.0.
Generally, the smaller $b^\mathrm{c}$, the higher the continuum 
$P_\mathrm{s}$, because there is less unpolarized flux from the clouds. 
For the dry planets, the clouds act as a depolarizing reflecting 
surface (cf.\ Fig.~\ref{fig:cloudfreelambP}), resulting in peaks in 
$P_\mathrm{s}$ across the absorption bands, where the clouds are 
less visible. 
For the ocean planets, the continuum $P_\mathrm{s}$ increases with 
decreasing $b^\mathrm{c}$, as the polarized flux from the glint that
penetrates the clouds increases. 
And while an optically thick cloud covering the glint leaves $P_\mathrm{s}$ 
featureless across the absorption band, an optically thinner cloud 
transmits the glint and creates a dip in $P_\mathrm{s}$.
In fact, the thin cloud ($b^\mathrm{c}$=0.1) covering the glint 
makes the dip in $P_\mathrm{s}$ even more apparent.

Figure~\ref{fig:h2o_fc} is similar to Fig.~\ref{fig:h2o_bc}, except
here the planets have patchy clouds (all with $b^\mathrm{c}=10$ at 550~nm), 
for different cloud coverage fractions $f_\mathrm{c}$, namely 0.25, 0.50, 
and 0.75. For all coverages, the glint is cloud-free.
It can be seen that increasing $f_\mathrm{c}$ decreases $P_\mathrm{s}$ 
outside the absorption bands, because it increases the unpolarized flux. 
However, even for a heavily cloudy planet ($f_\mathrm{c}=0.75$), an 
ocean will leave a dip in $P_\mathrm{s}$ when the glint is cloud-free. 

Figure~\ref{fig:h2o_hc}, finally, shows $P_\mathrm{s}$ for dry and ocean 
planets with patchy clouds ($f_\mathrm{c} = 0.50$) and a cloud-free glint,
but for clouds at different altitudes $h_\mathrm{c}$. 
While on the standard planets the clouds extend from 2 to 4~km, 
here, we have included curves for lower ($h_\mathrm{c}= 0-2$~km) as well
as higher ($h_\mathrm{c}= 8-10$~km) clouds.
Increasing the cloud altitude hardly affects $P_\mathrm{s}$ in the continuum
because at these long wavelengths, there is almost no Rayleigh 
scattering above the clouds anyway. The dip in $P_\mathrm{s}$, however, 
deepens as the amount of H$_2$O gas above the clouds and thus the absorption
optical thickness decreases, and the contribution of the
low polarization of the clouds increases inside the band. 

Figure~\ref{fig:h2o_analysis} convincingly shows that the depth and shape 
of the dip in $P_\mathrm{s}$ depend on the H$_2$O VMR, and on the optical 
thickness $b^\mathrm{c}$, coverage fraction $f_\mathrm{c}$, and altitude 
$h_\mathrm{c}$ of the clouds. However, in all our model computations, 
a dip in $P_\mathrm{s}$ instead of a peak only occurs in the presence of
an ocean. Such dips in $P_\mathrm{s}$ could thus reveal the 
presence of an ocean.

\begin{figure}[t!]
\centering
\includegraphics[width=0.99\linewidth]{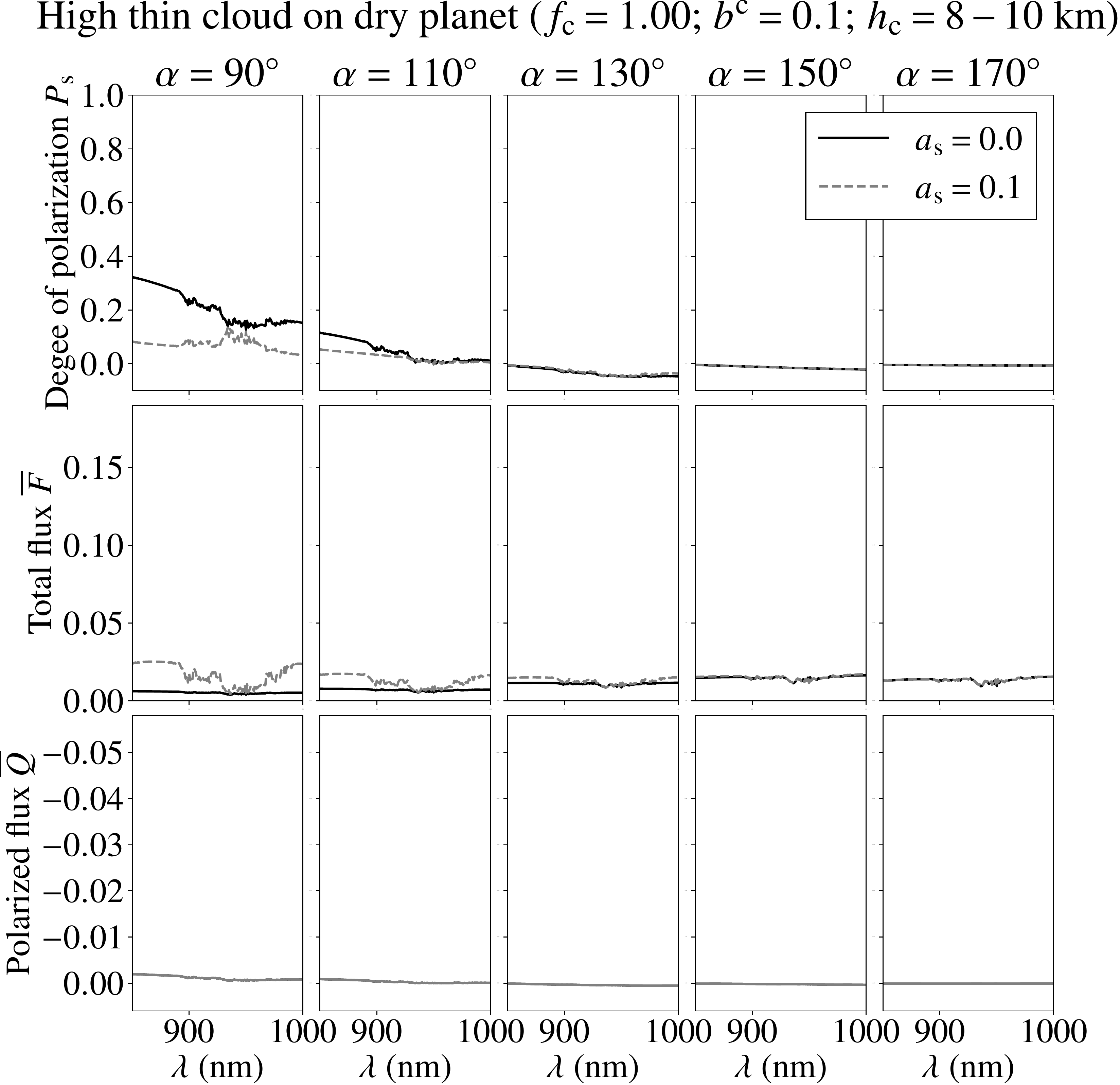}
\caption{The computed degree of polarization $P_\mathrm{s}$, total flux $\overline{F}$ and polarized flux $\overline{Q}$ of 
	     starlight reflected by dry planets with black surfaces,
	     thus with $a_\mathrm{s}=0.0$
	     (black solid lines), and dark surfaces, with $a_\mathrm{s}=0.1$  
	     (grey dashed line), and a high ($h_\mathrm{c}= 8-10$~km) and 
	     optically thin ($b^\mathrm{c}=0.1$) water cloud, presented at five planetary phase angles $\alpha$ from 90$^\circ$ to 170$^\circ$.}
\label{fig:h2oanalysis_bc01hc8}
\end{figure}

We note that dips in $P_\mathrm{s}$ of light that is locally reflected by 
an Earth-like gaseous atmosphere with a high aerosol layer above a black 
surface were found by \citet{Stam1999} (see their Fig.~13 for the effect 
of stratospheric H$_2$SO$_4$ aerosol on the shape of the O$_2$A band 
in $P_\mathrm{s}$). Also, \citet{Emdeetal2017} computed dips in 
$P_\mathrm{s}$ of light reflected by dry black surface planets 
at $\alpha=90^\circ$ with a high, thin ice cloud (see their Fig.~7). 
Indeed, we also find a dip in $P_\mathrm{s}$ between 850 and 1000~nm 
for a dry planet with a black surface at $\alpha=90^\circ$ if 
we place a thin ($b^\mathrm{c}=0.1$) layer of water cloud droplets 
between altitudes of 8 and 10~km, as shown in 
Fig.~\ref{fig:h2oanalysis_bc01hc8}. 
However, contrary to what we find for ocean planets, 
total flux $\overline{F}$ 
and polarized flux $\overline{Q}$ are extremely small in that case and 
the dip in $P_\mathrm{s}$ vanishes with increasing $\alpha$. 
This potentially false ocean signature in $P_\mathrm{s}$ also vanishes 
when a small amount of unpolarized flux is added to the signal, e.g.\ 
due to reflection by a non-black surface, e.g.\ with $a_\mathrm{s}=0.1$.

A detection of dips in the measured $P_\mathrm{s}$-spectrum of the sunlight that is reflected by the Earth when the glint is cloud-free would confirm our model predictions, and could already be done at a single phase angle. Assuming the depolarization by the lunar surface is spectrally smooth across gaseous absorption bands, those dips can be searched for in Earth-shine observations. \citet{Sterzik2012} and \citet{Sterzik2019} reported mainly peaks in Earth-shine spectra of $P_\mathrm{s}$ measured between 420 and 920 nm. 
In some $P_\mathrm{s}$-spectra of \citet{Sterzik2019}, dips can be
seen in the O$_2$A band, although they are not discussed.
Our simulations show that the near-infrared wavelength range could be more suitable to search for dips in $P_\mathrm{s}$, since those dips should be
deeper and spectrally wider than the dips at shorter wavelengths. \citet{MilesPaez2014} took Earth-shine measurements of $P_\mathrm{s}$ in the near-infrared and reported peaks across H$_2$O absorption bands near 930 and 1120 nm, which could suggest that at the time of the observations, 
a continent or cloud covered the glint. More spectropolarimetric Earth-shine measurements 
in the near-infrared over longer time periods, in which the cloud 
coverage of the glint can vary, are needed to confirm our predicted ocean signatures.

%%%%%%%%%%%%%%%%%%%%%%%%%%%%%%%%%%%%%%%%%%%%%%%%%%%%%%%%%%%%%%%%%%%%%%%%%%
% FIGURE 16
%%%%%%%%%%%%%%%%%%%%%%%%%%%%%%%%%%%%%%%%%%%%%%%%%%%%%%%%%%%%%%%%%%%%%%%%%%
%%%%%%%%%%%%%%%%%%%%%%%%%%%%%%%%%%%%%%%%%%%%%%%%%%%%%%%%%%%%%%%%%%%%%%%%%%

%-----------------------------------------------------------------------------
\section{Comparison with earlier work}
\label{sect_comparison}

%%%%%%%%%%%%%%%%%%%%%%%%%%%%%%%%%%%%%%%%%%%%%%%%%%%%%%%%%%%%%%%%%%%%%%%%%%
% FIGURE 17
%%%%%%%%%%%%%%%%%%%%%%%%%%%%%%%%%%%%%%%%%%%%%%%%%%%%%%%%%%%%%%%%%%%%%%%%%%
\begin{figure}[h!]
\centering
\includegraphics[width=0.9\linewidth]{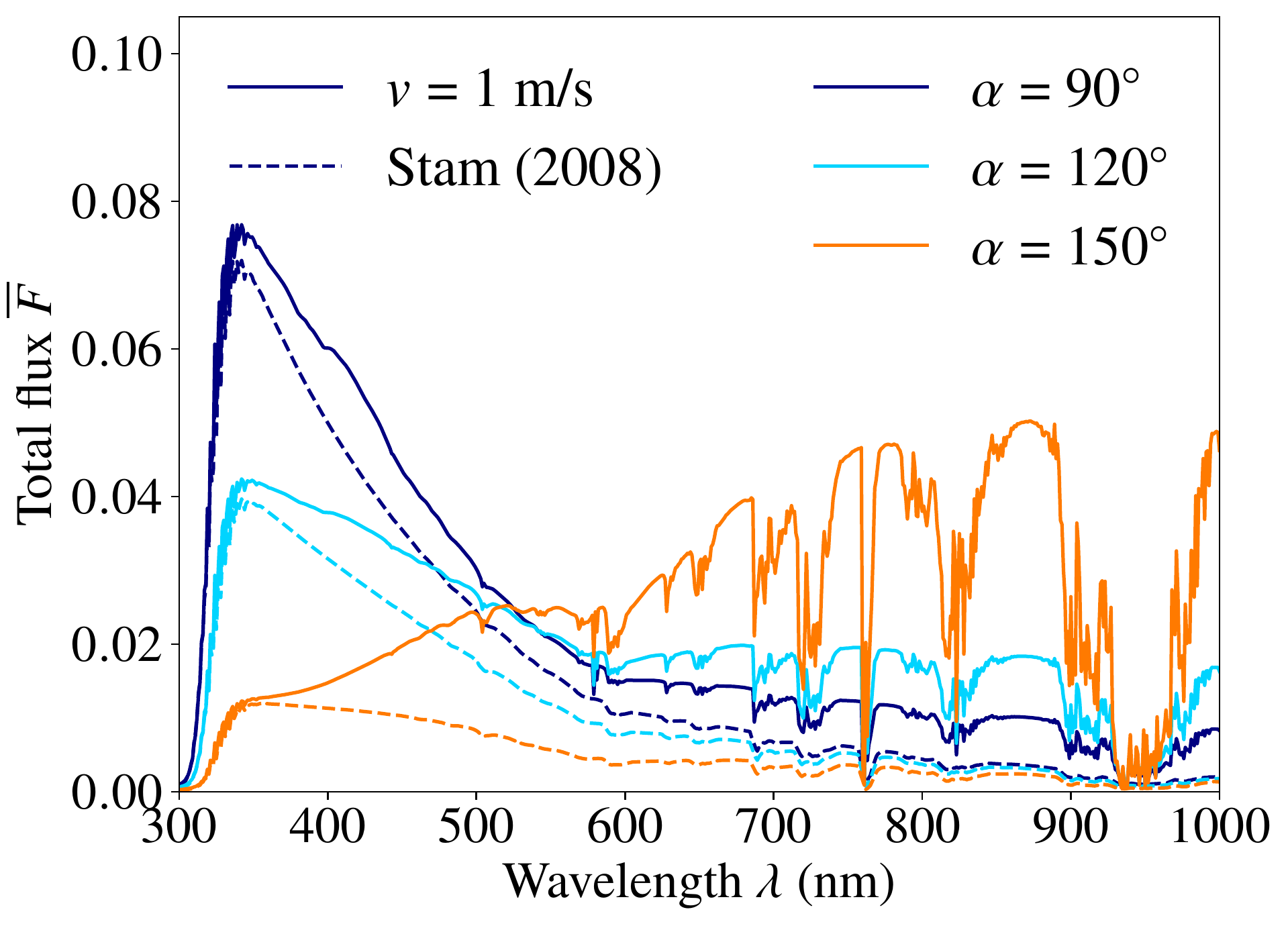} 
\vspace*{-0.1cm}
\hspace*{-0.2cm}
\includegraphics[width=0.92\linewidth]{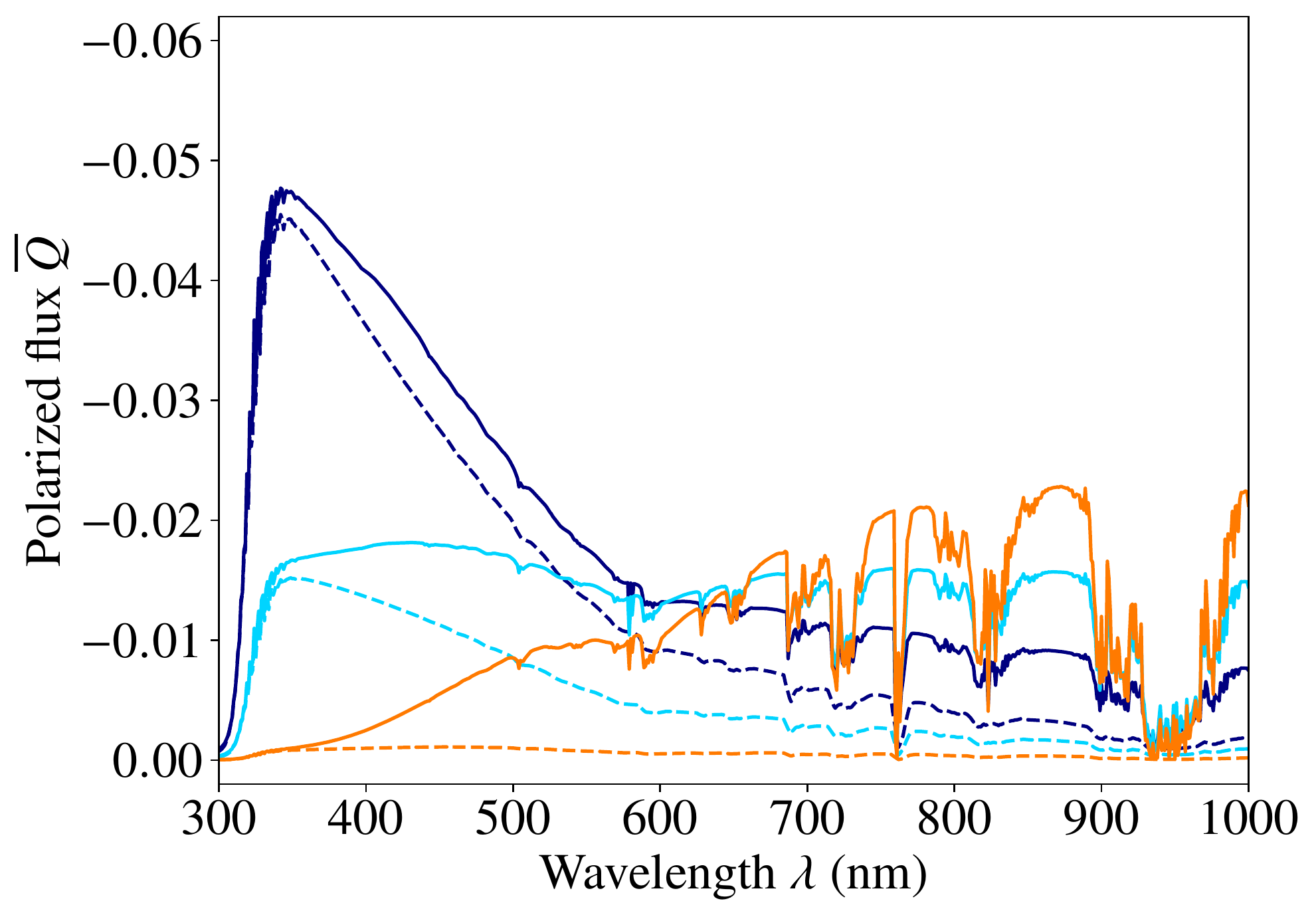} 
%\vspace*{0.2cm}
\hspace*{0.2cm}
\includegraphics[width=0.88\linewidth]{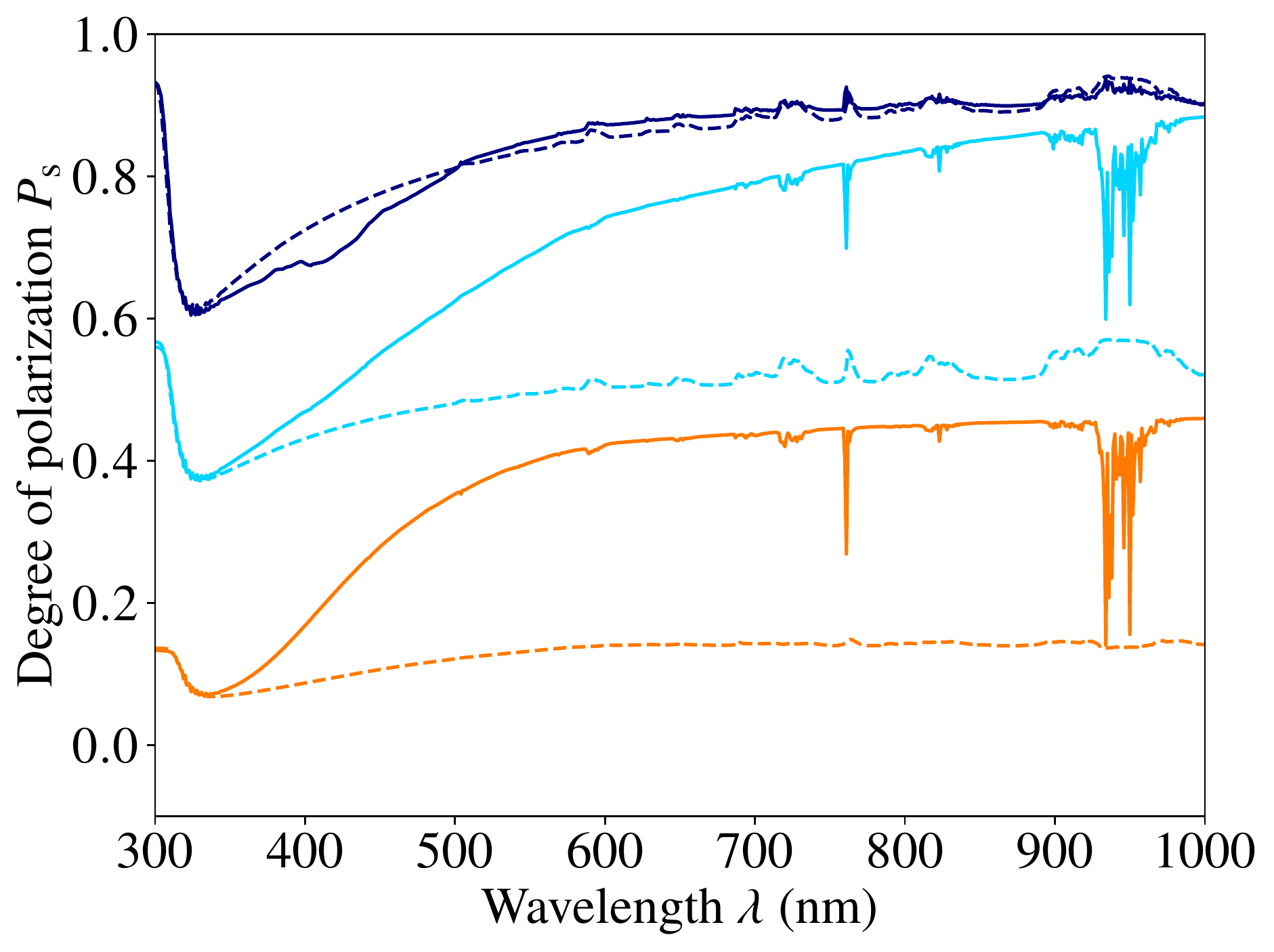} 
\caption{Comparison of our results (solid lines) to the results of 
         \citet{Stam08} (dashed lines) for the computed total flux 
         $\overline{F}$, polarized flux $\overline{Q}$, and degree of 
         polarization $P_\mathrm{s}$ of 
         light that is reflected by cloud-free ocean planets at 
         $\alpha=90^\circ$, $120^\circ$, and $150^\circ$. Our ocean 
         planet has a natural blue color and a glint (for $v=1$~m/s). 
         The ocean model of \citet{Stam08} is black and flat, and while
         its interface is a Fresnel-reflector, the lack of waves 
         results in an infinitely narrow glint.}
\label{fig:Stam08}
\end{figure}
%%%%%%%%%%%%%%%%%%%%%%%%%%%%%%%%%%%%%%%%%%%%%%%%%%%%%%%%%%%%%%%%%%%%%%%%%%

All our model computations show that $P_\mathrm{s}$ dips
instead of peaks in gaseous absorption bands only in the presence of
an ocean. Such dips in $P_\mathrm{s}$ could thus reveal an ocean.
These spectropolarimetric ocean signatures are absent in the computed
spectra of planets with oceans by \citet{Stam08},
because in that paper, the oceans had flat surfaces, not rough, 
wavy ones that broaden the glint and cast shadows. 
With a flat interface
and monodirectional incident starlight (thus when the distance between
the star and the planet is much larger than the radius of the planet),
the glint is infinitely narrow as it is the reflected direct starlight
and hence a mirror image of the star. This infinitely
narrow glint is lost upon integration over the planetary disk. 
Note that the diffuse skylight that is reflected
by a flat interface will still contribute to the planet's signal.
Another difference between the ocean model in this paper and the one
used in \citet{Stam08} is that our ocean has a natural blue color
and sea foam (see Sect.~\ref{sect_surface_model}),
while the ocean in \citet{Stam08} is black.

Figure~\ref{fig:Stam08} shows our computed $\overline{F}$,
$\overline{Q}$, and $P_\mathrm{s}$ for cloud-free ocean planets, 
similar as in Figs.~\ref{fig:cloudfreeoceanF}, \ref{fig:cloudfreeoceanQ} and~\ref{fig:cloudfreeoceanP},
except for $v=1$~m/s, compared to the results of
\citet{Stam08}, for $\lambda$ ranging from 300 to 1000~nm and at 
$\alpha= 90^\circ$, $120^\circ$, and $150^\circ$. At 90$^\circ$, the bump in $\overline{F}$ and the dip in $P_\mathrm{s}$ near 
$\lambda=400$~nm are absent in the spectra of \citet{Stam08} because
they are caused by the ocean's blue color. 
For $\alpha=120^\circ$ and 150$^\circ$, the bump and dip are
increasingly weaker in our spectra because of the longer average 
optical paths through the optically thick atmosphere at those short 
wavelengths, which decreases the amount of light that reaches the ocean.

At the longer wavelengths, where the atmospheric scattering 
optical thickness is small, and outside absorption bands, 
the ocean glint strongly increases our $\overline{F}$, $|\overline{Q}|$,
and $P_\mathrm{s}$ with increasing $\alpha$ from $90^\circ$ 
as compared to the results from \citet{Stam08}. This increase 
is caused by the increase of the local 
reflection angles on the ocean waves, and with that by the 
increase of Fresnel reflection and the glint brightness. 

\begin{figure}[h!]
\centering
\includegraphics[width=1\linewidth]{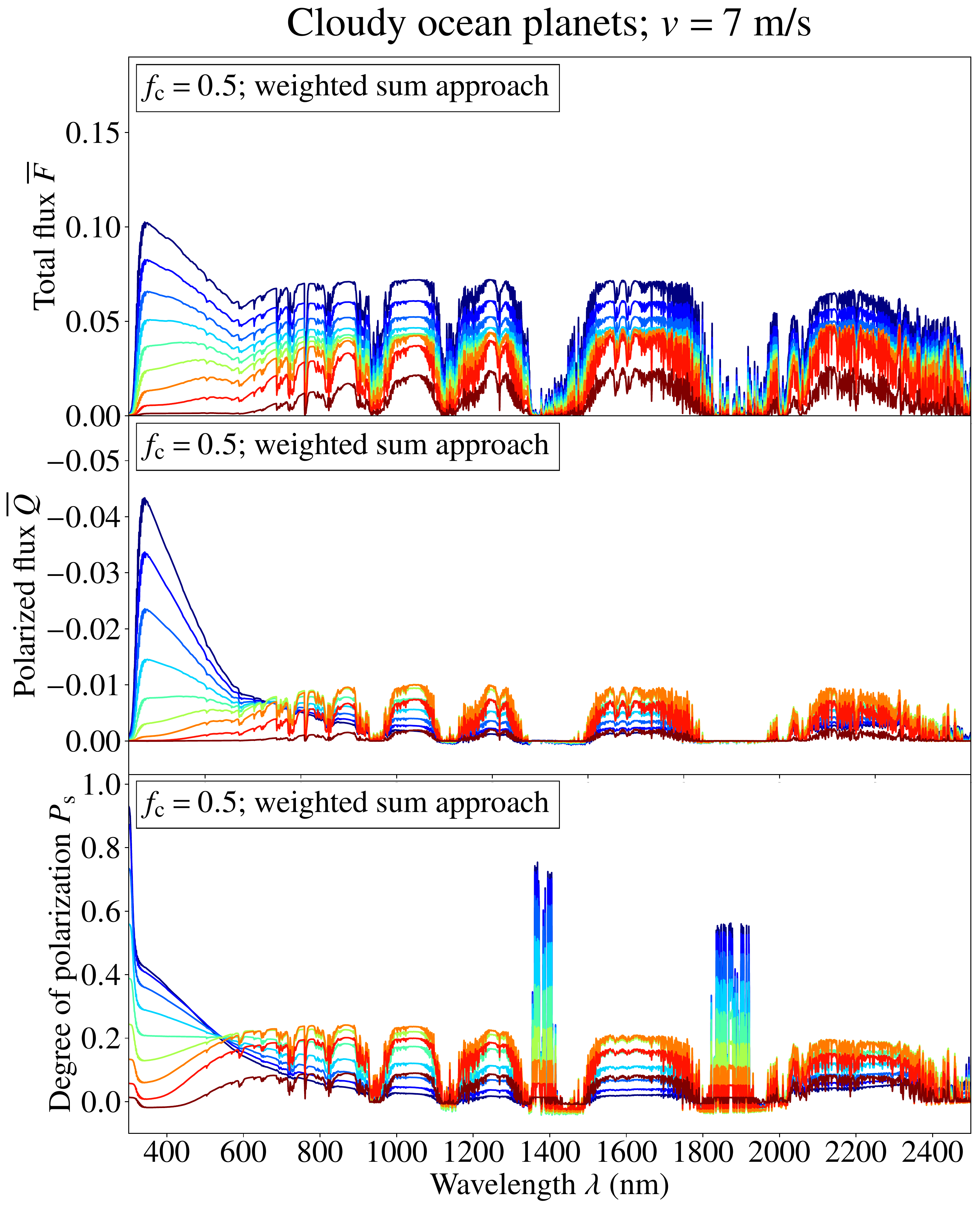}
\caption{Similar to Figs.~\ref{fig:cloudyoceanF}, \ref{fig:cloudyoceanQ},
         and \ref{fig:cloudyoceanP} for $f_\mathrm{c} = 0.5$, but then using a weighted sum of  $\overline{\textbf{F}}$ of a cloud-free ocean planet ($f_\mathrm{c}=0.0$) and  $\overline{\textbf{F}}$ of a completely cloudy ocean planet ($f_\mathrm{c}=1.0$), with weighting factors equal to 0.5.}
\label{fig:quasi}
\end{figure}

For accurate modelling of the signatures of oceans on exoplanets, 
the glint on the waves appears to be very important. 
The influence of the waves
depends significantly on the pattern of the clouds: if the glint is 
cloud-free, it leaves clear signatures in the spectra of 
$\overline{F}$, $\overline{Q}$, and $P_\mathrm{s}$.
If the glint is covered by clouds, the signature disappears,
unless the cloud patch is optically thin.
It is thus important to use horizontally inhomogeneous cloud patterns 
when modelling spectra of exoplanets with patchy clouds and oceans. 
Indeed, for this modelling, a weighted sum of a cloud-free and 
a completely cloudy planet to model partly cloudy planet signals
\citep[as used by e.g.][]{Stam08,Zugger2010,Zugger2011IR,Emdeetal2017} 
would fail to fully represent an actual planet signal. 
This is illustrated in Fig.~\ref{fig:quasi}, where we use a weighted 
sum approach to mimic an ocean planet with a cloud fraction $f_\mathrm{c}$ 
of 0.5. When comparing these curves with those in 
Figs.~\ref{fig:cloudyoceanF}, \ref{fig:cloudyoceanQ}, and 
\ref{fig:cloudyoceanP}, which pertain to an actual horizontally
inhomogeneous ocean planet with patchy clouds and the same 
$f_\mathrm{c}=0.5$ with a cloud-free glint, it is clear that the 
weighted sum planet has a weaker glint signature.  
Indeed, the weighted sum planet has a (polarized) flux due to the glint 
that is half of that of the horizontally inhomogeneous ocean planet 
with patchy clouds.

%%%%%%%%%%%%%%%%%%%%%%%%%%%%%%%%%%%%%%%%%%%%%%%%%%%%%%%%%%%%%%%%%%%%%%%%%%
% FIGURE 18
%%%%%%%%%%%%%%%%%%%%%%%%%%%%%%%%%%%%%%%%%%%%%%%%%%%%%%%%%%%%%%%%%%%%%%%%%%
%%%%%%%%%%%%%%%%%%%%%%%%%%%%%%%%%%%%%%%%%%%%%%%%%%%%%%%%%%%%%%%%%%%%%%%%%%

%-----------------------------------------------------------------------------
\section{Summary and conclusion}
\label{sect_conclusion}

We have presented the computed spectra of the total flux $\overline{F}$, 
the (linearly) polarized flux $\overline{Q}$, and the degree of (linear)
polarization $P_\mathrm{s}$ of starlight that is reflected by cloud-free 
and cloudy dry planets and ocean planets, from the ultraviolet (300~nm) 
to the infrared (2500~nm), at a 1~nm wavelength resolution and at 
planetary phase angles $\alpha$ ranging from $90^\circ$ to $170^\circ$. 
The atmospheres of our model planets are Earth-like and cloud-free, 
with patchy clouds, or fully cloudy. The clouds consist
of spherical, liquid water droplets. The ocean is composed 
of a Fresnel reflecting, wind-ruffled surface, with sea foam 
and shadows of waves, and below that a seawater body bounded below by 
a black surface.

Our results show that at wavelengths outside absorption bands, 
the ocean glint strongly increases $\overline{F}$ and $|\overline{Q}|$ 
with increasing $\alpha$ from $90^\circ$. The increase is caused by the
increase of the local reflection angles on the ocean waves, and with 
that by the increase of Fresnel reflected light, yielding a brighter 
glint. The increase is more apparent towards longer wavelengths, where 
the scattering optical thickness of the gaseous atmosphere is small. 
This wavelength dependence of the glint's visibility results in an 
intersection of the $\overline{F}$ and $|\overline{Q}|$-spectra 
for different phase angles at visible wavelengths. 

Our results show that the advantage of measuring $\overline{Q}$ 
instead of $\overline{F}$ when trying to detect an exo-ocean, is 
that only in the presence of Fresnel reflection, a significant 
polarized flux $|\overline{Q}|$ is expected at wavelengths larger 
than $\sim$~700~nm. According to \citet{Cowanfalseglint2012}, false positive ocean signatures 
could appear in $\overline{F}$ for planets with reflecting dry surfaces 
or polar caps. Our results indeed show that $\overline{F}$ 
can be nonzero and spectrally flat for dry planets with spectrally flat, 
bright surfaces and/or water clouds. 
Moreover, we find that in the near-infrared ($\lambda \gtrsim 700$~nm), 
a negative polarized flux $\overline{Q}$ (thus a direction of polarization 
that is perpendicular to the plane through the planet, the star, and the 
observer) only appears for an ocean planet with a visible glint. 
For all (cloudy and cloud-free) dry planets and for ocean planets that 
are completely covered by optically thick clouds, $\overline{Q}$ is near
zero or slightly positive. A significant amount of negative $\overline{Q}$ 
in the near-infrared, at phase angles up from $90^\circ$ appears to 
reveal the presence of an ocean.

In the spectra of the degree of polarization $P_\mathrm{s}$, we 
find high values of $P_\mathrm{s}$ outside absorption bands for ocean 
planets where the glint is cloud-free. This ocean signature is 
increasingly apparent with increasing wavelength, thus with decreasing 
atmospheric gaseous optical thickness. 
The high $P_\mathrm{s}$-values that result from the glint are caused 
by the shift of the maximum $P_\mathrm{s}$ from the single scattering 
Rayleigh peak near $\alpha=90^\circ$ to the Brewster angle at 
larger values of $\alpha$, as explained by \citet{Zugger2010} and 
\citet{TreesStam2019}.

The glint can cause the continuum-$P_\mathrm{s}$ to be higher 
than the usually relatively high $P_\mathrm{s}$-values from single 
Rayleigh scattering by gaseous molecules. 
Consequently, in gaseous absorption bands, where the glint signal
from the ocean surface is absorbed by the gas, and where $P_\mathrm{s}$
attains its value for single Rayleigh scattering, the 
$P_\mathrm{s}$-spectrum will show dips in the bands. 
According to our computations, such dips only occur for ocean planets
that are either cloud-free or that have patchy clouds that do not 
cover the glint.
For dry planets, $P_\mathrm{s}$ shows either no structure or peaks 
in absorption bands.

We specifically investigated the dips of $P_\mathrm{s}$ in the 
H$_2$O absorption bands between $\lambda =$ 850 and 1000~nm, and 
found that the depth and shape of the dips depend on the H$_2$O VMR, 
the cloud optical thickness (in case of a cloud-covered glint), 
the cloud altitude 
and the cloud coverage fraction (in case of a cloud-free glint). 
For all model planets, only in the presence of an ocean 
dips were observed.

Furthermore, our results show that the blue color of the water body of 
an ocean could only be considered for ocean detection at wavelengths 
shorter than $\sim$~550 nm, where the ocean reflection slightly increases
$\overline{F}$ and even more slightly $|\overline{Q}|$, which results in 
a shallow, broad dip in $P_\mathrm{s}$. 
At longer wavelengths, the light that enters the ocean is increasingly 
absorbed by the water body. When clouds contribute to the reflected light 
signal, the color signature of the ocean's water virtually disappears. 

The foam on the surface of our model oceans reflects incident light at all 
wavelengths between 300 and 2500~nm, although its albedo decreases 
with increasing wavelength. We only found a noticeable 
decrease of $P_\mathrm{s}$ due to the foam at the wind speed $v$ of 13~m/s
(which is considered to be a strong breeze on Earth),
where the foam covers 2.46\% of the ocean surface, and then 
at $\alpha=90^\circ$ and $100^\circ$, as at those phase angles, 
the optical path through the atmosphere is shorter than for 
the larger values of $\alpha$ that we consider. Of course, although not 
explicitly studied here, for even higher wind speeds and smaller phase 
angles, the foam signature is expected to be stronger, and with clouds 
covering the ocean, the signature is expected to be weaker.

The signature of the glint in the total and polarized fluxes and in the 
degree of polarization is not directly limited by the cloud coverage 
fraction $f_\mathrm{c}$, at it depends significantly on the horizontal 
cloud distribution: 
a cloud-free glint leaves clear signatures in the spectra, but if it is 
covered by a cloud patch, the signature disappears. 
In case of a random patchy cloud cover, the probability of a cloud-free 
glint is higher when the wind speed is lower such that the glint is 
narrower \citep[see][who generated 300 random patchy cloud patterns for each phase angle]{TreesStam2019}. The probability of a cloud-free glint of 
course decreases with increasing $f_\mathrm{c}$, but a (partly)
cloud-free glint at least part of the time is possible as long as 
there are gaps in the cloud deck. 
For future observations of Earth-like planets, multiple observations 
over time would increase the probability of catching a cloud-free glint,
improving the odds of a successful ocean detection.

\citet{TreesStam2019} already showed that a reddening of the degree of 
polarization and the polarized flux could reveal the presence of an ocean if consecutive measurements are taken at a range of phase angles. Between
observations at different phase angles, the optical properties of a planet can change, for example when weather systems evolve. The polarimetric glint signatures in the spectra of this paper could reveal the presence of an ocean with a measurement at a single phase angle. Dips in $P_\mathrm{s}$, and a negative $\overline{Q}$ in the near-infrared, 
can already be searched for at a planetary phase angle of $\alpha = 90^\circ$,
which would in principle be possible for any orbital inclination angle
\citep[see][]{Stam08}, and where the angular separation between the star and the planet is largest (which would be optimal for measuring the resolved 
planetary signal). This makes spectropolarimetry a strong tool for the 
detection of oceans on exoplanets, as the glint in the total flux, as
measured when polarization is ignored, is only 
distinctive in the most crescent phases.

%The spectropolarimetric ocean signatures reported in this paper are not 
%present in the computed spectra of planets with oceans by \citet{Stam08},
%because those oceans had flat Fresnel reflecting interfaces, not rough, 
%wavy interfaces that broaden the glint. The ocean 
%
%{\color{red}{the papers die hieronder worden genoemd gebruiken helemaal geen
%oceaanmodel ... zo lijkt het alsof ze dat wel gebruiken en dan aannemen
%dat de glint er niet is.}}
%The polarization signal from the ocean has therefore sometimes been 
%assumed to be small and even negligible in earlier Earth-like exoplanet 
%modelling \citep[e.g.][]{KaralidiStamHovenier2012}.  
%Note that the spectropolarimetric ocean signatures found in this paper also may influence the analyses of other Earth-like planet signatures,, horizontal inhomogeneities in general \citep{KaralidiandStam2012} and of the particles of water and ice clouds \citep{KaralidiStamHovenier2012}. 

%{\color{red}{hier nog iets zeggen over aardschijn en tijdsvariaties
%van patchy clouds: soms wel en soms geen glint te zien?}} 

%------------------------------------------------------------------------
\bibliographystyle{aa} % style aa.bst
\bibliography{paper.bib}

\end{document}